\documentclass[11pt]{article}

\usepackage{amsmath,amssymb}

\newcommand{\bea}{\begin{eqnarray}}
\newcommand{\eea}{\end{eqnarray}}
\newcommand{\de}{\mathrm{d}}

\newcommand{\ft}[2]{{\textstyle\frac{#1}{#2}}}

\renewcommand{\Im}{\operatorname{Im}}
\renewcommand{\Re}{\operatorname{Re}}

\newsavebox{\uuunit}
\sbox{\uuunit}
    {\setlength{\unitlength}{0.825em}
     \begin{picture}(0.6,0.7)
        \thinlines
        \put(0,0){\line(1,0){0.5}}
        \put(0.15,0){\line(0,1){0.7}}
        \put(0.35,0){\line(0,1){0.8}}
       \multiput(0.3,0.8)(-0.04,-0.02){10}{\rule{0.5pt}{0.5pt}}
     \end {picture}}

\allowdisplaybreaks

\numberwithin{equation}{section}

\begin{document}

\thispagestyle{empty}
{}

\hfill ITP-UU-12/20\\[-3ex]

\hfill Nikhef-2012-010\\

\vskip -3mm
\begin{center}
{\bf\LARGE
\vskip - 1cm
Non-holomorphic deformations of special geometry \\ [3mm]
and their applications}

\vspace{10mm}

{\large
{\bf Gabriel Lopes Cardoso$^{+}$\footnote{Lectures given by G.L.Cardoso at the 2011 Frascati School on Black Objects 
in Supergravity}},
{\bf Bernard de Wit$^{*,**}$
and Swapna Mahapatra$^\dagger$}
\vspace{1cm}

{\it 
$^+$
CAMGSD, Departamento de Matem\'atica\\ [1mm]
Instituto Superior T\'ecnico, Universidade T\'ecnica de Lisboa,
1049-001 Lisboa, Portugal \\ [2mm] 
$^*$
Nikhef Theory Group, Science Park 105, 1098 XG Amsterdam, The
  Netherlands \\[2mm]
  $^{**}$
Institute for Theoretical Physics, Utrecht University,
3584 CE Utrecht,\\
The Netherlands
\\[2mm]
  $^\dagger$ 
Physics Department, Utkal University, 
Bhubaneswar 751 004, India
}}

\vspace{5mm}

\texttt{gcardoso@math.ist.utl.pt, b.dewit@uu.nl, swapna@iopb.res.in}

\end{center}
\vspace{5mm}

\begin{center}
{\bf ABSTRACT}\\
\end{center}

The aim of these lecture notes 
is to give a pedagogical introduction to the subject of non-holomorphic
deformations of special geometry.  This subject was first introduced in the context of $N=2$ BPS black 
holes, but has a wider range of applicability.  
 A theorem is presented according to which an arbitrary point-particle Lagrangian can be formulated in terms of a complex function $F$,  whose features are analogous to those of 
the holomorphic function of special geometry. A crucial role is played by a symplectic vector that represents a complexification of the canonical variables, i.e. the  coordinates and canonical momenta. 
We illustrate the characteristic features of the theorem in the context of field theory
models with duality invariances.

The function $F$ may
depend on a number of external parameters that are not subject to duality transformations. 
We introduce duality covariant complex variables whose transformation rules under duality 
are independent of these parameters.
We express the real Hesse potential of 
$N=2$ 
supergravity in terms of the new variables and expand it in  powers of the external parameters.
Then  we relate this expansion to the 
one encountered in topological
string theory.

These lecture notes include exercises which are meant as a guidance to the reader.

\clearpage
\setcounter{page}{1}

\tableofcontents

\section{Introduction}

As is well known, an abelian $N=2$ supersymmetric vector
multiplet in four dimensions is described by a reduced chiral multiplet,
whose gauge covariant degrees of freedom include an (anti-selfdual) field strength $F_{\mu\nu}^-$ and a complex
scalar field $X$.  The Wilsonian effective Lagrangian for these vector
multiplets is encoded in a holomorphic function $F(X)$ which, when coupled to supergravity, is required to be homogeneous of degree two \cite{deWit:1984pk}. 
The abelian vector multiplets may be further coupled to (scalar) chiral multiplets that 
describe either additional dynamical fields or background fields. The function $F$ will then also
depend on holomorphic fields that reside in these chiral multiplets.  An example thereof is provided by the coupling
of vector multiplets to a conformal supergravity background.  
The multiplet that describes 
conformal supergravity
is the Weyl multiplet, and the chiral background is given by the square of it
\cite{Bergshoeff:1980is}.  In this case the function $F$, which now depends on the lowest component field of the chiral
background superfield, encodes the couplings of the vector multiplets to the square of
the Riemann tensor.  These couplings constitute a special class of higher-derivative couplings, 
namely, they depend on
the Riemann tensor but not on derivatives thereof.  In this paper we will only consider higher-derivative
couplings of this type, i.e. couplings that depend on field strengths but not on their derivatives.\footnote{In the language of the theorem
that will be presented in section \ref{sec:ubiquity}, this may be rephrased by saying that the Lagrangians we will consider depend
on coordinates and velocities, but not on accelerations.}
We refer to \cite{deWit:2010za} for a discussion
on other classes of higher-derivative couplings.
When higher-order derivative couplings are absent, we will
denote the function $F$  by $F^{(0)}(X)$, which then refers to a Wilsonian action that
is at most quadratic in space-time derivatives.

The abelian vector fields in these actions are subject to
electric/magnetic duality transformations under which the electric field strengths and
their magnetic duals are subjected to symplectic rotations. It is then
possible to convert to a different duality frame, by regarding half of
the rotated field strengths as the new electric field strengths and
the remaining ones as their magnetic duals. The latter are then
derivable from a new action. To ensure that the characterization of
the new action in terms of a holomorphic function remains preserved,
the scalars of the vector multiplets are transformed accordingly. This
amounts to rotating 
 the complex fields $X^I$ and the holomorphic
derivatives $F_I = \partial F/\partial X^I$ of the underlying function $F$ by the same symplectic
rotation as the field strengths and their dual partners
\cite{deWit:1984pk,Cecotti:1988qn}. 
Here the index $I$ labels
the vector multiplets (in supergravity it takes the values $I=0,1,\ldots,n$).
Thus, electric/magnetic
duality (which acts on the vector $(X^I,F_I)$), 
constitutes an equivalence transformation that relates
two 
Lagrangians (based on two different functions) and gives rise
to equivalent sets of equations of motion and Bianchi identities. A
subgroup of these equivalence transformations may constitute a
symmetry (an invariance) of the system.  For a duality transformation
to constitute a symmetry, the substitution $X^I \rightarrow {\tilde X}^I$
into $F_I$ must correctly induce the transformation $(X^I,F_I) \rightarrow 
(\tilde{X}^I,{\tilde F}_I)$ \cite{deWit:1996ix}.

At the Wilsonian level, 
when coupling the $N=2$ vector multiplets to supergravity, the scalar fields of the vector multiplets
parametrize a non-linear sigma-model whose geometry 
is called special geometry \cite{Strominger:1990pd}, a name that first arose in the study
of the geometry of the effective action of type-II string compactifications
on Calabi-Yau threefolds \cite{Cecotti:1988qn}.   The sigma-model space is a so-called
special-K\"ahler space,  
whose K\"ahler potential is \cite{deWit:1984pk}, 
\begin{equation}
  \label{eq:spec-K-pot}
  K(z,\bar z) = - \ln \left[\frac{\mathrm{i}\big(X^I\,\bar
      F^{(0)}_I - 
      \bar X^I\, F^{(0)}_I \big)}{\vert X^0\vert^2} \right]\,,
\end{equation}
where $F^{(0)}(X)$ is the holomorphic function that determines the 
supergravity action, which is quadratic in space-time
derivatives. Because $F^{(0)}(X)$ is homogeneous of second degree,
this K\"ahler potential depends only on the `special' holomorphic coordinates
$z^i=X^i/X^0$ and their complex conjugates, where $i=1,\ldots,n$, so
that we are dealing with a special-K\"ahler space of complex dimension
$n$. In view of the homogeneity, the symplectic rotations acting on
the vector $(X^I,F^{(0)}_I)$, induce corresponding
(non-linear) transformations on the special coordinates $z^i$. Up to a
K\"ahler transformation, the K\"ahler potential transforms as a
function under duality.

There actually exist various ways of defining special K\"ahler geometry.
Apart from its definition in terms of special holomorphic coordinates \cite{deWit:1984pk}, 
it can also be defined in a coordinate independent way \cite{Freed:1997dp}.
More recently, the formulation of special geometry in terms of 
special real instead of special holomorphic coordinates has been emphasized
\cite{LopesCardoso:2006bg,Ferrara:2006at,Cardoso:2010gc,VandenBleeken:2011ib,Mohaupt:2011aa,Mohaupt:2011ab}.  This formulation is based on
the real Hesse potential \cite{Hitchin:1999,Alek:1999,Cortes:2001}, which will play an important role below.

In order to pass from the Wilsonian effective action to the 
1PI low-energy effective action,  one needs to integrate over the massless modes of the model.
In the context of $N=2$ theories this induces non-holomorphic modifications in the
gauge and gravitational couplings of the theory that, at the Wilsonian level, are encoded in
the holomorphic function $F$.  An early example thereof is provided by the computation of 
the moduli dependence of string loop corrections to gauge coupling constants in heterotic
string compactifications \cite{Dixon:1990pc}.  These non-holomorphic modifications of the coupling functions
are crucial to ensure that the low-energy effective action possesses the expected duality symmetries.
This is therefore a generic feature of the low-energy effective action of $N=2$ models with duality symmetries.

Another context where these moduli dependent corrections play an important role is the one of BPS black hole solutions in $N=2$ models.
Their entropy should exhibit the duality symmetries of the underlying model, and this is achieved by taking
into account the non-holomorphic modifications of the low-energy effective action.  The need for non-holomorphic
modifications of the entropy was established 
in models with exact S-duality \cite{LopesCardoso:1999ur}, 
and their presence has been confirmed at
the semiclassical level from microstate counting
\cite{LopesCardoso:2004xf,Jatkar:2005bh}.
The fact that non-holomorphic modifications can be incorporated into the entropy of BPS black holes 
gave a first indication that the framework of special geometry can be consistently modified by a class of 
non-holomorphic deformations, to be described below.  This can be understood as follows.
The free energy of these BPS black holes
turns out to be given by a generalized version of the aforementioned Hesse potential
\cite{LopesCardoso:2006bg,Cardoso:2008fr,Cardoso:2010gc}.
The Hesse potential is related by a
Legendre transformation to the function $F$ that defines the effective
action, and thus it can be regarded as the associated `Hamiltonian'.
The Hamiltonian
transforms as a function under electric/magnetic duality transformations.
If the $N=2$ model under consideration has a duality symmetry, the Hamiltonian
will be invariant under symmetry transformations due to the presence of the aforementioned
non-holomorphic modifications.  Since the Hamiltonian is related to the function $F$ by an 
Legendre transformation, these non-holomorphic modifications will also be encoded in $F$.

This `Hamiltonian' picture of BPS black holes suggests that special geometry can be consistently modified
by a class of non-holomorphic deformations, whereby the holomorphic function $F(X)$ that
characterizes the Wilsonian action is replaced by a non-holomorphic function
\begin{equation}
F(X, \bar X) = F^{(0)}(X) + 2 \mathrm{i} \, \Omega (X, \bar X) \;,
\label{eq:F-0-Om}
\end{equation}
where $\Omega$ denotes a real (in general non-harmonic) function.  The Wilsonian limit is recovered by taking
$\Omega$ to be harmonic.  In section  \ref{sec:ubiquity} we show that the non-holomorphic deformations of special 
geometry described by \eqref{eq:F-0-Om} occur in a generic setting.
There we consider general point-particle Lagrangians (that depend on coordinates
and velocities)
and their associated Hamiltonians.  
We present a theorem that  shows that
the dynamics of these models can be reformulated in terms of a 
symplectic vector $(X, \partial F/\partial X)$ constructed out of a 
complex function $F$ of the form \eqref{eq:F-0-Om}, and whose real part
comprises the 
canonical variables of the associated Hamiltonian. 
We show that under duality transformations the transformed symplectic vector is again encoded
in a non-holomorphic function of the form \eqref{eq:F-0-Om}.
We illustrate the theorem with various field theory examples
with higher-derivative interactions.
We give a detailed discussion of these examples in order to illustrate the characteristic features of the theorem.
One example consists of the Born-Infeld Lagrangian for an abelian gauge  field, which we reformulate in the language
of the theorem based on \eqref{eq:F-0-Om}.  We subsequently promote the gauge coupling constant to a dynamical
field $S$ and discuss the duality symmetries of the resulting model.  We then turn to more general models 
with exact S- and T-duality and discuss 
the restrictions imposed on $\Omega$ by these symmetries.

The function $F$ in \eqref{eq:F-0-Om}
may depend on a number of external parameters which we denote by 
$\eta$. Under duality transformations, the symplectic vector  $(X, \partial F/\partial X)$  transforms into
 $(\tilde X, \partial \tilde{F}/\partial \tilde{X})$, while the parameters $\eta$ are inert. When expressing the
 transformed variables $\tilde X$ in  terms of the $X$, the relation will depend on $\eta$, i.e. $\tilde{X} = 
 \tilde{X}(X, \eta)$.
 In section \ref{sec:new-var}
we introduce covariant complex variables that constitute a complexification of the canonical variables of the Hamiltonian,
and whose duality transformation law is independent of $\eta$.
These variables ensure that when expanding the Hamiltonian 
in powers of the external parameters, the resulting expansion coefficients transform covariantly under duality
transformations. This expansion can also be studied by employing a modified derivative ${\cal D}_{\eta}$, which we 
construct.
The covariant variables introduced in this section have the same duality transformation properties
as the ones used
in topological string theory and can therefore be identified with the latter.  
A further indication of the relation with topological string theory
is provided by the generating function that relates the canonical variables of the Hamiltonian
to the covariant complex variables.  This generating function turns out to be the one that is used in the
wave function approach to perturbative topological string theory
\cite{Witten:1993ed,Dijkgraaf:2002ac,Verlinde:2004ck,Aganagic:2006wq,Gunaydin:2006bz} .

In section \ref{sec:hesse-top} we turn to 
supergravity models in the presence of higher-curvature interactions
encoded in the square of the Weyl superfield \cite{Bergshoeff:1980is,deWit:1996ix}.  We consider these models in an
$AdS_2 \times S^2$ background and compute the effective action in this background. This is first done at the level
of the Wilsonian effective action \cite{Sahoo:2006rp,Cardoso:2006xz}. 
Then we assume that the extension to the low-energy effective action
can be implemented by replacing the Wilsonian holomorphic function $F$ by the non-holomorphic function 
\eqref{eq:F-0-Om}.  Next, we perform a Legendre transformation of the low-energy effective action in this background
and obtain the associated `Hamiltonian', which takes the form of the aforementioned generalized Hesse potential.
Using the covariant complex variables introduced in section \ref{sec:new-var}, we expand
the associated Hesse potential (the Hamiltonian) and work out the first few iterations.
This reveals a systematic structure.  Namely, the Hesse potential decomposes into two classes of terms.
One class consists of combinations of terms, constructed
out of derivatives of $\Omega$,
that transform as functions under electric/magnetic duality. 
The
other class is constructed out of $\Omega$ and derivatives thereof.  Demanding this second 
class to also exhibit a proper behavior under duality transformations (as  a consequence of
the transformation behavior of the Hesse potential)
imposes 
restrictions on $\Omega$.  These restrictions are captured by a differential equation that 
constitutes 
half of the holomorphic anomaly equation 
encountered in the context of perturbative topological string theory.  The differential equation
is a consequence of the tension between
maintaining harmonicity of $\Omega$ and insisting
on a proper behavior under duality transformations \cite{deWit:1996ix}. We conclude section \ref{sec:hesse-top} with 
a brief discussion of open issues.

In the appendices we have collected various results, as follows.
Appendix \ref{integrab} discusses the transformation behavior under symplectic transformations 
of various holomorphic and anti-holomorphic derivatives of $F$. 
We use these expressions 
to give an alternative proof of the integrability of the resulting structures.
In addition, we show that when $F$ depends on an external parameter $\eta$, its derivative
$\partial_{\eta} F$ transforms as a function under symplectic transformations.
In appendix \ref{cov-der} we show that the modified derivative ${\cal D}_{\eta}$ of section \ref{sec:new-var}
acts as a covariant derivative for symplectic transformations.  This is done by
showing that when given a quantity $G(x, \bar x; \eta)$ that transforms
as a function under symplectic transformations, also ${\cal D}_{\eta} G$ transforms
as a function.  In appendix \ref{sec:top-string} we review the holomorphic anomaly equation of topological
string theory in the big moduli space.  
Appendix \ref{sec:funct-H-a-i-geq2} lists certain combinations that arise in the expansion of 
the Hesse potential in powers of $\eta$ and 
that transform as functions under electric/magnetic duality. 
In appendix \ref{sec:transf-der-om} 
we list the transformation properties of various derivatives of $\Omega$
under duality transformations using  the covariant variables of section \ref{sec:new-var}.

These lecture notes include exercises which we hope will constitute a guidance to the reader.

\section{Lecture I: Point-particle models and $F$-functions}
\label{sec:ubiquity}

We begin by considering a general point-particle Lagrangian 
that depends on coordinates $\phi$ and velocities $\dot \phi$.  
The associated Hamiltonian will depend on the canonical variables $\phi$ and $\pi$, where
$\pi$ denotes the canonical momentum.
After briefly
reviewing some of the salient features of the Hamiltonian description, such as canonical transformations in phase space,
we present a theorem that  shows that
the dynamics of these models can be reformulated in terms of a symplectic vector that is complex, and whose real part
comprises the 
canonical variables $(\phi, \pi)$.  This is achieved by introducing a complex function $F$ 
that depends on complex variables $x$, with the 
symplectic vector given by $(x, \partial F/
\partial x)$.  
This reformulation exhibits many of the special geometry features that are typical for $N=2$ supersymmetric systems. However, it also goes beyond the standard formulation of these systems
in that the function $F$ is of the form \eqref{eq:F-0-Om}, and hence non-holomorphic in general.

We illustrate the theorem with various field theory examples
with higher-derivative interactions.
We give a detailed discussion of these examples in order to illustrate the characteristic features of the theorem.
One example consists of the Born-Infeld Lagrangian for a Maxwell field, which we reformulate in the language
of the theorem.  
We subsequently promote the gauge coupling constant to a dynamical
field $S$ and discuss the duality symmetries of the resulting model.  We turn to more general models 
with exact S- and T-duality and discuss 
the restrictions imposed on $\Omega$ by these symmetries.

The reader not interested in the details of these examples may want to proceed to subsection \ref{sec-F-hom},
where we discuss 
the form of the Hamiltonian
when the function $F$ is such that it transforms homogeneously under a real rescaling of the variables involved.

\subsection{Theorem \label{sec:theor}}

Let us consider a point-particle model 
described by a Lagrangian ${L}$
with  $n$
coordinates $\phi^i$ and $n$ velocities $\dot \phi^i$.  The associated canonical momenta $\partial L /
\partial \dot \phi^i$
will be denoted by $\pi_i$.
The Hamiltonian ${H}$ of the system, which follows from $L$ by Legendre transformation,
\begin{equation}
{H}(\phi,\pi)= \dot \phi^i\, \pi_i - {L}(\phi,\dot \phi) \;,
\label{eq:leg-l-h}
\end{equation}
depends on $(\phi^i, \pi_i)$, which are called canonical variables, since they
satisfy the canonical Poisson bracket relations. The variables $(\phi^i, \pi_i)$
denote coordinates on 
a symplectic manifold called the classical phase 
space of the system.  
In these coordinates, the symplectic 2-form is $d \pi_i \wedge d \phi^i$.
This 2-form is preserved under canonical transformations of $(\phi^i,\pi_i)$ given by
\begin{equation}
  \label{eq:symplectic-pq}
  \begin{pmatrix} 
    \phi^i\\[2mm] \pi_i
  \end{pmatrix}
  \longrightarrow
  \begin{pmatrix} 
    \tilde \phi^i\\[2mm] \tilde \pi_i
  \end{pmatrix}
  =
  \begin{pmatrix} 
    U^i{}_j& Z^{ij}\\[2mm]  W_{ij} & V_i{}^j 
  \end{pmatrix}
  \begin{pmatrix} 
    \phi^j\\[2mm] \pi_j
  \end{pmatrix}\;,
\end{equation}
where $U, V, Z$ and $W$ denote $n \times n$ matrices that satisfy the relations 
\begin{eqnarray}
U^T \, V - W^T \, Z = V^T \, U - Z^T \, W = \mathbb{I} \;,\nonumber\\
U^T \, W = W^T \, U \;\;\;,\;\;\; Z^T \, V = V^T \, Z \;.
\label{eq:matrix-sympl}
\end{eqnarray}
These relations are precisely such that the transformation \eqref{eq:symplectic-pq}
constitutes an element of 
${\rm Sp} (2n, \mathbb{R})$.  This transformation leaves the Poisson brackets invariant. 
The Hamiltonian transforms
as a function under symplectic transformations, i.e. $\tilde{{H}} (\tilde \phi, \tilde \pi)
= {H} (\phi, \pi)$.  When the Hamiltonian is invariant under a subset of 
${\rm Sp} (2n, \mathbb{R})$ transformations, this subset describes a symmetry of the system.  This 
invariance is often called
duality invariance.
Observe that the Legendre transformation \eqref{eq:leg-l-h} also gives rise to the relation
$\partial {L}/\partial \phi^i = - \partial {H}/\partial \phi^i$ by virtue of 
$\pi_i = \partial {L}/ \partial \dot{\phi}^i$.

Now we present a theorem that states  that the Lagrangian
can be reformulated in terms of a complex function $F(x, \bar x)$
based on complex variables $x^i$, such that the 
canonical coordinates $(\phi^i, \pi_i)$ coincide with (twice) the real part of $(x^i, F_i)$, where
$F_i = \partial F(x, \bar x)/\partial x^i$.  
\\
{\bf Theorem}: 
Given a Lagrangian ${L}(\phi,\dot \phi)$ depending on $n$
coordinates $\phi^i$ and $n$ velocities $\dot \phi^i$, with corresponding
Hamiltonian ${H}(\phi,\pi)= \dot \phi^i\, \pi_i -{L}(\phi,\dot \phi)$,
there exists a description in terms of complex coordinates
$x^i=\tfrac12(\phi^i+\mathrm{i}\dot \phi^i)$ and a complex function
$F(x,\bar x)$, such that,
\begin{align}
  \label{eq:theorem-prop}
  2\, \mathrm{Re}\,x^i =&\, \phi^i\,,\nonumber\\
    2\, \mathrm{Re}\,F_i(x,\bar x) =&\, \pi_i\,, \quad\mbox{where}\;\; F_i =
    \frac{\partial F(x,\bar x)}{\partial x^i}\,.
\end{align}
The function $F(x,\bar x)$ is defined up to an anti-holomorphic function
and can be decomposed into a holomorphic and a purely imaginary
(in general non-harmonic) function,
\begin{equation}
  \label{eq:F(x)}
  F(x,\bar x) = F^{(0)}(x) + 2\mathrm{i} \Omega(x,\bar x)\,.
\end{equation}
The relevant equivalence transformations take the form, 
\begin{equation}
  \label{eq:ambiguity}
  F^{(0)}\to F^{(0)} + g(x)\,, \qquad \Omega\to \Omega- \mathrm{Im}\,
  g(x)\,, 
\end{equation}
which results in $F(x,\bar x)\to F(x,\bar x)+\bar g(\bar x)$.
The Lagrangian and Hamiltonian can then be expressed in terms of
$F^{(0)}$ and $\Omega$,
\begin{align}
  \label{eq:H-sympl}
  {L}=&\, 4 [\mathrm{Im}\, F -\Omega] \,,\nonumber \\
   {H} =&\, -\mathrm{i}(x^i \,\bar F_{\bar \imath} -\bar
  x^{\bar\imath} \,F_i)
  -4\,\mathrm{Im} [F-\tfrac12 x^i\,F_i] +4\, \Omega \nonumber\\
  =&\, -\mathrm{i}(x^i \,\bar F_{\bar \imath} -\bar
  x^{\bar\imath} \,F_i) -4\,\mathrm{Im}
  [F^{(0)}-\tfrac12 x^i\,F^{(0)}_i] -2(2\,\Omega
  -x^i\Omega_i -\bar x^{\bar\imath} \Omega_{\bar\imath}) \,,
\end{align}
with $F_i = \partial F/\partial x^i , F^{(0)}_i = \partial F^{(0)}/\partial x^i , 
\Omega_i = \partial \Omega / \partial x^i$, 
and similarly for ${\bar F}_{\bar \imath}, {\bar F}^{(0)}_{\bar \imath}$ and $\Omega_{\bar \imath}$.

Furthermore, a crucial observation is that the $2n$-vector $(x^i,F_i)$ 
denotes a complexification of the phase space coordinates
 $(\phi^i, \pi_i)$ that 
transforms precisely as 
$(\phi^i, \pi_i)$ under symplectic transformations, i.e.
\begin{equation}
  \label{eq:symplectic}
  \begin{pmatrix} 
    x^i\\[2mm] F_i(x,\bar x)
  \end{pmatrix}
  \longrightarrow
  \begin{pmatrix} 
    \tilde x^i\\[2mm] \tilde F_i(\tilde x,\bar{\tilde x})
  \end{pmatrix}
  =
  \begin{pmatrix} 
    U^i{}_j& Z^{ij}\\[2mm]  W_{ij} & V_i{}^j 
  \end{pmatrix}
  \begin{pmatrix} 
    x^j\\[2mm] F_j(x,\bar x)
  \end{pmatrix}\;.
\end{equation}
Hence, a $\mathrm{Sp}(2n,\mathbb{R})$ transformation of $(x^i,F_i)$ 
is a canonical transformation of ${H}(\phi, \pi)$.
The equations (\ref{eq:symplectic}) are, moreover, integrable: the
symplectic transformation yields a new function $\tilde F (\tilde x, \bar{\tilde{x}}) = 
\tilde F^{(0)}(\tilde x) + 2\mathrm{i} \tilde\Omega  (\tilde x, \bar{\tilde{x}})$, with 
 $\tilde\Omega$ real.\\
{\bf Proof:}
The proof of this theorem proceeds as follows. 
First we introduce the
$2n$-vector $(x^i,y_i)$, 
\begin{align}
  \label{eq:def-x-y}
  x^i = &\,  
    \tfrac12\Big(\phi^i +
  \mathrm{i}\frac{\partial {H}}{\partial \pi_i} \Big)\,,\nonumber\\
  y_i =&\, 
  \tfrac12\Big(\pi_i -
  \mathrm{i}\frac{\partial {H}}{\partial \phi^i} \Big)\,,
\end{align}
which is constructed out of two canonical pairs, one comprising the variables $(\phi^i, \pi_i)$ and 
the other one comprising derivatives of ${H}(\phi, \pi)$, namely
$( \partial {H} / \partial \pi_i, - \partial {H} / \partial \phi^i)$.  
Both pairs transform in the same way under canonical transformations \eqref{eq:symplectic-pq}.
Now we relate the vector  $(x^i,y_i)$ to the one given in \eqref{eq:theorem-prop}, and we 
show that  Lagrangian and the Hamiltonian can be expressed in terms
of a complex function $F(x, \bar x)$ as in \eqref{eq:H-sympl}.

The Legendre transformation \eqref{eq:leg-l-h} gives $\dot \phi^i 
= \partial {H} / \partial \pi_i$, where we used
$\pi_i = \partial {L}/ \partial \dot{\phi}^i$.
 This equation
establishes that 
the complex $x^i$ introduced in \eqref{eq:def-x-y}
coincide with the $x^i$ defined above \eqref{eq:theorem-prop}.
Then, expressing the Lagrangian in terms of  $x^i$ and $\bar x^{\bar\imath}$, gives
\begin{equation}
  \label{eq:x-derivative-L}
   \frac{\partial {L}(x,\bar
    x)}{\partial x^i} = -2\mathrm{i} y_i\,, 
\end{equation}
where we used 
the relation
$\partial {L}/\partial \phi^i = - \partial {H}/\partial \phi^i$
mentioned below \eqref{eq:matrix-sympl}.
Next we
write ${L}$ as the sum of a harmonic and a
non-harmonic function (which is always possible),
\begin{equation}
  \label{eq:n-n-harmonic-L}
  {L}= -2\mathrm{i}\big[F^{(0)}(x)-\bar F^{(0)}(\bar x)\big]
  +4\, \Omega(x,\bar x) \;.
\end{equation}
By introducing the combination
$F(x,\bar x) = F^{(0)}(x) +
2\mathrm{i} \Omega(x,\bar x)$, we observe that the relation \eqref{eq:x-derivative-L} can be concisely written as
$y_i=\partial F(x,\bar x) /  \partial x^i$, while the Lagrangian \eqref{eq:n-n-harmonic-L} becomes
${L} = 4 [\mathrm{Im}\, F -\Omega]$. 
Using this as well as \eqref{eq:def-x-y}, we obtain that
 the Hamiltonian ${H}(\phi,\pi) = \dot \phi^i\, \pi_i -{L}(\phi,\dot \phi) $ can be expressed as
 in \eqref{eq:H-sympl}.
\vskip 1mm
\noindent
{\sf Exercise 1: Verify that $H$ can be written as in \eqref{eq:H-sympl}.}
\vskip 1mm
\noindent
Thus, we have shown that the vector $(x^i, y_i)$ equals $(x^i, F_i)$, 
 and we have
established the validity of \eqref{eq:H-sympl}.

Now let us discuss the integrability of $(x^i, y_i )$ under canonical transformations.
The vector  $(x^i, y_i )$, given in \eqref{eq:def-x-y},
consists of two canonical pairs, and hence it transforms as in 
\eqref{eq:symplectic} under canonical transformations.
We denote the transformed variables by 
$(\tilde x^i, \tilde y_i )$.
The Hamiltonian
transforms as a function, i.e.
$\tilde{{H}} ({\rm \Re} \, \tilde x, {\rm Re} \, \tilde y) = {H} ({\rm Re} \, x, {\rm Re} \, y)$, as already mentioned.
Since we are dealing with a canonical transformation, 
the dual quantities $(\tilde x^i, \tilde y_i )$ and $\tilde H$ will satisfy the same relations as the original
quantities $(x^i, y_i )$ and $H$, so that we can apply the steps 
\eqref{eq:def-x-y}--\eqref{eq:n-n-harmonic-L}
to the dual quantities.
The dual variables $(\tilde x^i, \tilde y_i )$ have the decomposition given in \eqref{eq:def-x-y}, but now
in terms of the dual quantities.
The Lagrangian $\tilde L$ associated to $\tilde H$ is obtained by a  Legendre transformation of
 $\tilde H$, i.e. $\tilde L = \dot{\tilde{\phi}}^i \tilde{\pi}_i - \tilde H$.  Then, applying the steps 
given below \eqref{eq:def-x-y}
  to the dual Lagrangian shows that  
 $\tilde L = 4 [\mathrm{Im}\, \tilde F - \tilde \Omega]$, where $\tilde F$ is the sum of a holomorphic
function $\tilde F^{(0)}$ 
 and a real function $ \tilde\Omega$, i.e. 
 $\tilde F (\tilde x, \bar{\tilde{x}}) = 
\tilde F^{(0)}(\tilde x) + 2\mathrm{i} \tilde\Omega  (\tilde x, \bar{\tilde{x}})$.
This establishes that $(\tilde x^i, \tilde y_i)$ can be obtained from a new function $\tilde F$, and 
hence ensures the integrability of $(\tilde x^i, \tilde y_i)$ under symplectic transformations.

To complete the proof of the theorem, we need to discuss one more issue, namely
the decompositions of $F(x, \bar x)$ and  $\tilde F (\tilde x, \bar{\tilde{x}})$ and their
relation.
The decomposition of 
$F$ into 
$F^{(0)}$ and $\Omega$ suffers from the ambiguity \eqref{eq:ambiguity}, and so does the decomposition
of $\tilde F$.  Therefore, to be able to relate both decompositions, we need
to fix the ambiguity in the decomposition of $\tilde F$, once a decomposition of $F$ has been given.
To do so, we proceed as follows.

We consider a symplectic  transformation \eqref{eq:symplectic} which, as we just discussed,
yields a new function $\tilde F$.  Given a decomposition of $F$, we apply the same 
transformation
to the vector $(x^i, F_i^{(0)})$ alone, where $F^{(0)}_i = \partial F^{(0)} / \partial x^i$. 
This yields  
the vector $(\hat{x}^i, {\tilde F}_i^{(0)} (\hat x))$, as explained in 
appendix \ref{integrab}.
The transformed vector $(\hat x^i, \tilde F^{(0)}_i (\hat x))$
can be integrated, i.e. $\tilde F^{(0)}_i$ can be expressed as $\partial \tilde F^{(0)} (\hat x) / \partial \hat x^i$, where $\tilde F^{(0)} (\hat x)$ is uniquely determined up to a constant and up to terms linear in $\hat{x}^i$
(see \eqref{eq:F-0-transf})
\cite{deWit:1996ix}.  The expression for $\tilde F^{(0)} (\hat x)$ can be readily obtained by using 
that the combination $F^{(0)} - \tfrac12 \, x^i \, F^{(0)}_i$ transforms as a function under symplectic transformations, i.e.
$\delta \left(F^{(0)} - \tfrac12 \, x^i \, F^{(0)}_i \right) = \tfrac12 \left( \delta x^i \, F_i^{(0)} - x^i \,
\delta F^{(0)}_i \right) $. 
One obtains 
 $\tilde F^{(0)}(\hat x) =  \tfrac12 \, \hat x^i \, \tilde F^{(0)}_i (\hat x)  + F^{(0)} - \tfrac12 \, x^i 
 \, F^{(0)}_i$, up to a constant and up to terms linear in $\hat{x}^i$.
  Thus, to relate the decomposition of $\tilde F$ to the decomposition of $F$, we demand
that $\tilde F^{(0)}$ refers to the combination that
follows by applying a symplectic transformation to $(x^i, F_i^{(0)})$, as just described.
This in turn determines $\tilde \Omega = \tfrac14 [ \tilde L - 4 {\rm Im} \, \tilde F^{(0)}]$.
This completes the proof of the theorem.

We finish this subsection with a few comments.  First, we note that since both ${H}$ and 
$F^{(0)} - \tfrac12 \, x^i \, F^{(0)}_i$ transform as functions under symplectic transformations,
so does the following combination that appears in \eqref{eq:H-sympl},
\begin{equation}
2\,\Omega
  -x^i\Omega_i -\bar x^{\bar\imath} \Omega_{\bar\imath} \;.
  \label{eq:ex-sympl-func}
\end{equation}
Second, the transformation law of $2 \mathrm{i} \Omega_i = F_i - F_i^{(0)}$ under symplectic transformations
is determined by the transformation behavior
of $F_i$ and $F_i^{(0)}$, as described above. 
In appendix \ref{integrab} we give an equivalent expression for  $\tilde \Omega_i$ in terms of a power
series in derivatives of $\Omega$, see
\eqref{eq:tilde-omega}. The transformation law of $2 \mathrm{i} \Omega_{\bar \imath} = F_{\bar \imath}$, on the other hand, follows from the reality of $\tilde \Omega$, 
\begin{eqnarray}
\tilde{\Omega}_{\bar \imath} = (\overline{\tilde{\Omega}_i}  )       \;.
\label{eq:rel-omega-bari_cond}
\end{eqnarray}
Third, as mentioned in the introduction, 
the function $F(x, \bar x)$ may, in general, depend on a number of external parameters $\eta$
that are inert under symplectic transformations.
Without loss of generality, we may take 
$\eta$ to be solely encoded in $\Omega$ and, upon transformation, in  $\tilde \Omega$
(we can use the equivalence relation \eqref{eq:ambiguity} to achieve this).
In appendix \ref{integrab} we show that 
$\partial_{\eta} F = \partial F / \partial \eta$ transforms as a function under symplectic transformations
\cite{Cardoso:2008fr}.   We will return to this 
feature in subsection
\ref{sec-F-hom}.

Appendix \ref{integrab} also discusses the transformation behavior under symplectic transformations 
of various holomorphic and anti-holomorphic derivatives of $F$. 
We use these expressions 
to give an alternative proof of the integrability of \eqref{eq:symplectic}.

\subsection{Examples}

We now proceed to illustrate the features of the theorem discussed above in various models that have 
duality symmetries.  
To keep the discussion as transparent as possible in all cases, we 
consider the reduced Lagrangian that is obtained by restricting to spherically symmetric static configurations 
in flat spacetime.
The first model we consider
is the Born-Infeld model for an abelian gauge field, which has been known  
to have an ${\rm SO}(2)$ duality symmetry for a long time \cite{Schroed35}.  This symmetry may be enlarged to an
${\rm SL}(2, \mathbb{R})$ duality symmetry by coupling the system to a complex scalar field, called the dilaton-axion
field \cite{Gibbons:1995ap}.  This is the second model we consider.  
Then we turn 
to more general models 
with exact S- and T-duality and discuss 
the restrictions imposed on $\Omega$ by these symmetries.
We exhibit how the Born-Infeld-dilaton-axion
system fits into this class of models.  
Finally, we focus on the case when the 
function $F(x, \bar x)$ is taken to be homogeneous, 
and we discuss the form of the associated Hamiltonian.

\subsubsection{The Born-Infeld model
\label{sec:born-infeld-reduced}}

The Born-Infeld Lagrangian\footnote{We will use the notation $\mathcal{L}$ and $\mathcal{H}$ when
dealing with Lagrangian and Hamiltonian densities, respectively.}
 for an abelian gauge field in a spacetime with metric $g_{\mu \nu}$ is given by \cite{BornInf34}
\begin{equation}
  \label{eq:1}
  \mathcal{L}= -g^{-2} \left[\sqrt{\vert \det[g_{\mu\nu} +
        g\,F_{\mu\nu}] \vert } - \sqrt{\vert \det g_{\mu \nu} \vert } \right] \;.
\end{equation}
It depends on an external parameter $\eta = g^2$.
In the following we 
consider spherically symmetric static configurations in flat spacetime given by
\begin{align}
ds^2 =&\, - dt^2 + dr^2 + r^2 \left(d \theta^2 + \sin^2 \theta \, d\varphi^2 \right) \;, \nonumber\\
F_{rt} =&\, e(r) \;\;\;,\;\;\; F_{\theta\varphi} = p \, \sin \theta \;.
\label{eq:static-spher-sym}
\end{align}
Here, the $\theta$-dependence of $F_{\theta \varphi}$ is fixed by rotational invariance, and $p$
is constant by virtue of the Bianchi identity.
Evaluating \eqref{eq:1} for this configuration gives
\begin{equation}
  \label{eq:red-L_spher}
  \mathcal{L} = - g^{-2} r^2 \sin^2 \theta \, \left[\sqrt{| 1-g^2
      e^2(r)| }\;\sqrt{1+g^2 \, p^2 \, r^{-4}} -1\right]\,.
\end{equation}
Below we will rewrite \eqref{eq:red-L_spher} and bring it into the form \eqref{eq:H-sympl}.
Since this rewriting does not depend on the angular variables and since it applies to any 
$r$-slice, we integrate over the angular variables and 
pick the $r$-slice
 $4 \pi r^2 =1$, for convenience.  The resulting reduced Lagrangian reads,
 \begin{equation}
  \label{eq:red-L}
  \mathcal{L}(e, p) = - g^{-2} \left[\sqrt{1-g^2
      e^2}\;\sqrt{1+g^2 p^2} -1\right]\,,
\end{equation}
where we take $ g^2 e^2 < 1$.
\vskip 1mm
\noindent
{\sf Exercise 2:  Instead of flat spacetime, consider the $AdS_2 \times S^2$
line element $ds^2 = v_1 (- r^2 dt^2 + r^{-2} dr^2 )
+ v_2 (d \theta^2 + \sin^2 \theta \, d\varphi^2 )$, where $v_1$ and $v_2$ denote constants.
Show
that the resulting reduced Lagrangian takes the form \eqref{eq:red-L} after performing
a suitable rescaling of $g, e $ and $p$.}
\vskip 1mm

In the example \eqref{eq:red-L}, the role of the coordinate $\phi$ and of the velocity $\dot{\phi}$ introduced above
\eqref{eq:theorem-prop} is played by $p$ and $e$, respectively. The associated Hamiltonian ${\cal H}$ is obtained 
by Legendre transforming with respect to $\dot \phi = e$. The conjugate momentum $\pi$ is given by the electric charge $q$, so that
\begin{equation}
{\cal H}(p,q) = q \, e - \mathcal{L}(e,p) \;.
\label{eq:leg-H-L}
\end{equation}
Computing
\begin{equation}
  \label{eq:def-q-f}
  q=\frac{\partial\mathcal{L}}{\partial e} = e\;\sqrt{\frac{1+g^2
      p^2} {1-g^2 e^2}} \;\;\; , \;\;\;
  f \equiv\frac{\partial\mathcal{L}}{\partial p}= -p\; \sqrt{\frac{1-g^2
      e^2} {1+g^2 p^2} } \,,
\end{equation}
where we introduced $f$ for later convenience, and substituting in \eqref{eq:leg-H-L},
we obtain for the 
Hamiltonian,
\begin{equation}
  \label{eq:red-H}
  \mathcal{H} (p,q) =
  g^{-2}\left[\sqrt{1+g^2(p^2+q^2)}
    -1\right ] \,. 
\end{equation}
This Hamiltonian is manifestly invariant under ${\rm SO}(2)$ rotations of $p$ and $q$ and, in particular,
under the discrete symmetry that 
interchanges the electric and magnetic charges.  The external parameter $\eta = g^2$ is inert under these
transformations.
These rotations constitute the only continuous symmetry of the system \cite{Schroed35}.
Their infinitesimal form can be represented by an ${\rm Sp}(2, \mathbb{R})$-transformation 
\eqref{eq:matrix-sympl} with $U = V = 1$ and  $Z = - W =  - c$, where $c \in 
\mathbb{R}$.

Now, following the construction described in the previous subsection \ref{sec:theor},
we introduce 
a complex coordinate $x$ in terms of the coordinate $\phi = p$ and the velocity $\dot \phi = e$,
and a complex function $F(x, \bar x; g^2)$,
\begin{equation}
  \label{eq:complex}
  x= \tfrac12(p+\mathrm{i} e)\,,\qquad 
  F(x,\bar x; g^2) = F^{(0)}(x) + 
  2\mathrm{i}  \Omega(x,\bar x;g^2) \,,
\end{equation}
where
\begin{align}
  \label{eq:def-Omega}
     F^{(0)}(x) =&\, -\tfrac12\mathrm{i} \,x^2 \;, \nonumber\\
   \Omega(x,\bar x;g^2) =&\, \tfrac18 \, g^{-2} \left(\sqrt{1+
       g^2(x+\bar x)^2} - \sqrt{1+g^2(x-\bar x)^2}\right)^2 \,. 
\end{align}
The split into $F^{(0)}$ and $\Omega$ is done in such a way that 
$F^{(0)}$ will encode the contribution at the two-derivative
level (which corresponds to 
the term ${\cal L} \approx -\tfrac14 F_{\mu\nu}^2 +\mathcal{O}(g^2)$ in \eqref{eq:1}),
while $\Omega$ will encode the higher-derivative contributions.  
Indeed, with these definitions
the Lagrangian \eqref{eq:red-L} can be written as 
\begin{equation}
  \label{eq:lag-F}
  \mathcal{L} = 4[\mathrm{Im} F-\Omega] \,,
\end{equation}
in agreement with the first equation of \eqref{eq:H-sympl}. Next, using the first equation of \eqref{eq:def-q-f},
we establish
\begin{equation}
p= 2\,\mathrm{Re}\, x\,, \quad
  q= 2\, \mathrm{Re} \,F_x \,, 
  \label{eq:p-q-F}
\end{equation}
in accordance with \eqref{eq:theorem-prop}, where we recall that the conjugate momentum $\pi$
equals $q$.  Then, inserting \eqref{eq:p-q-F} into \eqref{eq:red-H} yields
\begin{eqnarray}
  \label{eq:4}
 \mathcal{H} = \mathrm{i} (\bar x\,F_{x}-x\,\bar F_{\bar x}) +
  4\,g^2\frac{\partial \Omega}{\partial g^2} \,,
 \end{eqnarray}
where $F_x = \partial F(x, \bar x; g^2) / \partial x$.  
This is in agreement with the second equation of \eqref{eq:H-sympl}, since $F^{(0)}$ satisfies
$F^{(0)} = \tfrac12 \, x F^{(0)}_x$, and $\Omega$ obeys the homogeneity relation
\begin{equation}
  \label{eq:Omega-homog}
 2 \, \Omega = x \, \Omega_x + {\bar x} \, \Omega_{\bar x} - 2 \, g^2 \, \frac{\partial \Omega}{\partial g^2} \;,
\end{equation} 
which is a consequence of the behavior of $\Omega$ under 
 the real scaling $x \rightarrow \lambda \, x$ and $g^{2}
\rightarrow \lambda^{-2} \, g^{2}$.
\\[1mm]
{\sf Exercise 3: Establish \eqref{eq:Omega-homog} by differentiating the relation
$\Omega(\lambda \, x, \lambda \, {\bar x} ; \lambda^{-2} \, g^2) = \lambda^2 \, \Omega (x, \bar{x}; g^2) $.}
\\[1mm]
{\sf Exercise 4: Verify \eqref{eq:lag-F}, \eqref{eq:p-q-F} and \eqref{eq:4}.}
\vskip 1mm

Rather than performing a Legendre transformation of $\mathcal{L}(e, p)$ with respect
to $e$, we may instead consider performing a Legendre transformation with respect to $p$.
The resulting quantity $\mathcal{S}(e, f)$ will then depend on the canonical
pair $(e,f)$, rather than on $(p,q)$. Using the expression for $f$ given in \eqref{eq:def-q-f},
we obtain
\begin{equation}
  \label{eq:S}
  \mathcal{S}(e,f) = f \, p -\mathcal{L} (e, p) =
  g^{-2}\left[\sqrt{1-g^2(e^2+f^2)} -1\right ] \,,
\end{equation}
which is invariant under ${\rm SO}(2)$ rotations of $e$ and $f$.
Next, we express $\mathcal{S}(e,f)$ in terms of  $x$ and $F_x$ introduced in \eqref{eq:complex}. First we establish
\begin{equation}
  \label{eq:2Im-f}
  f= 2\,\mathrm{Im} \,F_x \,, 
\end{equation}
so that
\begin{equation}
  \label{eq:x-Fx}
  x= \tfrac12(p+\mathrm{i} e)\,,\qquad F_x =\tfrac12(q+\mathrm{i} f)\,. 
\end{equation}
Then, using \eqref{eq:4}
and \eqref{eq:x-Fx}, 
we obtain\footnote{In the context of BPS black holes, $\mathcal{H}$ is the Hesse potential, and 
the double Legendre transform of $\mathcal{H}$
yields the entropy function \cite{Sen:2005wa,LopesCardoso:2006bg}.}
\begin{eqnarray}
  \label{eq:S-reexp}
  \mathcal{S} = f \, p - q \, e + \mathcal{H} 
  = - i \left( {\bar x} \, F_x - x \, {\bar F}_{\bar x} \right) + 4 \, g^2 \, \frac{\partial \Omega}{\partial g^2}\;.
\end{eqnarray}

Let us now return to the discussion about symplectic transformations alluded to below \eqref{eq:red-H}.
A  symplectic transformation \eqref{eq:symplectic}
may either constitute a symmetry (an invariance)
of the system or correspond to  a symplectic reparametrization
of the system giving rise to an equivalent set of equations of motion and Bianchi identities
\cite{deWit:1996ag}.
When a symplectic transformation describes a symmetry, a convenient method for verifying
this consists in 
performing the substitution $x^i \rightarrow \tilde{x}^i$ in the derivatives $F_i$, and checking that this 
correctly induces the symplectic transformation on $(x^i, F_i)$ \cite{deWit:1996ix}.

To elucidate this, 
let us consider a particular example, namely
the discrete symmetry
that interchanges the electric and magnetic charges.  It can be implemented by the transformation
 $(x,F_x) \rightarrow (F_x,-x)$, which operates on the canonical pairs $(p, q)$ and $(e, f)$
through \eqref{eq:x-Fx}.  
This constitutes a symplectic transformation \eqref{eq:matrix-sympl} with 
$U= V =0 \,,\, Z = 1 \,,\, W =-1$.
To verify that the transformation $x \rightarrow \tilde x = F_x$ 
correctly induces the transformation of $F_x$, we compute
\begin{equation}
  \label{eq:F-x}
  F_x= -\mathrm{i} x\;\frac{1+g^2(x^2-\bar x^2)}{\sqrt{1+
       g^2(x+\bar x)^2} \;\sqrt{1+g^2(x-\bar x)^2}} \,.
\end{equation}
Also, 
expressing $e$ in terms of $p$ and $q$ (by using the first relation of 
\eqref{eq:def-q-f}), 
we may express $x$ in terms of $p = 2 {\rm Re} \, x$ and $q = 2 {\rm Re} \,F_x$, 
\begin{equation}
  \label{eq:e-pq}
x =
  \tfrac12\left( p+\frac{\mathrm{i}q}{\sqrt{1+ g^2(p^2+ q^2)}}\right)\,.
\end{equation}
We leave the following exercise to the reader.
\\[1mm]
{\sf Exercise 5: Using \eqref{eq:F-x}, show that the transformation 
$x \rightarrow F_x$ induces the transformation
$F_x\rightarrow -x$ by inserting the former 
on the right hand side of $F_x$.  Similarly, 
using \eqref{eq:e-pq}, show that the transformation
 $({\rm Re} \, x,{\rm Re} \, F_x) \rightarrow ({\rm Re} \, F_x,- {\rm Re} \, x)$
 induces the transformation
$x \rightarrow F_x$.}
\vskip 1mm
Next, let us discuss an example of a symplectic transformation that does not constitute a symmetry 
of the system, but instead describes a reparametrization of it.  Namely, consider the
following transformation
of the canonical pair $(p, q)$, 
\begin{equation}
\begin{pmatrix}
p \\ q 
\end{pmatrix} = 
\begin{pmatrix}
2 {\rm Re} \, x   \\
2 {\rm Re} \, F_x 
\end{pmatrix}
\longrightarrow
 \begin{pmatrix}
\tilde{p} \\ \tilde{q} 
\end{pmatrix} = 
\begin{pmatrix}
2 {\rm Re} \, \tilde{x}   \\
2 {\rm Re} \, \tilde{F}_{\tilde{x}} 
\end{pmatrix}
=
\begin{pmatrix}
p + \alpha \, q  \\ q
\end{pmatrix} \;\;\;,\;\;\; \alpha \in \mathbb{R} \;.
\label{eq:transf-sympl-spef}
\end{equation}
This constitutes a symplectic transformation \eqref{eq:matrix-sympl} given by
$U= V =1 \,,\, Z = \alpha \,,\, W =0$.  Since, however, it does not represent an ${\rm SO}(2)$ rotation of
$p$ and $q$, it does not leave the Hamiltonian \eqref{eq:red-H} invariant.  To determine the new function
 $\tilde{F}(\tilde{x}, \bar{\tilde{x}}; g^2)$ associated with this reparametrization, we  
start on the Hamiltonian side and use the fact that $\mathcal{H}$ 
transforms as a function under
symplectic transformations.  Using \eqref{eq:transf-sympl-spef} this gives
\begin{equation}
 \label{eq:H-dual}
\tilde{\mathcal{H}} (\tilde p, \tilde q) = \mathcal{H} (p,q) =
 g^{-2}\left[\sqrt{1+g^2[(\tilde p - \alpha\tilde q)^2+{\tilde q}^2]}
   -1\right ] \,. 
\end{equation}
Now we determine the corresponding Lagrangian by Legendre
transformation,
\begin{equation}
 \label{eq:Lag-dual}
\tilde{\mathcal{L}}(\tilde{e}, \tilde{p}) = {\tilde e} \,{\tilde q} -\tilde{\mathcal{H}}(\tilde{p}, \tilde{q}) \;,
\end{equation}
where
\begin{equation}
  \label{eq:tilde-e}
  \tilde e = \frac{\partial \tilde{\mathcal{H}}}{\partial \tilde q} 
  = \frac{(1+\alpha^2)\tilde q - \alpha \, \tilde p}{\sqrt{
      1+g^2(1+\alpha^2)^{-1} [((1+\alpha^2)\tilde q -\alpha \, \tilde p)^2 +
     \tilde p^2]}} \;. 
\end{equation}
This yields, 
\begin{equation}
  \label{eq:e-tilde-expl} 
  \tilde q = \frac{\alpha\,\tilde p}{1+\alpha^2}  + \frac{\tilde
    e}{1+\alpha^2} \,  
  \sqrt{\frac{1+ \alpha^2+  g^2 \, {\tilde p}^2}{1+\alpha^2 - g^2 \,
      {\tilde e}^2}} \,, 
\end{equation}
which, when inserted in \eqref{eq:Lag-dual}, gives
\begin{equation}
  \label{eq:tilde-L}
  \tilde{\mathcal{L}} (\tilde e,\tilde p) = \frac{\alpha \,{\tilde e} \,
    {\tilde p}} {1+\alpha^2}  
  - g^{-2} \left[\frac{1}{1+\alpha^2}\, \sqrt{1+\alpha^2 - g^2 \,
      {\tilde e}^2}\;\sqrt{1+\alpha^2+ g^2 \, {\tilde p}^2} -1\right]\,. 
\end{equation}
In order to bring the Lagrangian $\tilde{\mathcal{L}}$ into the form $\tilde{\mathcal{L}} = 4 \left[
  \Im {\tilde F} - {\tilde \Omega} \right]$, as in \eqref{eq:H-sympl},
we express $\tilde{\mathcal{L}}$ in terms of the complex coordinate
\begin{equation}
{\tilde x} = \ft12 \left( \tilde p + \mathrm{i} \tilde e \right) \;,
\end{equation}
which is the transformed version of the coordinate $x$ introduced in \eqref{eq:complex}.
Then, we consider all the terms in $\tilde{\mathcal{L}}$ that are independent of $g^2$,
and we express them in terms of a function ${\tilde F}^{(0)} (\tilde x)$, as follows,
\begin{equation}
  \frac{1}{1 + \alpha^2} \left[ 
    \alpha \,{\tilde e} \,
    {\tilde p}  + \tfrac12 \left({\tilde e}^2 -  {\tilde p}^2 \right) \right]
   = 4 \, 
  \Im {\tilde F}^{(0)} (\tilde x) \;.
\end{equation}
This yields
\begin{equation}
 {\tilde F}^{(0)} (\tilde x)= \frac{\alpha - \mathrm{i}}{2\,(1+\alpha^2)} \,
  {\tilde x}^2  \;,
  \label{eq:tilF0-m}
  \end{equation}
up to a real constant.  
It represents the function that is obtained by applying the symplectic transformation 
\eqref{eq:transf-sympl-spef} to $F^{(0)}(x)$, as explained at the end of subsection
\ref{sec:theor}.
Next, we introduce the function 
\begin{equation} 
\tilde F (\tilde x, \bar{\tilde{x}}; g^2) =  {\tilde F}^{(0)} (\tilde x)
+  2 \mathrm{i} {\tilde \Omega}( \tilde x, 
  \bar{ \tilde x}; g^2) \;,
  \label{eq:F-dual-m}
 \end{equation}
 with $\tilde \Omega$ real, and we 
require it to satisfy $\tilde{\mathcal{L}} = 4 \left[
  \Im {\tilde F} - {\tilde \Omega} \right]$. This implies that all the 
$g^2$-dependent terms will be encoded in ${\tilde \Omega}( \tilde x, 
  \bar{ \tilde x}; g^2) $.  We obtain
\begin{equation}
  \label{eq:def-Omega_dual}
   {\tilde \Omega}(\tilde x,\bar{\tilde x};g^2) =
   \frac{1}{8(1+\alpha^2)\, g^2} 
   \left(\sqrt{1+\alpha^2 +  
       g^2({\tilde x}+\bar{\tilde x})^2} - \sqrt{1+\alpha^2+
       g^2(\tilde x-\bar{\tilde x})^2}\right)^2 \,.  
\end{equation}
This result gives the function $\tilde F$  associated with the reparametrization
\eqref{eq:transf-sympl-spef}.  We now check that it correctly reproduces the relation 
${\tilde q}= 2\, \mathrm{Re} \,{\tilde F}_{\tilde x}$, as required by
 \eqref{eq:transf-sympl-spef}.  We compute $\tilde F_{\tilde x}$  and obtain,
\begin{align}
  \label{eq:tilde-F-x}
  \tilde F_{\tilde x}=&\, \frac{\alpha\,\tilde
    x}{1+\alpha^2} \\
  &\, 
  -\frac{\mathrm{i}}{2\,(1+\alpha^2)} \left\{  
    (\tilde x-\bar{\tilde x}) \,  
  \sqrt{\frac{1+ \alpha^2+  g^2 (\tilde x+\bar{\tilde x})^2}{1+\alpha^2 +
        g^2 (\tilde x-\bar{\tilde x})^2}}
    + (\tilde x+\bar{\tilde x}) \,  
  \sqrt{\frac{1+ \alpha^2+  g^2 (\tilde x-\bar{\tilde x})^2}
    {1+\alpha^2 + g^2 (\tilde x+\bar{\tilde x})^2}}   \right \} \;. \nonumber
\end{align}
We leave the following exercise to the reader.
\\[1mm]
{\sf Exercise 6: Using \eqref{eq:tilde-F-x}, verify explicitly that
$2 \mathrm{Re} \,{\tilde F}_{\tilde x}$
equals \eqref{eq:e-tilde-expl}.
}
\vskip 1mm
Now we want to see how $\tilde{F}_{\tilde{x}}$ is related to $F_x$.  According to the discussion around
\eqref{eq:symplectic}, 
the symplectic transformation \eqref{eq:transf-sympl-spef} of the
canonical pair $({\rm Re} \, x,{\rm Re} \, F_x)$ induces a corresponding transformation of the vector
$(x,F_x)$,
\begin{equation}
\begin{pmatrix}
 \tilde{x}   \\
 \tilde{F}_{\tilde{x}} 
\end{pmatrix}
 = 
\begin{pmatrix}
x + \alpha \, F_x  \\ F_x
\end{pmatrix} \;.
\label{eq:trans-x-Fx}
\end{equation}
This is indeed the case, as 
can be verified explicitly by expressing the transformed variables $(\tilde p, \tilde e)$ in
terms of the original variables $(p, e)$ using \eqref{eq:def-q-f}, \eqref{eq:tilde-e} and
\eqref{eq:transf-sympl-spef},
\begin{equation}
{\tilde p} = p + \alpha\,e\;\sqrt{\frac{1+g^2
      p^2} {1-g^2 e^2}} \;\;\;,\; \;\;{\tilde e} = e -\alpha\, p
  \;\sqrt{\frac{1 - g^2  e^2} {1+g^2 p^2}} \;,
\end{equation}
and employing the relation
\begin{equation}
\frac{1+\alpha^2 - g^2 {\tilde e}^2} {1+ \alpha^2+ g^2 {\tilde p}^2} =  
\frac{1 - g^2 e^2} {1+g^2 p^2} \;.
\end{equation}
\\[1mm]
{\sf Exercise 7: Verify \eqref{eq:trans-x-Fx} explicitly using \eqref{eq:tilde-F-x}.}

\subsubsection{Including a dilaton-axion complex scalar field}

The Born-Infeld system discussed in the previous section possesses a continuous $SO(2)$ duality symmetry group.
It is possible to enlarge this duality symmetry group to 
${\rm Sp}(2, \mathbb{R})$ by coupling the abelian gauge field to a complex scalar field 
$S = \Phi + \mathrm{i} \, B $
\cite{Gibbons:1995ap}.  This is achieved by replacing $g \, F_{\mu \nu}$ in \eqref{eq:1}
with $ g \, \Phi^{1/2} \,  F_{\mu \nu}$ and adding a term $ B F_{\mu \nu } {\tilde F}^{\mu \nu} $ 
to the Lagrangian, as follows
\cite{Gibbons:1995ap}
\begin{equation}
  \label{eq:BI-dil-gen}
  \mathcal{L}= -g^{-2} \left[\sqrt{\vert \det[g_{\mu\nu} +
        g\,\Phi^{1/2} \,  F_{\mu\nu}] \vert } - \sqrt{\vert \det g_{\mu \nu} \vert } \right] 
        + \tfrac14 \, B \, F_{\mu \nu} {\tilde F}^{\mu \nu} \;.
\end{equation}
Then, the combined system of equations of motion and Bianchi identity for $F_{\mu \nu}$
 is invariant under ${\rm Sp}(2, \mathbb{R})$ transformations, provided that $S$ transforms in a suitable fashion.
The associated Hamiltonian will then be invariant under these transformations.  This will be discussed momentarily.
The coupling  $g \, \Phi^{1/2}$ replaces the gauge coupling constant with a dynamical field, customarily called
the dilaton field, while the term $ B F_{\mu \nu } {\tilde F}^{\mu \nu}$ introduces a scalar field degree of freedom
called the axion.  For this reason, $S$ is also called the dilaton-axion field.

As before, let us consider spherically symmetric static configurations of the form
\eqref{eq:static-spher-sym}.   Picking again the $r$-slice $4 \pi r^2 = 1$, for convenience, the reduced Lagrangian
is now given by
\begin{equation}
\mathcal{L} (e, p, \Phi, B) = - g^{-2} \left[\sqrt{1-g^2 \, \Phi \,
      e^2}\;\sqrt{1+g^2 \, \Phi \, p^2} -1\right] + B \, e \, p \,,
      \label{eq:red-lag-S-bS}
\end{equation}
where we take $g^2 \, \Phi \, e^2 < 1$.
This reduces to the previous one in \eqref{eq:red-L} when setting $S = 1$. 
 To obtain the associated Hamiltonian ${\cal H}$,
\begin{equation}
{\cal H}(p,q, \Phi, B) = q \, e - \mathcal{L} (e,p, \Phi, B) \;,
\label{eq:leg-H-L-S}
\end{equation}
we first compute $q=\partial\mathcal{L}/\partial e$,
\begin{equation}
  \label{eq:def-q-S}
  q=
  e \, \Phi \;\sqrt{\frac{1+g^2 \, \Phi \,
      p^2} {1-g^2 \, \Phi \, e^2}} +  B \, p \;.
 \end{equation}
Inverting this relation yields
\begin{equation}
e = \frac{q- B \, p }{ \sqrt{\Phi^2  + g^2 \, \Phi \, \left[ \Phi^2  \, p^2 + (q - B \, p )^2 \right] }} \;,
\end{equation}
and substituting in \eqref{eq:leg-H-L-S} gives 
\begin{equation}
  \label{eq:red-H-Phi-B}
  \mathcal{H} (p,q, \Phi, B) =
  g^{-2}\left[\sqrt{1+g^2[\Phi \, p^2+ \Phi^{-1} \, \left(q - B \, p \right)^2]}
    -1\right ] \,. 
\end{equation}
Then, expressing $\Phi$ and $B$ in terms of $S$ and $\bar S$ results in 
\begin{equation}
  \label{eq:red-H-S}
  \mathcal{H} (p,q, S, \bar S) =
  g^{-2}\left[\sqrt{1+ 2 \, g^2 \, \Sigma(p, q, S, \bar S)} 
      -1\right ] \,,
\end{equation}
where 
\begin{equation}
\Sigma (p, q, S, \bar S) = \frac{q^2 + \mathrm{i} p \,q (S - \bar S) + p^2 \, |S|^2}{S + \bar S} \;.
\label{eq:Sigma}
\end{equation}
\\[1mm]
{\sf Exercise 8: Verify \eqref{eq:red-H-S}.}
\\[1mm]
Now we are in position to discuss the invariance of the Hamiltonian under ${\rm Sp}(2, \mathbb{R})$ transformations.
Consider a general ${\rm Sp}(2, \mathbb{R})$ transformation of the canonical pair $(p,q)$ given by
\begin{equation}
\begin{pmatrix}
p \\ q 
\end{pmatrix}
\longrightarrow
\begin{pmatrix}
\tilde{p} \\ \tilde{q} 
\end{pmatrix}
=
\begin{pmatrix}
d & - c\\ - b & a 
\end{pmatrix}
\begin{pmatrix}
p \\ q 
\end{pmatrix} \;,
\label{eq:em-duality-pq}
\end{equation} 
where $a, b, c, d \in \mathbb{R}$ and $ad-bc=1$.  The latter ensures that the transformation belongs to
${\rm SL}(2, \mathbb{R}) \cong {\rm Sp}(2, \mathbb{R})$.  Then, $\Sigma$ given in \eqref{eq:Sigma}
is invariant under 
\eqref{eq:em-duality-pq} provided that $S$ transforms according to \cite{LopesCardoso:2006bg}
\begin{eqnarray}
S \rightarrow \frac{a S - \mathrm{i} b}{\mathrm{i} c S + d} \;.
\label{eq:transf-S}
\end{eqnarray}
This explains the role of $S$ in achieving duality invariance.
It should be noted that $S$ does not constitute an additional canonical variable, but instead describes 
a background field. The external parameter $g^2$ is inert under these transformations.
\\[1mm]
{\sf Exercise 9: Show that $\Sigma$ is invariant under the combined transformation \eqref{eq:em-duality-pq}
and \eqref{eq:transf-S}.}
\\[1mm]
We observe that $\mathcal{H}$ homogeneously as
$\mathcal{H} \rightarrow \lambda^2 \mathcal{H}$
under the real scaling $(p, q) \rightarrow \lambda (p, q) \,,\,
g^2 \rightarrow \lambda^{-2} \, g^2 \,,\, S \rightarrow S$, with $\lambda \in \mathbb{R}$.

Let us now return to the reduced Lagrangian \eqref{eq:red-lag-S-bS} and recast it in the form
${\cal L} = 4 \left[ {\rm Im} F - \Omega \right]$,
where again we introduce the complex variable
$x = \ft12 (p + \mathrm{i} e)$. The function $F$ will now depend on the two complex scalar fields $x$ an $S$, 
\begin{align}
F(x, {\bar x}, S, {\bar S}; g^2) = F^{(0)}(x, S)
+ 2 \mathrm{i} \Omega(x, {\bar x}, S,
{\bar S}; g^2) \;,
\label{eq:F-S-x}
\end{align}
and is determined as follows. The holomorphic function $F^{(0)}$ encodes all the contributions 
that are independent of $g^2$, while $\Omega$, which is real, accounts for all the terms in the reduced Lagrangian that
depend on $g^2$.  This yields,
\begin{align}
  \label{eq:def-Omega-dil}
  F^{(0)}(x, S) =&\,  - \ft12 \mathrm{i} \, S \, x^2 \;, \\
     \Omega(x,\bar x, S, {\bar S};g^2) =&\, \tfrac18 \,g^{-2} \left(\sqrt{1+\ft12
       g^2 \, (S + \bar S) \, (x+\bar x)^2} - \sqrt{1+\ft12 g^2 \, (S + \bar S) \,
       (x-\bar x)^2}\right)^2 \,. \nonumber
\end{align}
Observe that under the scaling of $(p,q)$ and $g^2$ discussed below \eqref{eq:transf-S}, $e$ scales as $e \rightarrow
\lambda e$, and hence
$x$ scales as $x \rightarrow \lambda x$.  This in turn implies that 
$F$ scales as $F \rightarrow
\lambda^2 F$.

{From} \eqref{eq:theorem-prop} we infer that the canonical pair $(p,q)$ is given by 
$(2{\rm Re} \, x, 2 {\rm Re} \, F_x)$.
According to the discussion around
\eqref{eq:symplectic}, 
the symplectic transformation \eqref{eq:em-duality-pq} of the
canonical pair $({\rm Re} \, x,{\rm Re} \, F_x)$ induces a transformation of the vector
$(x,F_x)$ given by
 $(x, F_x ) \rightarrow (d \, x - c \, F_x, a \, F_x - b \, x)$.  Since \eqref{eq:em-duality-pq} together with \eqref{eq:transf-S}
constitutes a symmetry of the model, the transformation of $F_x$ must be induced by the transformation of $x$ and $S$ upon substitution.  We leave it to the reader to verify this.
\\[1mm]
{\sf Exercise 10: Show that the transformation of $x$ and $S$ (given in \eqref{eq:em-duality-pq} and
 \eqref{eq:transf-S}, respectively)
induces the transformation
$F_x\rightarrow  a \, F_x - b \, x$ by substituting $x$ and $S$ with $\tilde x$ and $\tilde S$ in $F_x$.}
\vskip 1mm
The reduced Lagrangian \eqref{eq:red-lag-S-bS} describes the system on an $r$-slice $4 \pi r^2 =1$.
Another background leading to a similar reduced Lagrangian, and hence to a similar
description in terms of a function $F$, is provided by an $AdS_2 \times S^2$ spacetime.
\\[1mm]
{\sf Exercise 11:
Consider the Born-Infeld-dilaton-axion system in 
an $AdS_2 \times S^2$ background 
and show
that, after performing
a suitable rescaling of $g, e$ and $p$,
the resulting reduced Lagrangian is again encoded in \eqref{eq:def-Omega-dil}.}
\vskip 1mm

\subsubsection{Towards $N=2$ supergravity models}

In the Born-Infeld example discussed above, the duality symmetry of the model was enlarged by coupling it
to an additional complex scalar field $S$.  This feature is not an accident. 
In the context of $N=2$ supersymmetric models, it is well known
that the presence of complex scalar fields is crucial in order for the model
to have duality symmetries.  To explore this in more detail, let us broaden the discussion
and consider functions $F$ that depend on three complex scalar fields $Y^I$ (with $I = 0, 1 ,2$), 
as well as on an external
parameter $\eta$.  They will have the form
\begin{align}
F(Y, {\bar Y}; \eta) = - \ft12 \,  \frac{Y^1 (Y^2)^2}{Y^0} + 2 \mathrm{i} \Omega(Y,{\bar Y}; \eta) \;.
\label{eq:F-Y-eta}
\end{align}
The function $F$ describing the Born-Infeld-dilaton-axion system, given in \eqref{eq:F-S-x}, 
 is a special case of \eqref{eq:F-Y-eta}.  It is obtained by performing the identification
$S = -\mathrm{i} \, Y^1/Y^0$, $x =Y^2$ and $\eta = g^2$.  This identification is consistent with the
scaling properties of $x, S$ and $g^2$ 
discussed below \eqref{eq:transf-S}.  Namely, by assigning the uniform scaling behavior 
$Y^I \rightarrow \lambda \, Y^I$ to the $Y^I$, we reproduce the scalings of $x, S$ and $g^2$.
The function \eqref{eq:F-Y-eta} may, however, also describe other models, such as 
genuine $N=2$ supergravity models and should thus be viewed
in a broader context.  Depending on the chosen context, the external parameter $\eta$ will
have a different interpretation. Observe that in the description \eqref{eq:F-Y-eta} based on the $Y^I$, duality
transformations are represented by ${\rm Sp}(6, \mathbb{R})$ matrices (which are $6 \times 6$ matrices of the form
\eqref{eq:matrix-sympl})
acting on $(Y^I, F_I)$, 
where $F_I = \partial F(Y, \bar Y;\eta) / \partial Y^I$. The external parameter $\eta$ is inert under these transformations.

Let us now assume that 
a model based on \eqref{eq:F-Y-eta} has a symmetry associated with a subgroup of 
${\rm Sp}(6, \mathbb{R})$.  This will impose restrictions on
the form of $\Omega$ \cite{LopesCardoso:1999ur,Cardoso:2008fr}.
For concreteness, we take the symmetry to be an 
${\rm SL}(2, \mathbb{R}) \times {\rm SL}(2, \mathbb{R})$
subgroup. 
The first ${\rm SL}(2, \mathbb{R})$ subgroup acts as follows on 
$(Y^I, F_I)$, 
\begin{equation}
        \begin{array}{rcl}
      Y^0 &\to& d \, Y^0 + c\, Y^1 \;,\\
      Y^1 &\to& a \, Y^1 + b \, Y^0 \;,\\
      Y^2 &\to& d\, Y^2 - c \,F_2  \;,
    \end{array}
    \quad
    \begin{array}{rcl}
      F_0 &\to& a\,  F_0 -b\,F_1 \;, \\
      F_1 &\to& d \,F_1 -c\, F_0 \;,\\
      F_2 &\to& a \, F_2 -b\,  Y^2 \;,
    \end{array}
    \label{eq:electro-magn-dual}
\end{equation}
where $a, b, c, d$ are real parameters that satisfy $a d - b c = 1$. 
This symmetry is referred to as S-duality. Let us describe its action on two complex scalar fields
$S$ and $T$ that are given by the scale invariant combinations
$S = - \mathrm{i} Y^1/Y^0$ and $T = - \mathrm{i} Y^2/Y^0$. The field $S$ is the one we encountered
above. The S-duality transformation \eqref{eq:electro-magn-dual} acts as 
\begin{equation}
S \rightarrow \frac{ a S - \mathrm{i} b }{ \mathrm{i} c S + d} \;\;\;,\;\;\;
T \rightarrow T + \frac{ 2 \mathrm{i} \, c }{\Delta_{\rm S} \, (Y^0)^2} \, \frac{\partial \Omega}{\partial T} 
\;\;\;,\;\;\; Y^0 \rightarrow \Delta_{\rm S} \, Y^0\;,
\label{eq:S-dual-T-Y}
\end{equation}
where we view $\Omega$ as function of $S, T, Y^0$ and their complex conjugates, and where
\begin{equation}
\Delta_{\rm S} = d + \mathrm{i} c \, S \;.
\end{equation}
The second ${\rm SL}(2, \mathbb{R})$ subgroup is referred to as 
T-duality group. Here we focus on the T-duality transformation that, in the absence
of $\Omega$, induces the transformation $T \rightarrow 2/T$.
It is given by the following ${\rm Sp}(6, \mathbb{R})$ transformation, 
\begin{equation}
     \begin{array}{rcl}
      Y^0 &\to& F_1 \;,\\
      Y^1 &\to& - F_0  \;,\\
      Y^2 &\to&  Y^2  \;,
    \end{array}
    \quad
    \begin{array}{rcl}
      F_0 &\to& -Y^1 \;, \\
      F_1 &\to& Y^0 \;, \\
      F_2 &\to& F_2 \;,
    \end{array}
    \label{eq:T-dual-inv}
\end{equation}
and yields
\begin{equation}
  S\rightarrow S+ \frac2{\Delta_{\mathrm{T}}(Y^0)^2}
  \,\left[- Y^0 \frac{\partial\Omega}{\partial Y^0} + 
  T\frac{\partial\Omega}{\partial T} \right]  \;\;\;,\;\;\;
 T \rightarrow \frac{T}{\Delta_{\mathrm{T}}}  \;\;\;,\;\;\;
  Y^0 \rightarrow  \Delta_{\mathrm{T}} \, Y^0 \;,
  \label{eq:ST-T-full}
\end{equation} 
where
\begin{equation}
  \label{eq:Delta-T}
  \Delta_{\mathrm{T}} = \tfrac12 T^2 +\frac{2}{(Y^0)^2}
  \frac{\partial\Omega}{\partial S}\;. 
\end{equation}

As already mentioned below \eqref{eq:S-reexp},
when a symplectic transformation describes a symmetry of the system, a convenient method for verifying
this consists in 
performing the substitution $Y^I \rightarrow \tilde{Y}^I$ in the derivatives $F_I$, and checking that this 
substitution 
correctly induces the symplectic transformation of $F_I$.
This will impose restrictions on
the form of $F$, and hence also on $\Omega$.
Imposing that  S-duality \eqref{eq:electro-magn-dual} constitutes a symmetry of 
the model \eqref{eq:F-Y-eta}
results in the following 
conditions on
the transformation behavior of the derivatives of $\Omega$ \cite{Cardoso:2008fr}, 
\begin{align}
 \left( \frac{\partial \Omega}{\partial T} \right)^\prime_\mathrm{S}
=&\, \frac{\partial \Omega}{\partial T} \;, \nonumber\\
\left( \frac{\partial \Omega}{\partial S} \right)^\prime_\mathrm{S}  
=&\,
\Delta_{\rm S}{}^2 \,
\left( \frac{\partial \Omega}{\partial S} \right) + 
 \frac{\partial\left( \Delta_{\rm S}{}^2 \right)}{\partial S} \left[ - \ft12 Y^0 \, \frac{\partial \Omega}{\partial Y^0}   
 - \frac{\mathrm{i} c}{2 \Delta_{\rm S} \, (Y^0)^2} \left(\frac{\partial \Omega}{\partial T}\right)^2
 \right] \;, \nonumber\\
 \left(Y^0 \, \frac{\partial \Omega}{\partial Y^0} \right)^\prime_\mathrm{S}
=&\, Y^0 \, \frac{\partial \Omega}{\partial Y^0} + \frac{\mathrm{2i} c}{\Delta_{\rm S} \, (Y^0)^2} \left(\frac{\partial \Omega}{\partial T}\right)^2\;,
\label{eq:Oprime-O-rel}
\end{align}
while requiring \eqref{eq:T-dual-inv} to constitute a symmetry imposes the transformation behavior \cite{Cardoso:2008fr}
\begin{align}
  \label{eq:T-invariance}
  \left(\frac{\partial\Omega}{\partial S}\right)^\prime_\mathrm{T} =&\, 
  \frac{\partial\Omega}{\partial S} \;, \nonumber\\
  \left(\frac{\partial\Omega}{\partial T}\right)^\prime_\mathrm{T} =&\,
  \left(\Delta_{\mathrm{T}}  - T^2
  \right) \,\frac{\partial\Omega}{\partial T}
  + T\;Y^0 \frac{\partial\Omega}{\partial Y^0} \;,
  \nonumber\\ 
  \left(Y^0 \frac{\partial\Omega}{\partial Y^0}\right)^\prime_\mathrm{T} =&\,
  Y^0 \frac{\partial\Omega}{\partial Y^0} + 
  \frac{4}{\Delta_{\mathrm{T}}\,(Y^0)^2}
  \,\frac{\partial\Omega}{\partial S} 
  \left[- Y^0 \frac{\partial\Omega}{\partial Y^0} +
  T \frac{\partial\Omega}{\partial T }\right]\;. 
\end{align}
These equations allow for various classes of solutions.  For instance, if we only impose 
S-duality invariance, then an exact solution to the S-duality conditions \eqref{eq:Oprime-O-rel} is
\begin{equation}
\Omega (S, \bar S, Y^0, \bar Y^0; \eta) = \eta \left[ \ln Y^0 + \ln \bar Y^0 + \ln (S + \bar S) \right] \,,
\label{eq:sol-ln-Y0}
\end{equation}
which is invariant under \eqref{eq:S-dual-T-Y}.
If, on the other hand, we impose both S-duality and T-duality invariance, solutions to both
\eqref{eq:Oprime-O-rel} and \eqref{eq:T-invariance} may be constructed iteratively by assuming that $\Omega$ is analytic
in $\eta$ and power expanding in it, so that
\begin{equation}
\Omega (Y, \bar Y; \eta) = \sum_{n=1}^{\infty} {\eta}^{n} \, \Omega^{(n)} (Y, \bar Y) \;.
\label{ex:om-pseries}
\end{equation}
Then,
at order $\eta$, the differential equations \eqref{eq:Oprime-O-rel} reduce to
\begin{align}
 \left( \frac{\partial \Omega^{(1)}}{\partial T} \right)^\prime_\mathrm{S}
=&\, \frac{\partial \Omega^{(1)}}{\partial T} \;, \nonumber\\
\left( \frac{\partial \Omega^{(1)}}{\partial S} \right)^\prime_\mathrm{S} 
=&\,
\Delta_{\rm S}{}^2 \,
\left( \frac{\partial \Omega^{(1)}}{\partial S} \right) + 
 \frac{\partial
 \left( \Delta_{\rm S}{}^2 \right)}{\partial S} \left[ - \ft12 Y^0 \, \frac{\partial \Omega^{(1)}}{\partial Y^0}   \right] \;, \nonumber\\
 \left(Y^0 \, \frac{\partial \Omega^{(1)}}{\partial Y^0} \right)^\prime_\mathrm{S}
=&\, Y^0 \, \frac{\partial \Omega^{(1)}}{\partial Y^0} \;,
\label{eq:diff-Ome-1}
\end{align}
while the differential equations \eqref{eq:T-invariance} reduce to
\begin{align}
  \label{eq:T-invariance-Om1}
  \left(\frac{\partial\Omega^{(1)}}{\partial S}\right)^\prime_\mathrm{T} =&\, 
  \frac{\partial\Omega^{(1)}}{\partial S} \;, \nonumber\\
  \left(\frac{\partial\Omega^{(1)}}{\partial T}\right)^\prime_\mathrm{T} =&\, - \tfrac12  T^2\; \frac{\partial\Omega^{(1)}}{\partial T} + 
     T\;Y^0 \frac{\partial\Omega^{(1)}}{\partial Y^0} \;,
  \nonumber\\ 
  \left(Y^0 \frac{\partial\Omega^{(1)}}{\partial Y^0}\right)^\prime_\mathrm{T} =&\,  
  Y^0 \frac{\partial\Omega^{(1)}}{\partial Y^0} \;.
  \end{align}
Once a solution $\Omega^{(1)}$ to these equations has been found, the full expression  \eqref{ex:om-pseries}
 can be constructed by solving 
\eqref{eq:Oprime-O-rel}  and \eqref{eq:T-invariance}
iteratively starting from 
$\Omega^{(1)}$.

As an application, 
let us return to the Born-Infeld-dilaton-axion model \eqref{eq:def-Omega-dil} which, as we already mentioned, is a model
of the form
\eqref{eq:F-Y-eta} that scales as $F \rightarrow \lambda^2 \, F$ under $Y^I \rightarrow \lambda Y^I$ 
with $\lambda \in \mathbb{R}$ 
(see below \eqref{eq:def-Omega-dil}).
Let us first check that both S- and T-duality constitute invariances of the model. We recall that $x = Y^2$.
The 
S-duality transformation \eqref{eq:electro-magn-dual} precisely induces the transformations 
\eqref{eq:em-duality-pq} and \eqref{eq:transf-S}, since $(p, q) =  (2{\rm Re} \, x, 2 {\rm Re} \, F_x)$.
The T-duality transformation \eqref{eq:T-dual-inv} leaves $(x, F_x)$ invariant.  By expressing $\Omega$
given in 
\eqref{eq:def-Omega-dil} in terms of $S, T$ and $Y^0$ (and their complex conjugates), we see from
\eqref{eq:ST-T-full}
that also $S$ is invariant under this T-duality transformation, since $Y^0 \partial \Omega / \partial Y^0 = T \partial \Omega / \partial T$.  Consequently,
the Hamiltonian \eqref{eq:red-H-S} is also invariant under  \eqref{eq:T-dual-inv}.

Now consider expanding 
\eqref{eq:def-Omega-dil} in powers of $g^2$.  To first order we obtain 
\begin{equation}
\Omega^{(1)} = \ft18 \, |Y^0|^4 \, (S+\bar S)^2 \, |T|^4 \;.
\label{eq:om-1-g2}
\end{equation}
It is invariant under both \eqref{eq:electro-magn-dual} and \eqref{eq:T-dual-inv} to lowest order in $g^2$,
and it is straightforward to check that \eqref{eq:om-1-g2} indeed
satisfies the differential equations \eqref{eq:diff-Ome-1} and \eqref{eq:T-invariance-Om1}.
We note that under the aforementioned scaling $Y^I \rightarrow 
\lambda \, Y^I$, 
$\Omega^{(1)}$ 
scales as $\Omega^{(1)} \rightarrow \lambda^4 \, \Omega^{(1)}$.
This scaling behavior is thus very different from the one encountered in supergravity models, such as those considered in  
\cite{LopesCardoso:1999ur,Cardoso:2008fr}, where the function $F$ scaled homogeneously as
$F \rightarrow \lambda^2 \, F$, but the associated 
$\Omega^{(1)}$ did not scale at all.
This difference is due to the fact that in these models, the external parameter $\eta$ scaled as $\eta \rightarrow
\lambda^2 \, \eta$, while in the Born-Infeld-dilaton-axion model it scales as 
$\eta \rightarrow \lambda^{-2} \, \eta$.  

Thus, we see that the actual solutions to \eqref{eq:Oprime-O-rel}
and \eqref{eq:T-invariance} depend sensitively on the scaling behavior of the $Y^I$ and $\eta$.   For instance, 
the solution \eqref{eq:sol-ln-Y0} does not exhibit a homogeneous scaling behavior under
$Y^I \rightarrow \lambda Y^I$.
In the next subsection, we further analyze some of the consequences of this scaling behavior.


\subsection{Homogeneous $F(x, \bar x; \eta)$ \label{sec-F-hom}}

The theorem in subsection \ref{sec:theor} did not assume any homogeneity properties for $F$.
Here we will look at the case when $F$ is homogeneous of degree two and discuss some of its consequences.
As shown in the previous subsections, an example of a model with this feature is the
Born-Infeld-dilaton-axion system.

Let us consider a function $F(x, \bar x; \eta) = F^{(0)}(x) + 2 \mathrm{i} \Omega (x, \bar x; \eta)$
that depends on a real external parameter $\eta$, and let us discuss its behavior under the scaling
$x \rightarrow \lambda \, x \;,\; \eta \rightarrow \lambda^m \, \eta$ with $\lambda \in \mathbb{R}$.
We take $F^{(0)}(x)$ to be quadratic in $x$, so that $F^{(0)}$ scales as 
$F^{(0)}(\lambda\, x) = \lambda^2  \, F^{(0)}(x)$.  This scaling behavior can be extended to the full
function $F$ if we demand that the canonical pair $(\phi,\pi)$ given in \eqref{eq:theorem-prop}
scales uniformly as
$(\phi,\pi) \rightarrow \lambda \, (\phi,\pi)$.  Then we have
\begin{equation}
F(\lambda\, x, \lambda \, \bar x; \lambda^m \, \eta) = \lambda^2  \, F(x, \bar x; \eta) \;,
\label{eq:resc-F}
\end{equation}
which results in the homogeneity relation
\begin{equation}
2 \,F = x^i \, F_i + {\bar x}^{\bar \imath} \, F_{\bar \imath} + m \, \eta \, F_{\eta} \;,
\label{eq:homog-F}
\end{equation}
where $F_{\eta} = \partial F / \partial \eta$.
Inspection of \eqref{eq:def-x-y}
shows 
that the associated Hamiltonian $H$ scales with weight two as
\begin{equation}
{H} (\lambda \, \phi, \lambda \, \pi; \lambda^m \, \eta) = \lambda^2 \, {H} (\phi,\pi; \eta)\;,
\end{equation}
so that ${H}$ satisfies
the homogeneity relation,
\begin{equation}
2 \, {H} = 
\phi \, \frac{\partial { H}}{\partial \phi} + \pi \, 
\frac{\partial {H}}{\partial \pi}
+ m \, \eta \, \frac{\partial {H}}{\partial \eta} \;.
\label{eq:homog-H}
\end{equation}
Using \eqref{eq:def-x-y} as well as $y_i = F_i $, this can be written as 
\begin{equation}
\label{eq:H-hom-om}
{H} = \mathrm{i} \left( {\bar x}^{\bar \imath} \, F_i - x^i \, {\bar F}_{\bar \imath}
\right)  
+ \frac{m}{2} \, \eta \, \frac{\partial { H}}{\partial \eta} \;.
\end{equation}
Next, using that the dependence on $\eta$ is solely contained in $\Omega$, we obtain
\begin{equation}
\frac{\partial {H}}{\partial \eta}|_{\phi, \pi} = -  \frac{\partial {L}}{\partial \eta}|_{\phi, \dot \phi}
= - 4 \Omega_{\eta} \;,
\end{equation}
where $\Omega_{\eta} = \partial \Omega / \partial \eta$.
Thus, we can express \eqref{eq:H-hom-om} as
\begin{equation}
{H} = \mathrm{i} \left( {\bar x}^{\bar \imath} \, F_i - x^i \, {\bar F}_{\bar \imath}
\right)  
-  2 \, m \,
\eta \, \Omega_{\eta} \;.
\label{eq:hamil-sympl}
\end{equation}
This relation is in accordance with \eqref{eq:H-sympl} upon substitution of the homogeneity relations
$2 F^{(0)}(x) = x^i \, F^{(0)}_i $ and 
$2\,\Omega = x^i\Omega_i + \bar x^{\bar\imath} \Omega_{\bar\imath} + m \, \eta \, \Omega_{\eta} $
that follow from \eqref{eq:homog-F}. 

The Hamiltonian transforms as a function under symplectic transformations.  Since the first term in 
\eqref{eq:hamil-sympl} transforms as a function, it follows that $\Omega_{\eta}$ also transforms as a function.
This is in accordance with the general result quoted at the end of subsection \ref{sec:theor} which states that $\partial_{\eta}
F$ transforms as a function.

An application of the above is provided by the Born-Infeld-dilaton-axion system based on \eqref{eq:def-Omega-dil},
whose function $F$ scales according to \eqref{eq:resc-F} with $m = -2$ (in this example, 
$\eta = g^2$).

In certain situations, such as in the study of 
BPS black holes \cite{LopesCardoso:1998wt}, the discussion needs to be extended
to an external parameter $\eta$ that is complex, so that now we consider a function 
$F(x, \bar x; \eta, \bar \eta) = F^{(0)}(x) + 2 \mathrm{i} \Omega (x, \bar x; \eta, \bar \eta)$
 that scales as follows
(with $\lambda \in \mathbb{R}$),
\begin{equation}
F(\lambda \, x, \lambda \, \bar x; \lambda^m \, \eta, 
\lambda^m \, \bar \eta) = \lambda^2 \, F(x, \bar x; \eta, \bar \eta) \;.
\end{equation}
For instance, in the case of BPS black holes, $\eta$ is identified with $\Upsilon$,
which is complex and
denotes the (rescaled)
lowest component of the square of the Weyl superfield.  The extension to a complex $\eta$
results in the presence of an additional term
on the right hand side of \eqref{eq:homog-F} and \eqref{eq:homog-H},
\begin{align}
2 \, F =&\, x^i \, F_i + {\bar x}^{\bar \imath} \, F_{\bar \imath} + m  \left( \eta \, F_{\eta} + \bar \eta F_{\bar \eta} \right)
 \;, \nonumber\\
2 \, {H} =&\,
\phi \, \frac{\partial {H}}{\partial \phi} + \pi \, 
\frac{\partial { H}}{\partial \pi}
+ m  \left( \eta \, \frac{\partial {H}}{\partial \eta} + \bar \eta \, \frac{\partial {H}}{\partial \bar
\eta}
\right)\;,
\label{eq:homog-ceta}
\end{align}
and hence
\begin{equation}
{H} = \mathrm{i} \left( {\bar x}^{\bar \imath} \, F_i - x^i \, {\bar F}_{\bar \imath}
\right)  
+ \frac{m}{2} \left( \eta \, \frac{\partial {H}}{\partial \eta} + 
\bar \eta \, \frac{\partial {H}}{\partial \bar
\eta}
\right)\;.
\end{equation}
Then, since the dependence on $\eta$ and $\bar \eta$ is solely contained in $\Omega$, 
we obtain
\begin{equation}
{H} = \mathrm{i} \left( {\bar x}^{\bar \imath} \, F_i - x^i\, {\bar F}_{\bar \imath}
\right)  
-  2 \, m \left( 
\eta \, \Omega_{\eta} + \bar \eta \, \Omega_{\bar \eta} \right)\;.
\end{equation}
This is in accordance with \eqref{eq:H-sympl} upon substitution of the homogeneity relations
$2 F^{(0)}(x) = x^i \, F^{(0)}_i $ and 
$2\,\Omega = x^i\Omega_i + \bar x^{\bar\imath} \Omega_{\bar\imath} + m \, 
(\eta \, \Omega_{\eta} +  \bar \eta \, \Omega_{\bar \eta} )$
that follow from \eqref{eq:homog-ceta}.
The case of BPS black holes mentioned above corresponds to $m=2$ \cite{LopesCardoso:2006bg,Cardoso:2008fr}.

The above extends straightforwardly to the case of multiple real external parameters.

\section{Lecture II: Duality covariant complex variables \label{sec:new-var}}

As already discussed, the function $F(x, \bar x)$ may depend on a number of external parameters
$\eta$.  Under duality transformations \eqref{eq:symplectic}, 
the symplectic vector  $(x^i, F_i(x, \bar x))$  transforms into
 $(\tilde x^i, \tilde F_i (\tilde x, \bar{\tilde x}))$, while the parameters $\eta$ are inert. When expressing the
 transformed variables $\tilde x^i$ in  terms of the original $x^i$, 
 the relation will depend on $\eta$, i.e. $\tilde{x}^i = 
 \tilde{x}^i(x, \bar x, \eta)$.
 In this section 
we introduce duality covariant complex variables $t^i$ whose duality transformation law is independent of $\eta$.
These variables constitute a complexification of the canonical variables of the Hamiltonian and ensure
that when expanding the Hamiltonian 
in powers of the external parameters, the resulting expansion coefficients transform covariantly under duality
transformations. This expansion can also be organized by employing a suitable covariant derivative, which we 
construct.
The covariant variables introduced here have the same duality transformation properties
as the ones used
in topological string theory and can therefore be identified with the latter.

We begin by writing the Hamiltonian ${ H}$ given in \eqref{eq:H-sympl} in the form
\begin{align}
  \label{eq:Ham-sympl}
   {H} =
  -\mathrm{i}(x^i \,\bar F^{(0)}_{\bar\imath} -\bar x^{\bar\imath} \,F^{(0)}_i)
  -4\,\mathrm{Im} 
  [F^{(0)}-\tfrac12 x^i\,F^{(0)}_i] -2\left[ 2\,\Omega
  -(x^i - {\bar x}^{\bar\imath}) (\Omega_i - \Omega_{\bar\imath}) \right] \;,
\end{align}
where we made use of \eqref{eq:F(x)}.  
We take $\Omega(x, \bar x; \eta)$ to depend on a single real parameter $\eta$ that is inert under
symplectic transformations.
The discussion given below can be extended to the case of
multiple real external parameters in a straightforward manner.
For later convenience, we introduce the notation $\Omega_{\eta} = \partial \Omega / \partial \eta, F_{\eta j} = \partial^2 F/\partial \eta
\partial x^j$, etc.

The Hamiltonian \eqref{eq:Ham-sympl} is given in terms of complex fields $x^i$ and $\bar x^{\bar \imath}$ 
whose transformation law under duality depends on the external parameter $\eta$.  Now we define
complex variables $t^i$ whose transformation law does not depend on $\eta$,
as follows.  We introduce the complex vector $(t^i, F_i^{(0)} (t))$
and equate its real part with the vector comprising the canonical variables $(\phi^i, \pi_i)$ \cite{Cardoso:2010gc},
\begin{align}
  \label{eq:canon-t-rel}
 2 {\Re} \, t^i = &\ \phi^i \;, \nonumber\\
 2 {\Re} \,   F_i^{(0)} (t)  = &\, \pi_i \;.
  \end{align}
This definition ensures that the vector $(t^i, F_i^{(0)} (t))$ describes a complexification of 
$(\phi^i, \pi_i)$ that transforms in the same way as $(\phi^i, \pi_i)$ 
under duality transformations, namely as in 
\eqref{eq:symplectic-pq}. This yields the transformation law
  \begin{equation}
    \tilde{t}^i= U^i{}_j\,t^j + Z^{ij}  F^{(0)}_j(t) \;,
    \end{equation}
which, differently from the one for the $\tilde x^i$, is independent of $\eta$.  

Using \eqref{eq:theorem-prop}, the new variables $t^i$ are related to the 
$x^i$ by
\begin{align}
  \label{eq:x-t-rel}
  2 {\Re} \, t^i    =&\, 2 {\Re} \, x^i  \;, \nonumber\\
  2 {\Re} \,   F_i^{(0)} (t) =&\,  2 {\Re} \, F_i(x, \bar x; \eta)  \;.
  \end{align}

Now we consider the series expansion of ${H}$ in powers of $\eta$.
If the expansion is performed
keeping $x^i$ and $x^{\bar\imath}$ fixed, the resulting coefficients functions in 
the expansion do not have a nice behavior under
sympletic transformations because of the aforementioned dependence of $\tilde x^i$ on $\eta$.
This implies that the coefficient functions at a given order in $\eta$ will transform into
coefficient functions at higher order.  This can be avoided by performing an 
expansion in powers of $\eta$ 
keeping $t^i$ and $t^{\bar\imath}$ fixed instead.  We obtain
\begin{equation}
{H} =  \sum_{n=0}^{\infty} \frac{\eta^n}{n!} \; f^{(n)}(t, \bar t) \;,
\label{eq:exp-ham-newv}
  \end{equation}
where the coefficient functions
\begin{equation}
f^{(n)} =  \partial^{^n}_\eta{H}(t, {\bar t};\eta)\Big\vert_{\eta=0} 
\label{eq:coef-f}
\end{equation}
transform as functions under symplectic transformations, 
i.e. ${\tilde f}^{(n)}(\tilde t, \bar{\tilde t}) = f^{(n)} (t, \bar t)$.  Viewing them as 
as functions of ${\Re} \, t^i$ and of ${\Re} \,   F_i^{(0)} (t)$, we can re-express them in terms
of $x^i$ and ${\bar x}^{\bar \imath}$ using  \eqref{eq:x-t-rel}, as follows.
First we introduce 
a modified derivative ${\cal D}_{\eta}$ \cite{deWit:1996ix,deWit:1996ag} that has the feature that it
annihilates the canonical variables $(\phi^i, \pi_i)$, so that
\begin{equation}
{\cal D}_{\eta} \left( {\Re} \, x^i \right) = 0 \;\;\;,\;\;\; {\cal D}_{\eta} \left({\Re} \, F_i \right)=0 \;.
\label{eq:D-rex-ref}
\end{equation}
We then use ${\cal D}_{\eta}$ to expand ${H}$ 
in powers of $\eta$ while keeping
${\Re} \, x^i$ and ${\Re} \, F_i$ fixed,
\begin{equation}
{H} =  \sum_{n=0}^{\infty} \frac{\eta^n}{n!} \; {H}^{(n)} \;,
\label{eq:exp-hamil-cov}
  \end{equation}
  where the coefficient functions are given by
\begin{equation}
{H}^{(n)} = \mathcal{D}^{^n}_\eta {H}(x,\bar x;\eta)\Big\vert_{\eta=0}\,.
\label{eq:coef-H}
\end{equation}
By comparing  \eqref{eq:exp-ham-newv} with
 \eqref{eq:exp-hamil-cov}, we conclude that $f^{(n)} = { H}^{(n)}$, so that 
 the symplectic coefficient functions $f^{(n)}$ can be
 expressed as
 \begin{equation}
  \label{eq:dn-H--DF}
f^{(n)} = \partial^{^n}_\eta{H}(t, {\bar t};\eta)\Big\vert_{\eta=0} = \mathcal{D}^{^n}_\eta {H}(x,\bar x;\eta)\Big\vert_{\eta=0}\,.
\end{equation}
The modified derivative ${\cal D}_{\eta}$ used in the expansion is given by 
\begin{equation}
{\cal D}_{\eta} = \partial_{\eta} + \mathrm{i} \, {\hat N}^{ij} \left( F_{\eta j} + {\bar F}_{\eta \bar\jmath} \right) \left(\partial_i
- \partial_{\bar\imath} \right) \;,
\label{eq:cov-der-multiple}
\end{equation}
where ${\hat N}^{ij}$ denotes the inverse of 
\begin{equation}
\hat{N}_{ij} = - \mathrm{i} \left[ F_{ij} - {\bar F}_{\bar\imath \bar\jmath} - F_{i \bar\jmath} + {\bar F}_{\bar\imath j} \right] \;.
\label{eq:hat-N-ij}
\end{equation}
Using  \eqref{eq:F(x)}, the above can also be written as
\begin{equation}
{\cal D}_{\eta} = \partial_{\eta} - 2 \, {\hat N}^{ij} \left( \Omega_{\eta j} - {\Omega}_{\eta \bar\jmath} \right) \left(\partial_i
- \partial_{\bar\imath} \right) \;,
\label{eq:cov-der-multiple-Om}
\end{equation}
with
\begin{align}
\hat{N}_{ij} =&\, N_{ij} + 4 {\rm Re} \left(\Omega_{ij} - \Omega_{i \bar \jmath} \right) \;, 
\nonumber\\
N_{ij} =&\,
- \mathrm{i} \left[ F^{(0)}_{ij} - {\bar F}^{(0)}_{\bar\imath \bar\jmath} \right] \;.
\label{eq:hat-N-N}
\end{align}
Observe that $\hat{N}_{ij}$ is a real symmetric matrix.
\\[1mm]
{\sf Exercise 12: Verify \eqref{eq:D-rex-ref}.}
\vskip 1mm

We now give the first few terms in the expansion of $H$.
We choose to evaluate them using \eqref{eq:coef-H}.
Expanding $\Omega$ in a power series\footnote{Note that here we have
chosen a different normalization for the $\Omega^{(n)}$ compared to the one in \eqref{ex:om-pseries}.} in $\eta$,
\begin{equation}
\Omega(x, \bar x; \eta) = \sum_{n=1}^{\infty} \frac{\eta^n}{n!} \, \Omega^{(n)} (x, \bar x) \;,
\label{eq:new-nor-om}
\end{equation}
we obtain
\begin{align}
\label{eq:H-exp-comb}
  f^{(0)}=&\,
  -\mathrm{i}(x^i \,\bar F^{(0)}_{\bar\imath} -\bar x^{\bar\imath} \,F^{(0)}_i)
  -4\,\mathrm{Im} 
  [F^{(0)}-\tfrac12 x^i\,F^{(0)}_i] \;, \nonumber\\
  f^{(1)}=&\, - 4 \, \Omega^{(1)} \;, \nonumber\\
   f^{(2)} =&\, - 4 \left[ \Omega^{(2)} - 2 N^{ ij} \left(\Omega_i^{(1)} - 
  \Omega_{\bar\imath}^{(1)} \right) \left(\Omega_j^{(1)} - 
  \Omega_{\bar\jmath}^{(1)} \right) 
\right] \;,
  \nonumber\\
   f^{(3)} =&\, - 4 \left[ \Omega^{(3)} 
  - 6  N^{ ij} \left(\Omega^{(2)}_i - \Omega^{(2)}_{\bar\imath} \right)
 \left(\Omega^{(1)}_j- \Omega^{(1)}_{\bar\jmath} \right)  \right. \\
 &\,+12 N^{ik} N^{jl} \left(\Omega^{(1)}_{ij} - \Omega^{(1)}_{i \bar\jmath} + 
 {\rm c.c.} \right)  \left(\Omega^{(1)}_k- \Omega^{(1)}_{\bar k} \right)
  \left(\Omega^{(1)}_l- \Omega^{(1)}_{\bar l} \right)
 \nonumber\\
 &\, \left. + 4 \mathrm{i}\,
  N^{ip} N^{jl} N^{km} 
 \left(\Omega^{(1)}_i- \Omega^{(1)}_{\bar\imath} \right)
  \left(\Omega^{(1)}_j- \Omega^{(1)}_{\bar\jmath} \right)
  \left(\Omega^{(1)}_k- \Omega^{(1)}_{\bar k} \right)
  \left(F^{(0)}_{plm} + {\bar F}^{(0)}_{\bar p \bar l \bar m} \right)
  \right] \;.\nonumber
\end{align}
Observe that at any given order in $\eta$, there is no distinction between $x^i$ and $t^i$, so that
in \eqref{eq:H-exp-comb} we may replace $x^i$ everywhere by $t^i$.

The expansion \eqref{eq:exp-hamil-cov} yields expansion functions that are symplectic functions.  This implies that 
${\cal D}_{\eta}$ acts as a covariant derivative for symplectic transformations.  This can
be verified explicitely and is done in appendix \ref{cov-der}, where we show that
if a quantity $G(x, \bar  x; \eta)$ transforms as a function under symplectic transformations,
then so does ${\cal D}_{\eta} G$.
In  particular, applying ${\cal D}_{\eta}$ to ${ H}$ yields the relation
\begin{equation}
  \label{eq:d-H--DF}
  \partial_\eta{ H}(t, {\bar t};\eta) = \mathcal{D}_\eta {H}(x,\bar x;\eta)\,,
\end{equation}
where the right-hand side defines a symplectic function. 
More generally, applying multiple derivatives $\mathcal{D}_\eta^{^n}$ on
any symplectic function depending on $x^i$ and $\bar x^{\bar \imath}$, will
again yield a symplectic function. 
As an example, consider applying 
$\mathcal{D}_\eta$ and $\mathcal{D}_\eta^{^2}$ on \eqref{eq:Ham-sympl},
\begin{align}
  \label{eq:D-eta-H}
  \mathcal{D}_\eta{} {H} (x, \bar x; \eta)=&\, -4\,\partial_\eta \Omega(x,\bar
  x;\eta) \,, \nonumber\\
    \mathcal{D}_\eta^{^2} {H} (x, \bar x; \eta)
  =&\,- 4 \left[  \partial_\eta^{^2}\Omega -2\,\hat N^{ij} 
   \partial_\eta\omega_i
  \,\partial_\eta\omega_j
    \right]\,, 
\end{align}
where $\omega_i= \Omega_i-\Omega_{\bar \imath}$. 
According to the above, 
both these expressions transform as
functions under symplectic transformations. For the first expression
this is confirmed by the result \eqref{eq:trafo-F-eta-tilde}
which shows that $\partial_\eta\Omega$ transforms as a
function.
The second expression shows that, while
$\partial_\eta^{^2}\Omega$ does not transform as a function, there exists
a modification that can be included such that the result does again
transform as a function. Expressions like these were derived earlier
in a holomorphic setup
\cite{deWit:1996ix,deWit:1996ag}. 
Furthermore, we note that the 
differential operators $\mathcal{D}^i$, defined by
\begin{equation}
  \label{eq:D-I}
  \mathcal{D}^i = \hat N^{ij} \left( \frac{\partial}{\partial x^j} - 
    \frac{\partial}{\partial {\bar x}^{\bar \jmath}} \right) \,,
\end{equation}
are mutually commuting, and they also commute with $\mathcal{D}_\eta$,
\begin{equation}
  \label{eq:commuting}
  {}[\mathcal{D}^i,\mathcal{D}^j]= [\mathcal{D}^i, \mathcal{D}_\eta] =
  0 \,. 
\end{equation}

\noindent
{\sf Exercise 13: Verify \eqref{eq:commuting}.}
\\[1mm]
As already mentioned, 
it is possible to extend the above to the case of several independent real parameters 
$\eta, \eta', \eta'',\ldots$. In that case the additional operators,
$\mathcal{D}_{\eta'}$, etc., will also commute with the operators
considered in (\ref{eq:commuting}).

Obviously, when imposing the restriction $\eta =0$ on
the functions ${\cal D}_{\eta}^{^n} {H}$,  they
reduce to the expressions for the $f^{(n)}$ obtained in \eqref{eq:H-exp-comb}. This can be explicitly verified for the functions given in \eqref{eq:D-eta-H} by comparing them to
the expressions in \eqref{eq:H-exp-comb}.

Let us return to the relation \eqref{eq:canon-t-rel} and 
discuss it in the light of phase space variables.
As mentioned in subsection \ref{sec:theor},
we view $(\phi^i, \pi_i)$ as coordinates on a classical phase space equipped with the symplectic form
$d \pi_i \wedge d \phi^i$.  
Let us express the symplectic form in terms of the $t^i$ using \eqref{eq:canon-t-rel}, 
\begin{equation}
d \pi_i \wedge d\phi^i = \mathrm{i} \, N_{ij} \, dt^i \wedge d {\bar t}^{\bar \jmath} 
\;,
\end{equation}
with $N_{ij}$ given in \eqref{eq:hat-N-N}.  This relation may be interpreted as a canonical transformation
from variables $(\phi^i, \pi_i)$ to $(t^i, \bar t^{\bar \imath})$ which is generated by a function $S$
that depends on
half of all the coordinates.  We take $S$ to depend on $\phi^i$ and $t^i$.  We determine it 
in the linearized approximation by expanding
$N_{ij}$ around 
a background value $t^i_B$.  Performing the shift
\begin{equation}
 t^i \rightarrow t^i_B + t^i \;\;\;,\;\;\; {\bar t}^{\bar \imath} \rightarrow {\bar t}^{\bar \imath}_B 
 + {\bar t}^{\bar \imath} \;,
\end{equation}
and keeping only terms linear in the fluctuations $t^i$ and ${\bar t}^{\bar \imath}$, we obtain from 
\eqref{eq:canon-t-rel},
\begin{align}
\phi^i =&\,
t^i + {\bar t}^{\bar \imath }  \;\;\;, \nonumber\\
\pi_i =&\, F_{ij}^{(0)} (t_B) \, t^j  + {\bar F}_{\bar \imath \bar \jmath}^{(0)}({\bar t_B}) 
\, {\bar t}^{\bar \jmath}  \;\;\;,
\label{eq:potentials-eqs-back}
\end{align}
where we absorbed the fluctuation independent pieces into the
definition of $(\phi^i, \pi_i)$. Then, expressing $\pi_i$ in terms of $t^i$ and $\phi^i$, 
\begin{equation}
\pi_i = \mathrm{i} N_{ij} (t_B, {\bar t}_B) \, t^j  + {\bar F}_{\bar \imath \bar \jmath}^{(0)}({\bar t_B}) 
\, \phi^j  \;,
\label{eq:can-m-p-P}
\end{equation}
and introducing the combination
\begin{equation}
P_i = - \mathrm{i} \, N_{ij}(t_B, \bar{t}_B) \left(\phi^j - t^j \right) \;,
\end{equation}
yields
\begin{equation}
d \pi_i \wedge d\phi^i = \mathrm{i} \, N_{ij}(t_B, {\bar t}_B) \, dt^i \wedge d {\bar t}^{\bar \jmath} = d P_i \wedge d t^i \;.
\end{equation}
Hence, the 1-form $ \pi_i \, d\phi^i - P_i  \, d t^i$ is closed, so that locally, 
\begin{equation}
\pi_i \, d \phi^i - P_i \, dt^i = d S \;,
\end{equation}
where $S(\phi, t)$ is called the generating function of the canonical transformation.
Then, integrating this relation yields the following expression for the 
generating function $S(\phi, t; t_B, \bar{t}_B)$ \cite{Dijkgraaf:2002ac,Verlinde:2004ck,Aganagic:2006wq},
\begin{eqnarray}
 S(\phi, t; t_B, \bar{t}_B)  = \tfrac12 {\bar F}^{(0)}_{\bar \imath \bar \jmath}(\bar{t}_B) \, \phi^i \phi^j 
 + \mathrm{i} \, N_{ij} (t_B, \bar{t}_B) \,  \phi^i t^j - \tfrac12 \mathrm{i} \, N_{ij}(t_B, \bar{t}_B) \,  t^i t^j + c(t_B, \bar{t}_B) \;,
 \label{eq:Sf-back}
\end{eqnarray}
where $c$ denotes a background dependent integration constant.
Observe that $S(\phi,t; t_B, \bar{t}_B)$ is holomorphic in the fluctuation $t$ and non-holomorphic in the background
$t_B$.  The generating function $S(\phi, t; t_B, \bar{t}_B)$ plays a crucial role in the wave function approach to perturbative topological string theory.  This approach represents a concise framework 
\cite{Witten:1993ed,Dijkgraaf:2002ac,Verlinde:2004ck,Aganagic:2006wq,Gunaydin:2006bz} for deriving the 
holomorphic anomaly equation
of topological string theory \cite{Bershadsky:1993ta,Bershadsky:1993cx}, 
and will be reviewed in appendix \ref{sec:top-string}.

\section{Lecture III: The Hesse potential and the topological string }
\label{sec:hesse-top}

In the previous sections we showed that the dynamics of a general class of Lagrangians is encoded
in a non-holomorphic function $F$ of the form given in \eqref{eq:F-0-Om}.  
This function $F$
may depend on a number of external parameters $\eta$.  We expressed the associated 
Hamiltonian in terms of duality covariant complex variables and showed that in these variables, the expansion of 
the Hamiltonian in a power series in $\eta$ yields expansion coefficients that transform as functions under duality.
In this section we apply these techniques
to
supergravity models in the presence of higher-curvature interactions
encoded in the square of the Weyl superfield \cite{Bergshoeff:1980is,deWit:1996ix}. 
We consider these models in an 
$AdS_2 \times S^2$ background.  
The Hamiltonian \eqref{eq:H-sympl} associated to the reduced
Lagrangian is a (generalized) Hesse potential. The Hesse potential plays a central role in the formulation of special geometry in terms
of real variables \cite{Freed:1997dp,Hitchin:1999,Alek:1999,Cortes:2001}.
The external parameter $\eta$, which is now complex, is identified with 
the lowest component field of the square of the Weyl superfield.

We begin by reviewing the computation of the Wilsonian effective Lagrangian in an $AdS_2 \times S^2$ background 
\cite{Sahoo:2006rp,Cardoso:2006xz}
and
relate it to the presentation of section \ref{sec:ubiquity}.
We then generalize the discussion to the case of a function $F$ of type \eqref{eq:F(x)}
with a non-harmonic $\Omega$. We express the Hesse potential in terms of the aforementioned duality covariant complex variables,
and expand it in powers of $\eta$ and $\bar \eta$.
This reveals 
a systematic structure. 
Namely, the Hesse potential decomposes into two classes of terms.
One class consists of combinations of terms, constructed
out of derivatives of $\Omega$,
that transform as functions under electric/magnetic duality. 
The
other class is constructed out of $\Omega$ and derivatives thereof.  Demanding this second 
class to also exhibit a proper behavior under duality transformations (as  a consequence of
the transformation behavior of the Hesse potential)
imposes 
restrictions on $\Omega$.  These restrictions are captured by a differential equation that 
equals 
half of the holomorphic anomaly equation encountered in perturbative topological string theory.

\subsection{The reduced Wilsonian Lagrangian in an $AdS_2 \times S^2$ background}

We consider the coupling of $N=2$ vector multiplets 
to 
$N=2$ supergravity in the presence of higher-curvature interactions
encoded in the square of the Weyl superfield \cite{Bergshoeff:1980is,deWit:1996ix}.  We use the conventions of $N=2$ supergravity, 
whereby the vector multiplets are labelled by a capital index $I = 0, \dots, n$
(instead of the index $i$
used in the previous sections). The degrees of freedom of a vector multiplet include
an abelian gauge field and a complex scalar field, and these will thus carry an index $I$.  We denote the 
complex scalar fields
by $X^I$.  The square of the Weyl superfield has various component fields.
The highest component field contains the square of the anti-selfdual components of the Riemann tensor, while  
the lowest one, denoted by $\hat A$, equals the square of an anti-selfdual tensor field.
Below we will find it convenient to work with rescaled complex fields $Y^I$ and $\Upsilon$, which
are related to the $X^I$ and ${\hat A}$ by a complex rescaling \cite{LopesCardoso:1998wt}.

First we evaluate the Wilsonian effective Lagrangian of these models
on a field configuration consistent
with the ${\rm SO}(2,1) \times
{\rm SO}(3)$ isometry of an $AdS_2 \times S^2$ background.  The spacetime metric $g_{\mu \nu}$ and the
field strengths $F_{\mu\nu}{}^I$ of the abelian gauge fields are given by
\begin{eqnarray}
  \label{eq:general-fields}
  &&
  \mathrm{d}s^2{} = 
   v_1\Big(-r^2\,\mathrm{d}t^2 + \frac{\de r^2}{r^2}\Big) 
  + v_2 \Big(\de \theta^2 +\sin^2\theta \,\de\varphi^2\Big)\,,
  \nonumber\\ 
  &&
  F_{rt}{}^I = e^I\,,\qquad F_{\theta\varphi}{}^I = p^I\,
  \sin\theta \,.
\end{eqnarray}
The $\theta$-dependence of
$F_{\theta\varphi}{}^I$ is fixed by rotational invariance and the
$p^I$ denote the magnetic charges. 
The quantities $v_1, v_2, e^I$ and $p^I$ are all constant by virtue of the ${\rm SO}(2,1) \times
{\rm SO}(3)$ symmetry.

It is well-known \cite{Bergshoeff:1980is}
that the Wilsonian Lagrangian $\mathcal{L}$
is encoded in a holomorphic function
$F(X, \hat A)$, which is homogeneous of degree two under the scaling discussed in \eqref{eq:resc-F}, i.e. 
$F(\lambda X, \lambda^2 \hat A) = \lambda^2 \, F(X, \hat A)$. 
Evaluating the Wilsonian Lagrangian in the background \eqref{eq:general-fields}
and integrating over  $S^2$ \cite{Sen:2005wa},
\begin{equation}
  \label{eq:reduced-action}
  \mathcal{F} = \int \mathrm{d}\theta\,\mathrm{d}\varphi\;
  \sqrt{\vert g\vert} \, \mathcal{L} \,, 
\end{equation}
yields the reduced Wilsonian Lagrangian which depends on $e^I$ and $p^I$, on the rescaled fields
$Y^I$ and $\Upsilon$, and on $v_1$ and $v_2$ through the ratio $U = v_1/v_2$.

In the following, 
we will restrict to supersymmetric backgrounds, for simplicity, in which case $U=1$ and $\Upsilon = - 64$
\cite{LopesCardoso:1998wt}. Then, the reduced Wilsonian Lagrangian reads \cite{Sahoo:2006rp,Cardoso:2006xz},
\begin{align}
\mathcal{F} (e, p, Y, \bar Y; \Upsilon , \bar \Upsilon) =&\, - \tfrac18 \mathrm{i} \left( F_{IJ} - {\bar F}_{\bar I \bar J} \right)  \left(e^I e^J - p^I p^J \right)
- \tfrac14 \left(F_{IJ} + {\bar F}_{\bar I \bar J} \right) e^I p^J \nonumber\\
&\, + \tfrac12 \mathrm{i} e^I \left( F_I + F_{IJ} {\bar Y}^{\bar J} - {\rm h.c.} \right)
- \tfrac12 p^I \left(F_I - F_{IJ} {\bar Y}^{\bar J} + {\rm h.c.} \right) 
\nonumber\\
&\, + \mathrm{i} \left( F - Y^I F_I + \tfrac12 {\bar F}_{\bar I \bar J} Y^I Y^J - {\rm h.c.} \right) \;,
\end{align}
where $\Upsilon = \bar \Upsilon = - 64$ and
$F_I = \partial F / \partial Y^I \,,\, F_{IJ} = \partial^2 F / \partial Y^I \partial Y^J $, etc.
Introducing the complex scalar fields $x^I = \tfrac12 (p^I + \mathrm{i} e^I )$ of section \ref{sec:born-infeld-reduced} (see \eqref{eq:x-Fx}),
the reduced Lagrangian becomes a function of two types of complex scalar fields, namely the $x^I$ that incorporate
the electromagnetic information, and the moduli fields $Y^I$.

Now we recall that in an $AdS_2 \times S^2$ background the electro/magnetic quantities appearing in 
\eqref{eq:general-fields}
are related to the 
moduli fields $Y^I$.  When the background is supersymmetric, the relation takes the form
\cite{LopesCardoso:2000qm}
\begin{equation}
x^I = \mathrm{i} \bar{Y}^I \;.
\label{eq:x-Y}
\end{equation}
In the context of BPS black holes, 
the real part of this equation yields the magnetic attractor equation.
Then, using \eqref{eq:x-Y}, the reduced Wilsonian Lagrangian becomes equal to
\begin{equation}
{\cal F} (Y, \bar Y; \Upsilon, \bar \Upsilon) = - 2 \, {\rm Im} F(Y, \Upsilon) \;,
\label{eq:wilson-F}
\end{equation}
with $\Upsilon =  \bar \Upsilon = - 64$.
\\[1mm]
{\sf Exercise 14: Verify \eqref{eq:wilson-F}.}
\vskip 1mm

Let us reformulate the reduced Lagrangian \eqref{eq:wilson-F}, which is based on a holomorphic
functions $F(Y, \Upsilon)$,  in terms of the function $F(Y, \bar Y; \Upsilon, \bar \Upsilon) = F^{(0)}(Y) + 
2 \mathrm{i} \Omega(Y, \bar Y; \Upsilon, \bar \Upsilon)$ introduced in section \ref{sec:ubiquity}.
This is achieved by using the equivalence transformation
\eqref{eq:ambiguity}.  Writing the holomorphic function  $F(Y, \Upsilon)$ as  $F(Y, \Upsilon) = F^{(0)}(Y)
- g(Y, \Upsilon)$ and applying \eqref{eq:ambiguity}, we obtain $\Omega = - {\rm Im} \, g(Y, \Upsilon)$.
Thus, at the Wilsonian level, $\Omega$ is a harmonic function, and the reduced Lagrangian can be expressed as
\begin{align}
{\cal F} (Y, \bar Y; \Upsilon, \bar \Upsilon) =&\, - 2 \, \left[ {\rm Im} F^{(0)}(Y) +  
\Omega(Y, \bar Y; \Upsilon, \bar \Upsilon) 
\right] \nonumber\\
=&\,
- 2 \, \left[ {\rm Im} F(Y, \bar Y; \Upsilon, \bar \Upsilon) - \Omega(Y, \bar Y; \Upsilon, \bar \Upsilon) 
\right] \;,
\label{eq:wilson-Om}
\end{align}
with $\Upsilon = \bar \Upsilon = -64$.
Both $F^{(0)}$ and $\Omega$
are homogeneous functions of degree two, so that 
${\cal F} (\lambda Y, \lambda\bar Y; 
\lambda^2 \Upsilon, \lambda^2 \bar \Upsilon) = \lambda^2 \, {\cal F} (Y, \bar Y; \Upsilon, \bar \Upsilon)$.

The reduced Lagrangian \eqref{eq:wilson-Om} agrees with the one in \eqref{eq:H-sympl}, up to an overall
normalization factor of $-2$. In the following, we rescale \eqref{eq:wilson-Om} by this factor, so that
from now on
\begin{eqnarray}
{\cal F} (Y, \bar Y; \Upsilon, \bar \Upsilon) = 4 \, \left[ {\rm Im} F(Y, \bar Y; \Upsilon, \bar \Upsilon) - 
\Omega(Y, \bar Y; \Upsilon, \bar \Upsilon) 
\right] \;.
\label{eq:wilson-Om-resc}
\end{eqnarray}
Using \eqref{eq:x-Y}, we infer that 
$p^I = - \mathrm{i} \left(Y^I - {\bar Y}^I \right)$
and $e^I = Y^I + {\bar Y}^I$. According to \eqref{eq:theorem-prop}, on the other hand, the real part of $Y^I$ plays the role of the
canonical variable $\phi^I$, so that we have $\phi^I = e^I$.  We 
may thus view $\cal F$ as a function of $p^I$ and $\phi^I$, 
 and consider its Legendre transformation 
either with respect to $p^I$ or with respect to $\phi^I$.  Performing the Legendre transformations with
respect to the $p^I$, i.e. ${\cal H} =  {\cal F} - p^I \, \pi_I $, results in 
\begin{equation}
\pi_I = \frac{\partial \cal{F}}{\partial p^I} = F_I + {\bar F}_{\bar I} \;,
\end{equation}
and hence
\begin{align}
  \mathcal{H} =&\,
  \mathrm{i} \left[Y^I \,\bar F_{\bar I} -\bar
  Y^{\bar I} \,F_I \right]
  + 2\left[2\,\Omega
  -Y^I\Omega_I -\bar Y^{\bar I} \Omega_{\bar I} \right] \nonumber\\
=&\,  
  \mathrm{i}\left[Y^I \,\bar F^{(0)}_{\bar I} -\bar Y^{\bar I} \,F^{(0)}_I \right]
    +2\left[ 2\,\Omega
  -(Y^I - {\bar Y}^{\bar I}) (\Omega_I - \Omega_{\bar I}) \right] \;,
  \label{eq:hesse-Y}
\end{align}
which is the analogue of the Hamiltonian \eqref{eq:H-sympl} (up to an overall sign difference in the definition of both
quantities).
In the context of BPS black holes, $\cal H$ denotes the BPS free energy of the black hole. 
When viewed
as a function of $\phi^I$ and $\pi_I$, ${\cal H}(\phi, \pi)$ is called the Hesse potential.
\\[1mm]
{\sf Exercise 15: Verify \eqref{eq:hesse-Y}.}
\vskip 1mm

On the other hand, performing the Legendre transformations with
respect to the $\phi^I$,
 i.e. ${\cal S} =  {\cal F} - \phi^I \, q_I $, results in 
\begin{equation}
q_I = \frac{\partial \cal{F}}{\partial \phi^I}  = - \mathrm{i} \left( F_I -  {\bar F}_{\bar I} \right) \;,
\label{eq:q-attrac}
\end{equation}
and hence
\begin{align}
  \mathcal{S}  &=\, -
    \mathrm{i} \left[Y^I \,\bar F_{\bar I} -\bar
  Y^{\bar I} \,F_I \right]
  + 2\left[2\,\Omega
  -Y^I\Omega_I -\bar Y^{\bar I} \Omega_{\bar I} \right] \nonumber\\
&=\,   - 
  \mathrm{i}(Y^I \,\bar F^{(0)}_{\bar I} -\bar Y^{\bar I} \,F^{(0)}_I)
    +2\left[ 2\,\Omega
  -(Y^I + {\bar Y}^{\bar I}) (\Omega_I + \Omega_{\bar I}) \right] \;.
  \label{eq:entr-BPS}
\end{align}
In the context of  BPS black holes, \eqref{eq:q-attrac} is the electric attractor equation, and ${\cal S}$
denotes the black hole entropy when viewed as function of $p^I$ and $q_I$ \cite{LopesCardoso:1998wt}. 
\\[1mm]
{\sf Exercise 16: Verify \eqref{eq:entr-BPS}.}
\vskip 1mm

The entropy 
${\cal S}$ can be obtained from the 
Hesse potential
by a double Legendre transformation with respect to $(\phi^I, \pi_I)$ \cite{LopesCardoso:2006bg}, i.e.
\begin{equation}
{\cal S}(p,q) = {\cal H}(\phi, \pi) + \pi_I \, p^I - \phi^I \, q_I 
\end{equation}
with $p^I = - \partial {\cal H} / \partial \pi_I$ and $q_I = \partial {\cal H} / \partial \phi^I$.

\subsection{The reduced low-energy effective action in an $AdS_2 \times S^2$ background}

When passing from the Wilsonian to the low-energy effective action, non-holomorphic terms
emerge that are crucial for maintaining duality invariances
\cite{Dixon:1990pc}, and that therefore need to be incorporated
into the framework of the previous subsection.  In the following, we assume that these terms can 
be incorporated into $\Omega$ by giving up the requirement that $\Omega$ is harmonic.
We take the reduced low-energy effective Lagrangian
and the associated Hesse potential to be given by 
\eqref{eq:wilson-Om-resc} and \eqref{eq:hesse-Y}, respectively, 
but now based on a non-harmonic $\Omega$.

The Hesse potential \eqref{eq:hesse-Y} is given in terms of complex scalar fields $Y^I$ and $\bar Y^I$.
Under duality transformations, the scalar fields $Y^I$ transform into $\tilde Y^I = \tilde Y^I (Y, \bar Y, \Upsilon,
\bar \Upsilon)$
(and similarly for the $\bar{\tilde{Y}}^I$),
as discussed in section \ref{sec:new-var}.  In order to obtain expansion coefficients that have a proper
behavior under duality when expanding ${\cal H}$ in powers of $\Upsilon$ and $\Upsilon$, 
we first express ${\cal H}$
in terms of the duality covariant complex coordinates introduced in section \ref{sec:new-var}.
This can be achieved by iteration,
and the result for the Hesse potential in the new coordinates then takes the form of an infinite power
series in terms of $\Omega$ and its derivatives.  We explicitly evaluate the first terms in this expansion
up to order $\Omega^5$. This suffices for appreciating the general structure of the full result.
The actual calculations are rather cumbersome, and we have relegated some relevant material to appendices 
\ref{sec:funct-H-a-i-geq2} and \ref{sec:transf-der-om}.
The expression for the Hesse potential, given in \eqref{eq:Hesse-decomp}, consists of a sum of contributions $\mathcal{H}^{(a)}_i$, each of which
transforms as a function under symplectic transformations.  The function $\mathcal{H}^{(1)}$ is the only one that contains $\Omega$,
while all the other $\mathcal{H}^{(a)}_i$ contain derivatives of $\Omega$.  Using that $\mathcal{H}^{(1)}$  transforms as a function under symplectic
transformations, we determine the transformation law of $\Omega$, which is given in \eqref{eq:Omega-tilde4}.  In the following, we present a
detailed derivation of these results.  We suppress the superscript in $F^{(0)}$ for the most part, for simplicity.

The Hesse potential $\mathcal{H}$ is defined in terms of the real variables $(\phi^I, \pi_I)$, whose definition depends on the full
effective action.  These may be expressed in terms of the duality covariant variables introduced in \eqref{eq:canon-t-rel}, and
which will be denoted by $\mathcal{Y}^I$ in the following.  Inspection of \eqref{eq:x-t-rel} shows that
these new variables are such that they
coincide precisely with the fields $Y^I$
that one would obtain from $(\phi^I,\pi_I)$ by using only the
lowest-order holomorphic function $F^{(0)}$, 
\begin{align}
  \label{eq:new-Yfields-Y}
  2\,\mathrm{Re}\,\mathcal{Y}^I =&\, \phi^I = 2\,\mathrm{Re}\, Y^I
  \,,\nonumber\\ 
    2\,\mathrm{Re}\,F_I^{(0)}(\mathcal{Y}) =&\, \pi_I =
    2\,\mathrm{Re}\,F_I(Y,\bar Y;\Upsilon,\bar\Upsilon) 
    \;.
\end{align}
Since the relation between the new variables and the real variables $(\phi^I, \pi_I)$
depends only on $F^{(0)}$, their duality transformations will not
depend on the the details of the full effective action. 
Under symplectic transformations they transform
according to,
\begin{equation}
  \label{eq:duality-new-Y}
  \tilde{\mathcal{Y}}^I= U^I{}_J\mathcal{Y}^J  + Z^{IJ}\,F^{(0)}_J(\mathcal{Y}) =
  \mathcal{S}_0{}^I{}_J(\mathcal{Y}) \, \mathcal{Y}^J\,, 
\end{equation}
where 
\begin{equation}
  \label{eq:def-S-0}
  \mathcal{S}_0{}^I{}_J(\mathcal{Y})=  U^I{}_J +
  Z^{IK}\,F^{(0)}_{KJ}(\mathcal{Y})\,.   
\end{equation}

At the two-derivative level, where $\Omega=0$, we have $\mathcal{Y}^I=Y^I$,
but in higher orders the relation between these moduli is complicated
and will depend on $\Omega$. Hence we write $\mathcal{Y}^I=Y^I +\Delta
Y^I$, where $\Delta Y^I$ is purely imaginary. Writing $F= F^{(0)} +
2\mathrm{i}\Omega$, we will express \eqref{eq:duality-new-Y} in terms
of $F^{(0)}$ and $\Omega$, so that we can henceforth suppress the
superscript in $F^{(0)}$. Hence, in the following, $F$ will denote a
holomorphic homogeneous function of degree two. Therefore it is not
necessary to make a distinction between holomorphic and
anti-holomorphic derivatives of this function. The equations
(\ref{eq:new-Yfields-Y}) can then be written as,
\begin{align}
  \label{eq:Ycal-fields}
  &
  F_I(\mathcal{Y}-\Delta Y)+\bar F_I(\bar{\mathcal{Y}}+\Delta Y)-
  F_I(\mathcal{Y})-\bar  F_I(\bar{\mathcal{Y}}) = \nonumber \\
  &
  - 2\mathrm{i}\,
  \left[\Omega_I(\mathcal{Y}-\Delta Y,\bar{\mathcal{Y}}+\Delta Y )-
  \Omega_{\bar I}(\mathcal{Y}-\Delta Y,\bar{\mathcal{Y}}+\Delta Y)\right]
   \,. 
\end{align}
Upon Taylor expanding, this equation will lead to an infinite power
series in $\Delta Y^I$. Retaining only the term of first order in
$\Delta Y^I$ shows that it is proportional to the first derivative of
$\Omega$. Proceeding by iteration will then lead to an expression for
$\Delta Y^I$ involving increasing powers of $\Omega$ and its
derivatives taken at $Y^I=\mathcal{Y}^I$. Here it suffices to give the
result of this iteration up to fourth order in $\Omega$, 
\begin{align}
  \label{eq:Delta-3-order}
  \Delta Y^I = &\, 2\,(\Omega^I-\Omega^{\bar I})\nonumber\\
   &\, -2\mathrm{i} (F+\bar F)^{IJK} 
   (\Omega_J-\Omega_{\bar J}) (\Omega_K-\Omega_{\bar K}) 
   - 8\,\mathrm{Re}(\Omega^{IJ}-\Omega^{I\bar J})\,(\Omega_J-\Omega_{\bar J})
   \nonumber\\
   &\, 
   +\tfrac4{3}\mathrm{i} \left[(F-\bar F)^{IJKL} +3\mathrm{i} (F+\bar
       F)^{IJM} (F+\bar F)_M{}^{KL}\right]\nonumber\\
     &\qquad \qquad
      \times(\Omega_J-\Omega_{\bar  J}) (\Omega_K-\Omega_{\bar K})
     (\Omega_L-\Omega_{\bar L}) \nonumber    \\ 
   &\, 
   + 8\mathrm{i} \left[ 2\,(F+\bar F)^{IJ}{}_K \mathrm{Re}
   (\Omega^{KL}-\Omega^{K\bar L})+ \mathrm{Re}
   (\Omega^{IK}-\Omega^{I\bar K})  (F+\bar F)_{K}{}^{JL} \right]\nonumber\\
   &\qquad \qquad 
    \times (\Omega_J-\Omega_{\bar J})(\Omega_L-\Omega_{\bar L})
    \nonumber  \\
   &\, 
   + 32\,\mathrm{Re}
   (\Omega^{IJ}-\Omega^{I\bar J})\, \mathrm{Re}
   (\Omega_{JK}-\Omega_{J\bar K})\, (\Omega^K-\Omega^{\bar K})  \nonumber\\
   &\, 
   + 8\mathrm{i}\,\mathrm{Im}(\Omega^{IJK} -2\,\Omega^{IJ\bar K} +
   \Omega^{I\bar J\bar K} ) (\Omega_{J}-\Omega_{\bar J})\,
   (\Omega_K-\Omega_{\bar K})  +\mathcal{O}(\Omega^4)\;.
\end{align}
Here indices have been raised by making use of $N^{IJ}$, which denotes
the inverse of 
\begin{equation}
  \label{eq:def-N}
  N_{IJ}= 2\,\mathrm{Im} F_{IJ} \,, 
\end{equation}
where we stress that all the derivatives of $F$ and $\Omega$ are taken
at $Y^I=\mathcal{Y}^I$ and $\bar Y^I=\bar{\mathcal{Y}}^I$.

Furthermore, we obtain the following expression for the Hesse potential \eqref{eq:hesse-Y},
\begin{align}
  \label{eq:Hesse-Y-calY}
  \mathcal{H}(\mathcal{Y},\bar{\mathcal{Y}}) =&\,- \mathrm{i}[
  \bar{\mathcal{Y}}^I F_{I} (\mathcal{Y}) -\mathcal{Y}^I\bar F_{I}
  (\bar{\mathcal{Y}}) ]
  +4\,\Omega(\mathcal{Y},\bar{\mathcal{Y}}) \nonumber \\
  &\, -\mathrm{i}\Big[ \mathcal{Y}^I ( F_I(Y)- F_I(\mathcal{Y})) +
  \Delta
  Y^I F_{I} (Y)  -\mbox{h.c.}\Big]  \nonumber\\
  &\, + 4\Big[\Omega(Y,\bar Y)-\Omega(\mathcal{Y},\bar{\mathcal{Y}}) +
  \Delta Y^I \left(\Omega_I(Y,\bar Y) -\Omega_{\bar I}(Y,\bar
    Y)\right) \Big]   \,. 
\end{align}
Here we made use of (\ref{eq:Ycal-fields}) at an intermediate stage of
the calculation. Again this result
must be Taylor expanded upon writing $Y^I=\mathcal{Y}^I-\Delta Y^I$ and
$\bar Y^I=\bar{\mathcal{Y}}^I+\Delta Y^I$. The last two lines of
(\ref{eq:Hesse-Y-calY}) then lead to a power series in $\Delta Y$,
starting at second order in the $\Delta Y$,
\begin{align}
  \label{eq:Hesse-DY-calY}
  \mathcal{H}(\mathcal{Y},\bar{\mathcal{Y}}) =&\,- \mathrm{i}[
  \bar{\mathcal{Y}}^I F_{I} (\mathcal{Y}) -\mathcal{Y}^I\bar F_{I}
  (\bar{\mathcal{Y}}) ]
  +4\,\Omega(\mathcal{Y},\bar{\mathcal{Y}}) \nonumber \\
  &\, -N_{IJ}\Delta Y^I\Delta Y^J -\tfrac23\mathrm{i} (F+\bar F)_{IJK}
  \Delta Y^I\Delta Y^J\Delta Y^K \nonumber \\
  &\,  - 4\,\mathrm{Re}(\Omega_{IJ}-\Omega_{I\bar J}) \Delta Y^I\Delta
  Y^J + \tfrac14\mathrm{i} (F-\bar F)_{IJKL} 
  \Delta Y^I\Delta Y^J\Delta Y^K \Delta Y^L  \nonumber\\
  &\, +\tfrac83\mathrm{i}\, \mathrm{Im}(\Omega_{IJK}-3\Omega_{IJ\bar
    K}) \Delta Y^I\Delta Y^J \Delta Y^K  +\cdots\;. 
\end{align}
Inserting the result of the iteration (\ref{eq:Delta-3-order}) into
the expression above leads to the following expression for the Hesse
potential, up to terms of order $\Omega^5$,
\begin{align}
  \label{eq:Hesse-Y-calY-exp}
  \mathcal{H}(\mathcal{Y},\bar{\mathcal{Y}}) =&\,- \mathrm{i}[
  \bar{\mathcal{Y}}^I F_{I} (\mathcal{Y}) -\mathcal{Y}^I\bar F_{I}
  (\bar{\mathcal{Y}}) ]
  +4\,\Omega(\mathcal{Y},\bar{\mathcal{Y}}) \nonumber \\
  &\, -4\, \hat N^{IJ}  \omega_I\,\omega_J +\tfrac83\mathrm{i}
  (F+\bar F)_{IJK} \hat N^{IL}\hat N^{JM}\hat N^{KN}
  \omega_L\,\omega_M\,\omega_N \nonumber \\
  &\, - \tfrac43\mathrm{i} [(F-\bar F)_{IJKL}+ 3\mathrm{i}(F+\bar
  F)_{IJR}\hat N^{RS}(F+\bar F)_{SKL} ] \nonumber\\
  &\qquad \times \hat N^{IM}\hat
  N^{JN}\hat N^{KP} \hat N^{LQ} 
  \omega_M\,\omega_N\,\omega_P\,\omega_Q   \nonumber\\
  &\, -\tfrac{32}3\mathrm{i}\, \mathrm{Im}(\Omega_{IJK}-3\Omega_{IJ\bar
    K}) \hat N^{IL}\hat N^{JM}\hat N^{KN}
  \omega_L\,\omega_M\,\omega_N   +\mathcal{O}(\Omega^5)\;,
\end{align}
where $\omega_I=\Omega_I-\Omega_{\bar I}$, and where we also made use
of $\hat N^{IJ}$, which is the inverse of the real, symmetric matrix
$\hat N_{IJ}$ given in \eqref{eq:hat-N-N}, namely 
\begin{equation}
  \label{eq:hat-N}
  \hat N_{IJ}= 
  N_{IJ} +4\,\mathrm{Re}(\Omega_{IJ}-\Omega_{I\bar J}) \,.
\end{equation}
Upon expanding $\hat N^{IJ}$ we straightforwardly determine the
contributions to the Hesse potential up to fifth order in $\Omega$,
\begin{align}
  \label{eq:Hesse-1-4}
  \mathcal{H}=&\, \mathcal{H}\vert_{\Omega=0} + 4\,\Omega -
  4\,N^{IJ}(\Omega_I\Omega_J +\Omega_{\bar I}\Omega_{\bar J}) +
  8\,N^{IJ} \Omega_I\Omega_{\bar J} \nonumber\\
  &\, 
  +16\,\mathrm{Re}(\Omega_{IJ}-\Omega_{I\bar J}) N^{IK}N^{JL}
    \big(\Omega_K\Omega_L +\Omega_{\bar K}\Omega_{\bar
      L}-2\,\Omega_K\Omega_{\bar L} \big) \nonumber\\
    &\, 
    - \tfrac{16}3
  (F+\bar F)_{IJK}  N^{IL} N^{JM} N^{KN} \,\mathrm{Im}
  (\Omega_L\Omega_M\Omega_N - 3\,\Omega_L\Omega_M\Omega_{\bar N})
  \nonumber \\ 
  &\ - 64 N^{IP} {\rm Re} \left(\Omega_{PQ} - \Omega_{P \bar Q} \right) N^{QR}
  {\rm Re} \left(\Omega_{RK} - \Omega_{R \bar K} \right) N^{KJ}
  \left(\Omega_I \Omega_{J} + \Omega_{\bar I} \Omega_{\bar J} - 2
    \Omega_I \Omega_{\bar J} \right) 
  \nonumber\\ 
  &\ + 64 (F+\bar F)_{IJK}  N^{IL} N^{JM} N^{KP} \,\mathrm{Re}
  \left(\Omega_{PQ} - \Omega_{P \bar Q} 
  \right) N^{QN} \,\mathrm{Im}
  (\Omega_L\Omega_M\Omega_N - 3\,\Omega_L\Omega_M\Omega_{\bar N}) 
   \nonumber\\
  &\,   - \tfrac83\mathrm{i} [(F-\bar F)_{IJKL}+ 3\mathrm{i} (F+\bar
  F)_{R(IJ} N^{RS}(F+\bar F)_{KL)S} ]
   N^{IM}
  N^{JN} N^{KP} N^{LQ} \nonumber\\
 &\, \qquad \qquad \qquad \qquad \qquad 
 \times \mathrm{Re} \left(\Omega_M \Omega_N \Omega_P \Omega_Q - 4
   \Omega_M \Omega_N \Omega_P \Omega_{\bar Q}+ 
  3 \Omega_M \Omega_N \Omega_{\bar P} \Omega_{\bar Q} \right) \nonumber\\
   &\, +\tfrac{64}3 \, \mathrm{Im}(\Omega_{IJK}-3\Omega_{IJ\bar
    K})  N^{IL} N^{JM}  N^{KN}
  \,\mathrm{Im}
  (\Omega_L\Omega_M\Omega_N - 3\,\Omega_L\Omega_M\Omega_{\bar N}) 
     +\mathcal{O}(\Omega^5)\;.
\end{align}
We stress once more that this expression is taken at
$Y^I=\mathcal{Y}^I$. 

The expression \eqref{eq:Hesse-1-4} gives the Hesse potential in terms of the duality
covariant variables $\mathcal{Y}^I$ and $\bar{\mathcal{Y}}^I$, up to order $\Omega^5$.
It takes a rather complicated form,
even at this order of approximation. Nevertheless, it will turn out
that there is some systematics here. First of all, the Hesse potential
\eqref{eq:Hesse-1-4} transforms as a function under duality transformations
acting on the fields $\mathcal{Y}^I$.  This in turn enables one to
determine how $\Omega$ should transform. Clearly, when $\Omega=0$ the
Hesse potential transforms manifestly as a function. In general the
transformation behaviour of $\Omega$ must be rather complicated in
view of the non-linear dependence of the Hesse potential on
$\Omega$. To evaluate this transformation, we have to perform yet
another iteration procedure.

To demonstrate how this iteration proceeds, let us have a look at the first
few steps. Consider the expression \eqref{eq:Hesse-1-4} at first order in $\Omega$.
At this order, $\Omega$ must transform as a function, since both $\mathcal{H}$
and $\mathcal{H}\vert_{\Omega=0}$ transform as functions.
This implies that
\begin{align}
  \label{eq:transf-Omega1}
  \tilde\Omega(\tilde{\mathcal{Y}}, \tilde{\bar{{\mathcal{Y}}}}) =&\,
    \Omega(\mathcal{Y}, \bar{{\mathcal{Y}}}) \,,\nonumber\\
    \tilde\Omega_I(\tilde{\mathcal{Y}}, \tilde{\bar{{\mathcal{Y}}}})
      =&\, [\mathcal{S}_0^{-1}]^J{}_I(\mathcal{Y}) \,\Omega_J(\mathcal{Y},
      \bar{{\mathcal{Y}}}) \,.  
\end{align}
Now consider the terms of order $\Omega^2$ in \eqref{eq:Hesse-1-4}.
Applying the transformation given in the second line of \eqref{eq:transf-Omega1}
to these terms and demanding $\mathcal{H}$ to transform as a function, shows
that the result given in the first line of \eqref{eq:transf-Omega1}
must be modified to 
\begin{equation}
  \label{eq:second-order variation-Omega}
  \tilde \Omega=  \Omega - \mathrm{i}\big( \mathcal{Z}_0^{IJ}
  \,\Omega_I \Omega_J - \bar{\mathcal{Z}}_0^{\bar I\bar J}
  \,\Omega_{\bar I} \Omega_{\bar J} \big) +\mathcal{O}(\Omega^3)
  \,,  
\end{equation}
which in turn gives rise to the following result for derivatives
of $\Omega$,
\begin{align}
  \label{eq:transf-der-Omega2}
  \tilde\Omega_I=&\, [\mathcal{S}_0^{-1}]^J{}_I \Big[ \Omega_J
  +\mathrm{i} F_{JKL} \,\mathcal{Z}_0^{KM}\Omega_M\,\mathcal{Z}_0^{LN}
  \Omega_N -2\mathrm{i} \Omega_{JK}\mathcal{Z}_0^{KL} \Omega_L
  +2\mathrm{i} \Omega_{J\bar K}\bar{\mathcal{Z}}_0^{\bar K\bar L}
  \Omega_{\bar L} \Big]\nonumber\\
  &\,
  +\mathcal{O}(\Omega^3)\,,\nonumber \\
  \tilde\Omega_{IJ}=&\, [\mathcal{S}_0^{-1}]^K{}_I
  [\mathcal{S}_0^{-1}]^L{}_J  \Big[ \Omega_{KL} -F_{KLM}
  \,\mathcal{Z}_0^{MN} \Omega_N  \Big] +\mathcal{O}(\Omega^2)
  \,,\nonumber\\ 
  \tilde\Omega_{I\bar J}=&\, [\mathcal{S}_0^{-1}]^K{}_I
  [\bar{\mathcal{S}}_0^{-1}]^{\bar L}{}_{\bar J} \,\Omega_{K\bar
    L}  +\mathcal{O}(\Omega^2) \,,
\end{align}
where the symmetric matrix $\mathcal{Z}_0^{IJ}$ is defined by\footnote{This quantity was first defined in 
\cite{deWit:1996ix}.  It appeared later in \cite{Aganagic:2006wq}, where it was denoted by $\Delta$.}
\begin{equation}
  \label{eq:def-calZ-0}
  \mathcal{Z}_0^{IJ} = [\mathcal{S}_0^{-1} ]^I{}_K\, Z^{KJ}\,.
\end{equation}
Here we made use of the relations,
\begin{align}
  \label{eq:dual-tr-Omega}
  [\mathcal{S}_0^{-1}]^I{}_K \,[\bar{\mathcal{S}}_0]^{\bar K}{}_{\bar
    J} =&\, \delta^I{}_J -\mathrm{i}\mathcal{Z}_0^{IK}N_{KJ}\,,
  \nonumber\\ 
  \tilde N_{IJ} =&\,  [\mathcal{S}_0^{-1}]^K{}_I
  [\bar{\mathcal{S}}_0^{-1}]^{\bar L}{}_{\bar J}
  \,N_{KL}\,,\nonumber\\ 
  \delta\mathcal{Z}_0^{IJ} =&\, - \mathcal{Z}_0^{IK}\,\delta F_{KL}
  \,\mathcal{Z}_0^{LJ}\,,
\end{align}
which are independent of $\Omega$.

This iteration can be continued by including the terms of order
$\Omega^3$, making use of \eqref{eq:transf-der-Omega2} for derivatives
of $\Omega$, to obtain the expression for $\tilde\Omega$ up terms of
order $\Omega^4$. In the next iterative step one then derives the
effect of a duality transformation on $\Omega$ up to terms of order
$\Omega^5$. Before presenting this result, we wish to observe that
terms transforming as a proper function under duality, will not
contribute to this result. This is precisely what already happened to
the $\Omega$-independent contribution to the Hesse potential, which
decouples from the above equations. As it turns out there 
actually exists an infinite set of contributions to the Hesse potential
that transform as functions under duality. By separating those from
\eqref{eq:Hesse-1-4}, we do not change the transformation behaviour of
$\Omega$, but we can extract certain functions from the Hesse potential
in order to simplify its structure. We obtain
\begin{align}
  \label{eq:Hesse-decomp}
  \mathcal{H} =&\, \mathcal{H}^{(0)} + \mathcal{H}^{(1)} +
  \mathcal{H}^{(2)} + \big(\mathcal{H}^{(3)}_1 +
  {\mathcal{H}}^{(3)}_2 + \mathrm{h.c.}\big) +  \mathcal{H}^{(3)}_3 +
  \mathcal{H}^{(4)}_1  +\mathcal{H}^{(4)}_2 +\mathcal{H}^{(4)}_3 \nonumber\\
  &\, +\big(\mathcal{H}^{(4)}_4+ \mathcal{H}^{(4)}_5 +
  \mathcal{H}^{(4)}_6 + \mathcal{H}^{(4)}_7 + \mathcal{H}^{(4)}_8 + \mathcal{H}^{(4)}_9 +
  \mathrm{h.c.}\big)
  \ldots\,,  
\end{align}
where the $\mathcal{H}^{(a)}_i$ are certain expressions to be defined
below, whose leading term is of order $\Omega^a$. For higher values of
$a$ it turns out that there exists more than one functions with the
same value of $a$, and those will be labeled by $i=1,2,\ldots$. 
Of all the combinations $\mathcal{H}^{(a)}_i$ appearing in \eqref{eq:Hesse-decomp},
$\mathcal{H}^{(1)}$ is the only that contains $\Omega$, while
all the other combinations contain derivatives of $\Omega$.  Obviously,
$\mathcal{H}^{(0)}$ equals,
\begin{equation}
  \label{eq:Hesse-0}
  \mathcal{H}^{(0)} = - \mathrm{i}[
  \bar{\mathcal{Y}}^I F_{I} (\mathcal{Y}) -\mathcal{Y}^I\bar F_{I}
  (\bar{\mathcal{Y}}) ]\,,
\end{equation}
whereas $\mathcal{H}^{(1)}$ at this level of iteration is given by,
\begin{align}
  \label{eq:Hesse-1}
  \mathcal{H}^{(1)}=&\, 4\,\Omega - 4\,N^{IJ}(\Omega_I\Omega_J
    +\Omega_{\bar I}\Omega_{\bar J})
    \nonumber\\
    &\, +16\,\mathrm{Re} \Big[ (\Omega_{IJ}) (N \Omega)^I (N \Omega)^J
    \big]  + 16 \Omega_{I \bar J} \, (N \Omega)^I ( N \bar \Omega)^J
    \nonumber\\ 
    &\, - \tfrac{16}3 \,\mathrm{Im} \left[ F_{IJK} (N \Omega)^I (N
      \Omega)^J (N \Omega)^K
    \right] \nonumber\\
    &\, - \tfrac43 \mathrm{i} \Big[ \left(F_{IJKL} + 3 \mathrm{i}
      F_{R(IJ} N^{RS} F_{KL)S} \right) (N \Omega)^I (N \Omega)^J (N
    \Omega)^K (N \Omega)^L
     - {\rm h.c.} \Big] \nonumber\\
     &\, 
    - \tfrac{16}3 \left[ \Omega_{IJK} (N \Omega)^I (N \Omega)^J (N
      \Omega)^K  + {\rm h.c.}  \right] \nonumber\\
      &\, 
    - 16\left[ \Omega_{IJ\bar K} (N \Omega)^I (N \Omega)^J (N \bar \Omega)^K
      + {\rm h.c.}     \right] \nonumber\\
      &\ - 16 \mathrm{i} \,\Big[ F_{IJK} N^{KP} \,\Omega_{PQ}  (N
      \Omega)^I (N \Omega)^J (N \Omega)^Q 
      - {\rm h.c.} \Big]      \nonumber\\
     & \, - 16 \Big[ (N \Omega)^P  \, \Omega_{PQ} \, N^{QR}
     \Omega_{RK} \, (N \Omega)^K  
           + {\rm h.c.} \Big] \nonumber\\
       & \, - 32 \Big[ (N \Omega)^P \, \Omega_{PQ} \, N^{QR} 
         \Omega_{R \bar K} \, (N \bar \Omega)^K  + {\rm h.c.} \Big] \nonumber\\
& \, - 16 \Big[ (N \Omega)^P \, \Omega_{P \bar Q} \, N^{QR} 
      \Omega_{\bar R K} \, (N \Omega)^K 
       + {\rm h.c.} \Big]\nonumber\\
        &\ - 16 \mathrm{i} \,\Big[ F_{IJK} N^{KP} \,\Omega_{P\bar Q}  
             (N \Omega)^I ( N \Omega)^J (N \bar \Omega)^Q
            - {\rm h.c.} \Big]  + \mathcal{O}(\Omega^5)\,. 
\end{align}
Here we have used the notation $(N \Omega)^I = N^{IJ} \Omega_J \,,\, 
(N \bar\Omega)^I = N^{IJ} \Omega_{\bar J}$.   The symmetrization 
$F_{R(IJ} N^{RS} F_{KL)S}$ is defined with a symmetrization factor
$1/(4!)$.

The expressions for the higher-order functions $\mathcal{H}^{(a)}_i$ with
$a=2,3,4$ are given in  appendix
\ref{sec:funct-H-a-i-geq2}. Each of these higher-order functions transforms as a
function under symplectic transformations.  Demanding $\mathcal{H}^{(1)}$ to also transform as 
a function under these transformations determines the transformation behavior of $\Omega$.
Proceeding as already
explained below \eqref{eq:transf-Omega1} we obtain for the transformation law of $\Omega$
(up to order $\Omega^5$),
 \begin{align}
  \label{eq:Omega-tilde4}
  \tilde \Omega=&\, \Omega - \mathrm{i}\big( \mathcal{Z}_0^{IJ}
  \,\Omega_I \Omega_J - \bar{\mathcal{Z}}_0^{\bar I\bar J}
  \,\Omega_{\bar I} \Omega_{\bar J} \big)\nonumber \\
  &\, +\tfrac23\big( F_{IJK} \,\mathcal{Z}_0^{IL}\Omega_L
  \,\mathcal{Z}_0^{JM}\Omega_M \,\mathcal{Z}_0^{KN}\Omega_N +
  \mathrm{h.c.}\big) 
  \nonumber\\
  &\, -2 \big(\Omega_{IJ} \, \mathcal{Z}_0^{IK}\Omega_K
  \mathcal{Z}_0^{JL}\Omega_L  +\mathrm{h.c.}\big)
  +4\,\Omega_{I\bar J} \,
  \mathcal{Z}_0^{IK}\Omega_K\,  \bar{\mathcal{Z}}_0^{\bar J\bar
    L}\Omega_{\bar L}  \nonumber\\
    &\,+ \Big[ - \tfrac{\mathrm{i}}{3} F_{IJKL} (\mathcal{Z}_0 \Omega)^I (\mathcal{Z}_0 \Omega)^J (\mathcal{Z}_0 \Omega)^K
    (\mathcal{Z}_0 \Omega)^L \nonumber\\
     &\, \qquad + \tfrac{4 \mathrm{i}}{3} \Omega_{IJK} (\mathcal{Z}_0 \Omega)^I (\mathcal{Z}_0 \Omega)^J (\mathcal{Z}_0             
     \Omega)^K \nonumber\\
     & \, \qquad + \mathrm{i}\, F_{IJR} \, {\cal Z}_0^{RS} \, F_{SKL} \, ({\cal Z}_0 \Omega)^I ({\cal Z}_0 \Omega)^J 
      ({\cal Z}_0 \Omega)^K ({\cal Z}_0 \Omega)^L  \nonumber\\
      & \, \qquad - 4 \mathrm{i} \Omega_{IJ \bar K} \,  (\mathcal{Z}_0 \Omega)^I \,  (\mathcal{Z}_0 \Omega)^J
       (\bar{\mathcal{Z}}_0 \bar \Omega)^K \nonumber\\
       & \qquad  - 4 \mathrm{i} \, F_{IJK} {\cal Z}_0^{KP} \,\Omega_{PQ}  \,({\cal Z}_0 \Omega)^I ({\cal Z}_0 \Omega)^J \
       ({\cal Z}_0 \Omega)^Q \nonumber\\
        &\qquad  + 4 \mathrm{i} \, F_{IJK} \mathcal{Z}_0^{KP} \,\Omega_{P\bar Q}  
             (\mathcal{Z}_0 \Omega)^I ( \mathcal{Z}_0 \Omega)^J (\bar{\mathcal{Z}}_0\bar \Omega)^Q
            \nonumber\\
     & \, \qquad + 
      4 \mathrm{i} \, (\mathcal{Z}_0 \Omega)^P \, \Omega_{PQ} \, \mathcal{Z}_0^{QR} \left(\Omega_{RK} \,   
            (\mathcal{Z}_0\Omega)^K 
      -   2 \Omega_{R \bar K} \, (\bar{\mathcal{Z}}_0 \bar \Omega)^K \right)  \nonumber\\
      & \, \qquad 
      - 4 \mathrm{i}  \, (\mathcal{Z}_0 \Omega)^P \, \Omega_{P \bar Q} \, \bar{\mathcal{Z}}_0^{\bar Q \bar R} 
      \Omega_{\bar R K} \, (\mathcal{Z}_0 \Omega)^K 
            + {\rm h.c.} \Big]  + \mathcal{O}(\Omega^5)\,. 
    \end{align}
The transformation laws of the derivatives of $\Omega$, such as those in \eqref{eq:transf-der-Omega2},
are summarized in appendix \ref{sec:transf-der-om}.

The transformation law
\eqref{eq:Omega-tilde4}, which is entirely encoded in ${\cal Z}_0$ and in $\bar{\cal Z}_0$,
suggest a systematic pattern, which we now explore.  First we observe
that \eqref{eq:Omega-tilde4} simplifies when taking $\Omega$
to be harmonic both in
$\mathcal{Y}^I$ and $\Upsilon$,
\begin{equation}
\Omega ( \mathcal{Y}, \bar{\mathcal{Y}}; \Upsilon, \bar \Upsilon) = f(\mathcal{Y}, \Upsilon) + {\rm h.c.} \;.
\label{eq:harm-Om-calY}
\end{equation}
We obtain
\begin{align}
  \label{eq:Omega-tilde4-harmonic}
  \tilde \Omega=&\, \Omega + \Big[ - \mathrm{i} \mathcal{Z}_0^{IJ}
  \,\Omega_I \Omega_J \nonumber \\
  &\,\qquad  +\tfrac23  F_{IJK} \,\mathcal{Z}_0^{IL}\Omega_L
  \,\mathcal{Z}_0^{JM}\Omega_M \,\mathcal{Z}_0^{KN}\Omega_N 
  \nonumber\\
  &\, \qquad -2 \, \Omega_{IJ} \, \mathcal{Z}_0^{IK}\Omega_K
  \mathcal{Z}_0^{JL}\Omega_L  \nonumber \\
     &\, \qquad  - 
     \tfrac{\mathrm{i}}{3} F_{IJKL} (\mathcal{Z}_0 \Omega)^I (\mathcal{Z}_0 \Omega)^J (\mathcal{Z}_0 \Omega)^K
    (\mathcal{Z}_0 \Omega)^L \nonumber\\
     &\, \qquad + \tfrac{4 \mathrm{i}}{3} \, \Omega_{IJK} (\mathcal{Z}_0 \Omega)^I (\mathcal{Z}_0 \Omega)^J (\mathcal{Z}_0             
     \Omega)^K \nonumber\\
     & \, \qquad + \mathrm{i}\, F_{IJR} \, {\cal Z}_0^{RS} \, F_{SKL} \, ({\cal Z}_0 \Omega)^I ({\cal Z}_0 \Omega)^J 
      ({\cal Z}_0 \Omega)^K ({\cal Z}_0 \Omega)^L  \nonumber\\
             & \qquad  - 4 
             \mathrm{i} \, F_{IJK} {\cal Z}_0^{KP} \,\Omega_{PQ}  \,({\cal Z}_0 \Omega)^I ({\cal Z}_0 \Omega)^J \
       ({\cal Z}_0 \Omega)^Q \nonumber\\
       & \, \qquad + 
      4 \mathrm{i} \, \mathcal{Z}_0^{IP} \, \Omega_{PQ} \, \mathcal{Z}_0^{QR} \,
      \Omega_{RK} \, (\mathcal{Z}_0 \Omega)^K \, \Omega_I 
       + {\rm h.c.} \Big]  + \mathcal{O}(\Omega^5)\,,
    \end{align}
 which shows that $\tilde \Omega$ also is harmonic.  Hence, the harmonicity of $\Omega$ is preserved under
symplectic transformations.
  The transformation law 
 \eqref{eq:Omega-tilde4-harmonic} has a certain resemblance with the one encountered 
in the context of perturbative topological string theory, where ${\cal Z}_0^{IJ}$ plays the role of a propagator
\cite{Aganagic:2006wq}.  The relation with topological string theory will be discussed below.
Next, inserting \eqref{eq:harm-Om-calY} into \eqref{eq:Hesse-1}, we find that $\mathcal{H}^{(1)}$ is also almost harmonic, i.e.
it equals 
the real part of a function that contains only
purely holomorphic derivatives of $F$ and $\Omega$, contracted
with the non-holomorphic tensor $N^{IJ}$,
\begin{align}
  \label{eq:Hesse-1-harm}
\mathcal{H}^{(1)}  
  =&\, \Big[ 4\,f(\mathcal{Y}, \Upsilon)  - 4\,N^{IJ} \Omega_I\Omega_J
    \nonumber\\
    &\, +8  (\Omega_{IJ}) (N \Omega)^I (N \Omega)^J
    + \tfrac{8}3 \mathrm{i} \,  F_{IJK} (N \Omega)^I (N
      \Omega)^J (N \Omega)^K \nonumber\\
    &\, - \tfrac43 \mathrm{i}  \left(F_{IJKL} + 3 \mathrm{i}
      F_{R(IJ} N^{RS} F_{KL)S} \right) (N \Omega)^I (N \Omega)^J (N
    \Omega)^K (N \Omega)^L \nonumber\\
     &\, 
    - \tfrac{16}3  \Omega_{IJK} (N \Omega)^I (N \Omega)^J (N
      \Omega)^K   
     - 16 \mathrm{i} \, F_{IJK} N^{KP} \,\Omega_{PQ}  (N
      \Omega)^I (N \Omega)^J (N \Omega)^Q    \nonumber\\
     & \, - 16  (N \Omega)^P  \, \Omega_{PQ} \, N^{QR}
     \Omega_{RK} \, (N \Omega)^K  + {\rm h.c.} \Big]
                   + \mathcal{O}(\Omega^5)\,. 
\end{align}
Thus, when $\Omega$ is of the form \eqref{eq:harm-Om-calY}, $\mathcal{H}^{(1)}$ is given in terms of the real
part of a function that is holomorphic in $\Upsilon$.  Moreover,  since
$N^{IJ}$ is homogeneous of degree zero, this function is homogeneous of degree two in 
$\mathcal{Y}^I$ and homogeneous of degree zero in $\bar{\mathcal{Y}}^I$.

Let us now elucidate the relation of $\mathcal{H}^{(1)}$ given in \eqref{eq:Hesse-1-harm}
with topological string theory. We write 
$\mathcal{H}^{(1)}$ as
 \begin{equation}
\mathcal{H}^{(1)} = h(\mathcal{Y}, \bar{\mathcal{Y}}, \Upsilon) + {\rm h.c.} \;,
\label{eq:H1-harm-Ups}
\end{equation}
and we 
consider two expansions of $h(\mathcal{Y},\bar{\mathcal{Y}}, \Upsilon)$, namely one in powers of $\Omega$ and the other
one in powers of $\Upsilon$. First we consider the expansion in powers of $\Omega$.  Expanding 
$h$ as
\begin{equation}
h = \sum_{g=1}^{\infty} h^{(g)} 
\label{eq:exp-h}
\end{equation}
and comparing with \eqref{eq:Hesse-1-harm}, we obtain
\begin{align}
h^{(1)} =&\, 4 f \;\;\;,\;\;\; h^{(2)} = - 4 N^{IJ} \Omega_I \Omega_J \;, \nonumber\\
h^{(3)} =&\, 8 \Omega_{IJ} (N \Omega)^I (N \Omega)^J + \tfrac83 \mathrm{i} F_{IJK}
(N \Omega)^I (N \Omega)^J (N \Omega)^K \;,\nonumber\\
h^{(4)} =&\,   - \tfrac43 \mathrm{i}  \left(F_{IJKL} + 3 \mathrm{i}
      F_{R(IJ} N^{RS} F_{KL)S} \right) (N \Omega)^I (N \Omega)^J (N
    \Omega)^K (N \Omega)^L \nonumber\\
     &\, \, 
    - \tfrac{16}3  \Omega_{IJK} (N \Omega)^I (N \Omega)^J (N
      \Omega)^K   
      - 16 \mathrm{i} \, F_{IJK} N^{KP} \,\Omega_{PQ}  (N
      \Omega)^I (N \Omega)^J (N \Omega)^Q    \nonumber\\
     &\, \, - 16  (N \Omega)^P  \, \Omega_{PQ} \, N^{QR}
     \Omega_{RK} \, (N \Omega)^K  \;,
     \label{eq:hg-234}
\end{align}
where
$(N \Omega)^I = N^{IJ}  \, f_J$. 
This shows that all the $h^{(g)}$ are non-holomorphic in $\mathcal{Y}^I$ with the exception of 
$h^{(1)}$.
Using these expressions, one finds by direct calculation that the following relation holds,
\begin{eqnarray}
\partial_{\bar I} h^{(g)} = \tfrac14 \mathrm{i} \, {\bar F}_{\bar I}{}^{JK} \, \sum_{r=1}^{g-1} \partial_J h^{(r)} \,
\partial_K h^{(g-r)} \;\;\;,\;\;\; g\geq 2 \;,
\label{eq:anom-sugra}
\end{eqnarray}
where $ {\bar F}_{\bar I}{}^{JK}  = {\bar F}_{\bar I \bar L \bar M} N^{LJ} N^{MK}$. 
\\[1mm]
{\sf Exercise 17: Verify \eqref{eq:anom-sugra} for $g=2,3$.}
\\[1mm]
Equation \eqref{eq:anom-sugra} 
captures the $\bar{\mathcal{Y}}^I$-dependence of $h^{(g)}$ (for $g\geq 2$).  This dependence 
is a consequence of requiring $\mathcal{H}^{(1)}$
to have a proper behavior under 
symplectic transformations \cite{deWit:1996ix}.  
The differential equation \eqref{eq:anom-sugra} resembles the holomorphic anomaly equation
of perturbative topological string theory.
The latter arises in a specific 
setting, 
namely in the study of the non-holomorphicity of the 
genus-$g$ topological free energies $F^{(g)}$ \cite{Bershadsky:1993cx}. 
To exhibit the relation with the holomorphic anomaly equation, we turn to the second expansion and 
expand both 
$f(\mathcal{Y}, \Upsilon)$ and  $h(\mathcal{Y},\bar{\mathcal{Y}}, \Upsilon) $ in powers
of $\Upsilon$, 
\begin{align}
 f(\mathcal{Y}, \Upsilon) =&\, -  \tfrac12
 \mathrm{i} \,  \sum_{g=1}^{\infty} \Upsilon^g \, f^{(g)} (\mathcal{Y}) \;, \nonumber\\
h(\mathcal{Y}, \bar{\mathcal{Y}}, \Upsilon) =&\,  - 2 \mathrm{i}
\sum_{g=1}^{\infty} \Upsilon^g \, F^{(g)} (\mathcal{Y}, \bar{\mathcal{Y}}) \;.
\label{eq:exp-f-F}
\end{align}
Then we obtain 
\begin{align}
F^{(1)} (\mathcal{Y})  =&\, f^{(1)} (\mathcal{Y}) \;\;\;,\;\;\; F^{(2)} (\mathcal{Y}, \bar{\mathcal{Y}})
= f^{(2)} (\mathcal{Y}) + \tfrac12 \mathrm{i} \, N^{IJ} F^{(1)}_I F^{(1)}_J \;, \nonumber\\
F^{(3)}(\mathcal{Y}, \bar{\mathcal{Y}}) =&\,  f^{(3)} (\mathcal{Y}) + \mathrm{i} \,  N^{IJ} f^{(2)}_I F^{(1)}_J
-
\tfrac12 F^{(1)}_{IJ} (N F^{(1)})^I (N F^{(1)})^J 
\nonumber\\
&\, 
- \tfrac{1}{6} \mathrm{i} F_{IJK}
(N F^{(1)})^I (N F^{(1)})^J (N F^{(1)})^K \;,\nonumber\\
F^{(4)} (\mathcal{Y}, \bar{\mathcal{Y}}) =&\,   f^{(4)} (\mathcal{Y}) 
 + \mathrm{i} \,  N^{IJ} f^{(3)}_I F^{(1)}_J 
 +  \tfrac12 \mathrm{i} \, N^{IJ} f^{(2)}_I f^{(2)}_J \nonumber\\
&\,  -
\tfrac12 f^{(2)}_{IJ} (N F^{(1)})^I (N F^{(1)})^J 
-
 F^{(1)}_{IJ} (N f^{(2)})^I (N F^{(1)})^J 
  \nonumber\\
  &\,
  - \tfrac{1}{2} \mathrm{i} F_{IJK}
(N f^{(2)})^I (N F^{(1)})^J (N F^{(1)})^K  \nonumber\\
&\, +  \tfrac{1}{24}  \left(F_{IJKL} + 3 \mathrm{i}
      F_{R(IJ} N^{RS} F_{KL)S} \right) (N F^{(1)})^I (N F^{(1)})^J (N
    F^{(1)})^K (N F^{(1)})^L \nonumber\\
     &\, 
    - \tfrac{1}{6}  \mathrm{i} \, F^{(1)}_{IJK} (N F^{(1)})^I (N F^{(1)})^J (N
      F^{(1)})^K   \nonumber\\
      &\, + \tfrac{1}{2} \, F_{IJK} N^{KP} \,F^{(1)}_{PQ}  (N
      F^{(1)})^I (N F^{(1)})^J (N F^{(1)})^Q    \nonumber\\
     &\, - \tfrac{1}{2} \mathrm{i} \,  (N F^{(1)})^P  \, F^{(1)}_{PQ} \, N^{QR}
     F^{(1)}_{RK} \, (N F^{(1)})^K  \;,
     \label{eq:hg-234-ups}
\end{align}
where $(N F^{(1)})^I = N^{IJ} \, F^{(1)}_J$ and $(N f^{(2)})^I = N^{IJ} \, f^{(2)}_J$.
Observe that all the $F^{(g)}$ are non-holomorphic except $F^{(1)}$.
Using these expressions, one again finds by direct calculation, 
\begin{eqnarray}
\partial_{\bar I} F^{(g)} = \tfrac12  \, {\bar F}_{\bar I}{}^{JK} \, \sum_{r=1}^{g-1} \partial_J F^{(r)} \,
\partial_K F^{(g-r)} \;\;\;,\;\;\; g\geq 2 \;.
\label{eq:anom-sugra-ups}
\end{eqnarray}
This is similar to \eqref{eq:anom-sugra}, except that now the relation holds order by order in $\Upsilon$,
whereas \eqref{eq:anom-sugra} holds order by order in $\Omega$.  
Both expansions are, nevertheless, related.  Namely, taking $f$ in \eqref{eq:exp-f-F} to consist of only 
$f^{(1)}$, the expansion \eqref{eq:hg-234-ups}
 coincides with the expansion \eqref{eq:hg-234}.

Summarizing, we have found the following. 
 When expressing the Hesse potential, which is a symplectic function, in terms of the duality
covariant complex variables \eqref{eq:new-Yfields-Y}, we obtain an infinite set of contributions $\mathcal{H}^{(a)}_i$,
all of which transform as functions under symplectic transformations. 
One of them, namely $\mathcal{H}^{(1)}$, has a structure that arises in topological string theory.
$\mathcal{H}^{(1)}$ 
 is the only contribution that contains $\Omega$, while all the other combinations contain derivatives
 of $\Omega$.  When $\Omega$ is taken to be harmonic in all the variables (i.e. in both $\mathcal{Y}^I$
and $\Upsilon$), the resulting $\mathcal{H}^{(1)}$
is given in terms of the real
part of a function that is holomorphic in $\Upsilon$, 
homogeneous of degree two in 
$\mathcal{Y}^I$ and homogeneous of degree zero in $\bar{\mathcal{Y}}^I$.
Then, expanding  $\mathcal{H}^{(1)}$ in powers of $\Upsilon$ yields expansion
functions $F^{(g)}$, given in \eqref{eq:hg-234-ups}, that
transform as functions under symplectic transformations.  The $F^{(g)}$ are all non-holomorphic, with the exception of 
$F^{(1)}$, and the non-holomorphicity is governed by \eqref{eq:anom-sugra-ups}.  This differential equation
equals half of the holomorphic anomaly equation of perturbative topological string theory, which reads \cite{Grimm:2007tm}
\begin{eqnarray}
\partial_{\bar I} F^{(g)} = \tfrac12 \,  {\bar F}_{\bar I}{}^{JK} \,
\left(D_J \partial_K F^{(g-1)} + 
 \sum_{r=1}^{g-1} \partial_J F^{(r)} \,
\partial_K F^{(g-r)} \right) \;\;\;,\;\;\; g\geq 2 \;,
\label{eq:holom-top-string}
\end{eqnarray}
where $D_L V_M = \partial_L V_M + \mathrm{i} N^{PI} F_{ILM} V_P$. 
This is the holomorphic anomaly equation 
in the so-called big moduli space \cite{Grimm:2007tm}, and 
its derivation is reviewed in appendix \ref{sec:top-string} following \cite{Aganagic:2006wq}.
In the context of topological string theory, the $F^{(g)}$ 
denote free energies that arise in the perturbative expansion of the 
topological free energy $F_{\rm top} $ in powers of the topological string coupling $g_{\rm top}$,
i.e. $F_{\rm top} = \sum_{g=0}^{\infty}
g^{2g -2}_{\rm top} \, F^{(g)}$.  Whereas $F^{(0)}$ is holomorphic (it only depends on 
$\mathcal{Y}$), all the higher $F^{(g)}$
(with $g \geq 1$) are non-holomophic. For $g \geq 2$ this non-holomorphicity is captured by 
\eqref{eq:holom-top-string}.

The fact that the first term on the right hand side of 
\eqref{eq:holom-top-string} is missing in \eqref{eq:anom-sugra-ups} is due to the holomorphic nature of
the expansion function $F^{(1)}$ appearing in \eqref{eq:hg-234-ups}.
Were it to be non-holomorphic, it would induce a modification of the relation \eqref{eq:anom-sugra-ups}.
The required  modification arises by replacing the holomorphic quantity $F_I^{(1)} = f^{(1)}_I$ 
with the non-holomorphic combination $F_I^{(1)} = f^{(1)}_I + \tfrac12 \, \mathrm{i} \, F_{IJK} \, N^{JK}$
(see \eqref{eq:F1-nh-hat}).

\vskip 5mm

\subsection*{Acknowledgements}
We acknowledge helpful discussions with
Thomas Mohaupt, Ashoke Sen and Marcel Vonk. The work of G.L.C. is
partially supported by the Center for Mathematical Analysis, Geometry
and Dynamical Systems (IST/Portugal), as well as 
by Funda\c{c}\~{a}o para a Ci\^{e}ncia e a Tecnologia
(FCT/Portugal) through grants CERN/FP/116386/2010 and PTDC/MAT/119689/2010.
The work of B.d.W. is supported 
by the ERC Advanced Grant no. 246974, {\it Supersymmetry: a window
to non-perturbative physics}.  G.L.C. would like to thank Stefano Bellucci for
the invitation to lecture at BOSS2011.
We would like to thank each other's institutions for hospitality
during the course of this work.
S.M. would also like to thank 
Hermann Nicolai and the members of Quantum 
Gravity group at the Max Planck Institut f\"ur Gravitationsphysik (AEI, Potsdam),
where part of this work was carried out, for the warm hospitality.

\appendix

\section{Symplectic reparametrizations 
\label{integrab}}

In subsection \ref{sec:theor} we introduced the $2n$-vector $(x^i, F_i)$ and discussed its behavior under
symplectic transformations.  Here we consider derivatives of $F_i$ and show how they transform under
symplectic transformations.  We use the resulting expressions 
to give an alternative proof of integrability of the equations \eqref{eq:symplectic}.
In addition, we show that $\partial_{\eta} F$ transforms as a function under symplectic transformations.

We begin by recalling some of the elements of subsection \ref{sec:theor}.
The $2n$-vector $(x^i, F_i)$ is constructed using
$F(x, \bar x)  = F^{(0)} (x)  + 2 \mathrm{i}  \Omega (x, \bar x)$.  Under symplectic transformations, it transforms
as,
\begin{align}
  \label{eq:dual-F0-Omega} 
  \tilde x^i=&\, U^i{}_j\,x^j + Z^{ij}  [F^{(0)}_j(x)  +2\mathrm{i}
  \Omega_j(x,\bar x)]\,,\nonumber\\ 
  \tilde F_i (\tilde x,\bar{\tilde x}) =&\, V_i{}^j\,[F^{(0)}_j(x)
  +2\mathrm{i} \Omega_j(x,\bar x)] + W_{ij} \,x^j \,,
\end{align}
where 
$U, V, Z$ and $W$ are the  $n \times n$ submatrices \eqref{eq:matrix-sympl}
that define a symplectic transformation belonging to ${\rm Sp} (2n, \mathbb{R})$. 
Without loss of generality, we decompose $\tilde F_i $ as 
\begin{equation}
 \tilde F_i (\tilde x,\bar{\tilde x}) = \tilde F^{(0)}_{i}(\tilde x)  + 2\mathrm{i}
  \tilde\Omega_i(\tilde x,\bar{\tilde x}) \;.
  \label{decom-tilde-om}
  \end{equation}
This decomposition, which a priori is arbitrary,
can be related to  the decomposition of
$F_i  = F_i^{(0)}  + 2 \mathrm{i}  \Omega_i $ in the following way.
 The symplectic transformation \eqref{eq:dual-F0-Omega} 
is specified by the matrices $U, V, W$ and $Z$.  Consider 
applying the same transformation (specified by these matrices)
to the vector $(x^i, F_i^{(0)})$ alone.  This yields  
the vector $(\hat{x}^i, {\tilde F}_i^{(0)} (\hat x))$, which is expressed in terms of 
$ {\hat x}^i = \tilde x^i - 2\mathrm{i}  Z^{ij}  \Omega_j(x,\bar x)$ instead of $\tilde x^i$,
\begin{align}
  \label{eq:dual-F0-hat}
  \hat x^i=&\, U^i{}_j\,x^j + Z^{ij}  F^{(0)}_j(x) 
  \,,\nonumber\\ 
  \tilde F^{(0)}_{i}(\hat x)   =&\, V_i{}^j\,F^{(0)}_j(x) + W_{ij} \,x^j \,.
\end{align}
Thus, by demanding that $\tilde F_i^{(0)}$  follows
from the same symplectic transformation applied on $F^{(0)}_i$ alone, we 
relate the decomposition of $\tilde F_i$ to the decomposition of $F_i$.
Then, the second equation of \eqref{eq:dual-F0-Omega} can be written as
\begin{align}
  \label{eq:tilde-omega}
  \tilde\Omega_{i}(\tilde x,\bar{\tilde x}) =&\,
  V_i{}^j\,\Omega_j(x,\bar x) - \tfrac12\mathrm{i}  [\tilde
  F^{(0)}_{i}(\hat x)
  - \tilde F^{(0)}_{i}(\tilde x)] \\
  =&\,  V_i{}^j\,\Omega_j(x,\bar x) \nonumber\\
  &\, + 
  \tfrac12\mathrm{i}  \sum_{m=1}^{\infty}  \frac{(2\mathrm{i})^m}{m!} 
  Z^{j_1k_1}\Omega_{k_1}(x,\bar x)\cdots Z^{j_mk_m}\Omega_{k_m}(x,\bar
  x)  \; \tilde   F^{(0)}_{i j_1\cdots j_m} \nonumber
   (\hat x) \,, 
\end{align}
where the $\tilde   F^{(0)}_{i j_1\cdots j_m } (\hat x)$ denote 
multiple derivatives of $\tilde F^{(0)}_i(\tilde x)$ evaluated at $\hat x$.
The right-hand side of \eqref{eq:tilde-omega}
can be written entirely in terms of functions of $x$ and $\bar x$,
upon expressing 
$\tilde   F^{(0)}_{i j_1\cdots j_m } (\hat x)$
in terms of derivatives of $F^{(0)}_i(x)$ using \eqref{eq:dual-F0-hat}.
We give the first few
derivatives,
\begin{align} 
  \label{eq:F-0-der}
  \tilde F^{(0)}_{ij} (\hat x) =&\,(V_i{}^l F^{(0)}_{lk} + W_{ik})\,
      [\mathcal{S}_0^{-1}]^k{}_j \,,\\
   \tilde F^{(0)}_{ijk} (\hat x) =&\,   [\mathcal{S}_0^{-1}]^l{}_i
   \,[\mathcal{S}_0^{-1}]^m{}_j\,  [\mathcal{S}_0^{-1}]^n{}_k
   \,F^{(0)}_{lmn} \,,\nonumber \\
   \tilde F^{(0)}_{ijkl} (\hat x) =&\,   [\mathcal{S}_0^{-1}]^m{}_{i}
   \,[\mathcal{S}_0^{-1}]^n{}_{j}\,  [\mathcal{S}_0^{-1}]^p{}_k
   \,[\mathcal{S}_0^{-1}]^q{}_{l} \Big[F^{(0)}_{mnpq} - 3\, F^{(0)}_{r(mn}
   \mathcal{Z}_0^{rs} \, F^{(0)}_{pq)s} \Big]\,, \nonumber
\end{align}
where we used the definitions
\begin{align}
  \label{eq:def-hat-S0}
  \mathcal{S}_0^{\,i}{}_j =&\, U^i{}_j +Z^{ik} F^{(0)}_{kj}\,, \nonumber\\
  \mathcal{Z}_0^{ij} =&\, [\mathcal{S}_0^{-1}]^i{}_k \, Z^{kj} \;.
    \end{align}
Let us consider the first expression of \eqref{eq:F-0-der}.  While $F^{(0)}_{ij}$ is 
manifestly symmetric
in $i,j$, this appears not to be the case for $\tilde F^{(0)}_{ij}$.  However, using the properties
\eqref{eq:matrix-sympl} of the matrices $U, V, W$ and $Z$, it follows that 
 $\tilde F^{(0)}_{ij}$ is symmetric in $i,j$.    Using this, we obtain
 \begin{equation}
\tilde{F}^{(0)}_{ij} (\hat{x}) \, Z^{jk} = V_i{}^k - [{\cal S}_0^{-1,T}]_i{}^k \;.
\label{tildeF-V}
\end{equation}
\vskip 1mm
\noindent
{\sf Exercise 18:  Verify \eqref{tildeF-V} by 
computing $V^T {\cal S}_0$.}
\vskip 1mm
\noindent
The symmetry of $\tilde F^{(0)}_{ij}$ implies that
${\tilde F}^{(0)}_i (\hat{x})$ can be integrated, i.e.
${\tilde F}^{(0)}_i (\hat{x}) = \partial {\tilde F}^{(0)} ( \hat{x}) / \partial \hat{x}^i$, with
$\tilde{F}^{(0)} (\hat{x})$ given by the well-known expression \cite{deWit:1996ix},
\begin{align}
\tilde{F}^{(0)} (\hat{x}) =&\, F^{(0)}(x) - \tfrac12 x^i F_i^{(0)} + \tfrac12 (U^T W)_{ij} \, x^i x^j
+ \tfrac12 (U^T V + W^T Z )_{i}{}^j x^i F_j^{(0)} \nonumber\\
&\,+ \tfrac12 (Z^T V)^{ij} F_i^{(0)} F_j^{(0)} \;,
\label{eq:F-0-transf}
\end{align}
up to a constant and up to terms linear in $\hat{x}^i$.

In addition to \eqref{eq:def-hat-S0}, 
we will also need 
the combinations
$ \mathcal{S}$ and $ \hat{\mathcal{S}}$ given in \eqref{eq:def-cal-S} and \eqref{eq:def-hat-S} below, which
are related to $\mathcal{S}_0$ by
\begin{align}
  {\mathcal{S}}^i{}_j =&\, 
    \mathcal{S}_0^{\,i}{}_j  
  +2\mathrm{i} Z^{ik} \Omega_{kj} \,,\nonumber \\
  \hat{\mathcal{S}}^i{}_j =&\, \mathcal{S}_0^{\,i}{}_j +Z^{ik}
  \big[2\mathrm{i}\Omega_{kj} -4\, \Omega_{k\bar l}
  \bar{\mathcal{Z}}^{\bar l\bar m}  \Omega_{\bar m j} \big]\,, \nonumber\\
   {\cal Z}^{ij} =&\, [ {\cal S}^{-1}]^i{}_k \, Z^{kj} \;.
  \label{eq:Z-Z0}
\end{align}
Observe that the matrices $\mathcal{Z}_0$, $\mathcal{Z}$ and 
$ \hat{\cal Z} = \hat{\cal S}^{-1}  \,  Z$ are symmetric matrices 
 by virtue of the fact that 
  $Z U^T$ is a symmetric matrix
  \cite{deWit:1996ix}.

Next we consider the transformation behavior of the derivatives $F_{ij} = \partial F_i / \partial x^j$ 
and $F_{i \bar \jmath} = \partial F_i / \partial {\bar x}^{\bar \jmath}$.
First we observe that
\begin{equation}
\frac{\partial \tilde x^i}{\partial x^j} \equiv {\cal S}^i{}_j = U^i{}_j + Z^{ik} F_{kj} \,,
\;\;\;\; \frac{\partial \tilde x^i}{\partial \bar x^{\bar \jmath}} \equiv Z^{ik} F_{k \bar \jmath} \;.
\label{eq:def-cal-S}
\end{equation}
Applying the chain rule to \eqref{eq:dual-F0-Omega} yields the relation
\begin{equation}
F_{ij} \rightarrow {\tilde F}_{ij} = \left( V_i{}^l \hat{F}_{lk} + W_{ik} \right)  [ \hat{\cal S}^{-1} ]^k{}_j \;,
\label{eq:f-tilde}
\end{equation}
where  $\tilde{F}_{ij} = \partial \tilde{F}_i / \partial \tilde{x}^j $ and 
\begin{align}
\hat{F}_{ij} =&\, F_{ij} - F_{i \bar k} \,{\cal Z}^{\bar k \bar l} \, \bar{F}_{\bar l j} = 
F^{(0)}_{ij}
  +2\mathrm{i}\Omega_{ij} -4\, \Omega_{i\bar k} \, \bar{\mathcal{Z}}^{\bar
     k\bar l}  \, \Omega_{\bar l j} \,,
\nonumber\\
\hat{\mathcal{S}}^i{}_j =&\,  U^i{}_j + Z^{ik} \hat{F}_{kj} \;.
      \label{eq:def-hat-S}
\end{align}
\vskip 1mm
\noindent
{\sf Exercise 19: Derive \eqref{eq:f-tilde} by differentiating 
the second equation of \eqref{eq:dual-F0-Omega} 
with respect to either $x$ or $\bar{x}$. Then combine the two
resulting equations to arrive at \eqref{eq:f-tilde}.}\vskip 1mm
\noindent
Then, using the first equation of 
\eqref{eq:F-0-der} as well as \eqref{tildeF-V} in \eqref{eq:f-tilde}
yields,
\begin{align}
  \label{eq:holo-holo-Omega}
  \tilde\Omega_{ij}(\tilde x,\bar{\tilde x}) =&\, \tfrac12 \mathrm{i}
  \big[\tilde F^{(0)}_{ij}(\tilde x) 
  -\tilde F^{(0)}_{ij}(\tilde x^k-2\mathrm{i}Z^{kl}\Omega_l(x,\bar x))
   \big] \\
  &\,+[\hat{\mathcal{S}}^{-1}]^k{}_i\,[\hat{\mathcal{S}}^{-1}]^l{}_j 
  \Big[\Omega_{kl} + 2\mathrm{i}\Omega_{k\bar
    m}\bar{\mathcal{Z}}^{\bar m\bar n} \Omega_{\bar n l}  \nonumber\\
  &\, \qquad \qquad \qquad \qquad +2\mathrm{i}(\Omega_{km} + 2\mathrm{i}\Omega_{k\bar
    p}\bar{\mathcal{Z}}^{\bar p\bar r} \Omega_{\bar r m}) \mathcal{Z}_0^{mn}   
   (\Omega_{nl} + 2\mathrm{i}\Omega_{n\bar q} \bar{\mathcal{Z}}^{\bar
     q\bar s}  \Omega_{\bar s l})\Big] \,, \nonumber
\end{align}
which is symmetric by virtue of the symmetry of ${\tilde F}^{(0)}_{ij}$, 
$\Omega_{ij}$, ${ \cal Z}^{mn}$ and ${\cal Z}_0^{mn}$.

Subsequently we derive the following result from
(\ref{eq:tilde-omega}) \cite{Cardoso:2008fr}, 
\begin{equation}
  \label{eq:holo-nonholo-Omega}
  \tilde\Omega_{i\bar\jmath} =
  [\hat{\mathcal{S}}^{-1}]^k{}_i\,[{\bar{\mathcal{S}}}^{-1}]^{\bar
    l}{}_{\bar\jmath}\, \Omega_{k\bar l}
  =[{\mathcal{S}}^{-1}]^k{}_i[\bar{\hat{\mathcal{S}}}^{-1}]^{\bar 
    l}{}_{\bar\jmath}\, \Omega_{k\bar l}  \,.
\end{equation}
\vskip 1mm
\noindent
{\sf Exercise 20: Deduce \eqref{eq:holo-nonholo-Omega} by 
taking the first line of \eqref{eq:tilde-omega} and differentiating it with respect
to $\bar x$. Use the relation \eqref{tildeF-V} in the form
\begin{equation}
V_i{}^j = [{\cal S}_0^{-1,T}]_i{}^k + (V_i{}^l F_{lk}^{(0)} + W_{ik} ) {\cal Z}_0^{kj} \;,
\end{equation}
together with \eqref{eq:Z-Z0}.}\vskip 1mm
\noindent
The relation \eqref{eq:holo-nonholo-Omega}
 establishes that $\tilde\Omega_{i\bar\jmath}  =  
\overline{({\tilde{\Omega}}_{j \bar\imath})}.$  Using this as well as \eqref{eq:rel-omega-bari_cond},
and recalling that
$\tilde\Omega_{i\bar\jmath} = \partial \tilde\Omega_i/\partial {\bar{\tilde x}}^{ \bar \jmath}$,
we obtain $\tilde\Omega_{i\bar\jmath} =
\overline{({\tilde{\Omega}}_{j \bar\imath})} = \overline{(\partial \tilde\Omega_j/\partial {\bar{\tilde x}}^{ \bar \imath})} = \partial (\overline{\tilde{{\Omega}_{j}}})/\partial {\tilde x}^{i}
= \partial \tilde\Omega_{\bar \jmath}/\partial {\tilde x}^{i} \equiv \tilde\Omega_{\bar\jmath i}
$.
This, together with the symmetry of $\tilde\Omega_{ij}$, 
ensures the integrability of
\eqref{eq:dual-F0-Omega}, as follows.

We consider the 1-form $\tilde A = \tilde \Omega_i \, d {\tilde x}^i + \tilde \Omega_{\bar \imath}
\, d \bar{\tilde x}^{\bar \imath}$, which is real by virtue of 
$\tilde{\Omega}_{\bar \imath} = (\overline{\tilde{{\Omega}_i} }) $.
Its field strength reads 
$\tilde {\mathfrak{F}} = d \tilde A = \tilde \Omega_{ij} \, d {\tilde x}^j \wedge
\, d {\tilde x}^i + \left( \tilde \Omega_{i \bar \jmath} 
- \tilde \Omega_{\bar \jmath i} \right) 
\, d \bar{\tilde x}^{\bar \jmath}
\wedge d \tilde x^i + \tilde \Omega_{\bar \imath \bar \jmath} \, d \bar{\tilde x}^{\bar \jmath}
\wedge d
\bar{\tilde x}^{\bar \imath}$.  Then, using $\tilde \Omega_{ij} = \tilde \Omega_{ji}$ as well as 
$\tilde \Omega_{i \bar \jmath} = \tilde \Omega_{\bar \jmath i}$, we conclude that $\tilde{\mathfrak{F}}=0$,
which establishes  that locally $\tilde A = d \tilde \Omega$, with a real $\tilde \Omega$.

Hence we conclude that 
the equations 
\eqref{eq:dual-F0-Omega} are integrable and the decomposition \eqref{eq:F(x)}
is preserved, i.e. the transformed  $2n$-vector $(\tilde x^i, \tilde F_i)$
is constructed from a new function $\tilde F (\tilde x, \bar{\tilde x}) =  
\tilde{F}^{(0)} (\tilde x) + 2 \mathrm{i} \tilde
\Omega(\tilde x, \bar{\tilde x})$ with a real $\tilde
\Omega(\tilde x, \bar{\tilde x})$.  This was established in subsection  \ref{sec:theor}
by relying on the Hamiltonian.

Next, let us assume that the function $F$ depends on a auxiliary
real parameter $\eta$ that is inert under symplectic transformation,
i.e. $F(x, \bar x; \eta)$, 
 and let us consider partial derivatives with respect to
it.  A little calculation shows that $\partial_\eta F_i$ transforms in
the following way,
\begin{equation}
  \label{eq:auxiliary-eta}
  \partial_\eta \tilde F_i = [\hat {\cal S}{}^{-1}]^j{}_i \left[
  \partial_\eta F_j - F_{j\bar k} \,\bar{\mathcal{Z}}{}^{\bar k\bar l}
  \,\partial_\eta \bar F_{\bar l}\right] \,,
\end{equation}
where $\tilde x$ and $\bar{\tilde x}$ are kept fixed under the
$\eta$-derivative in $\partial_\eta \tilde F_i(\tilde x,\tilde{\bar
  x};\eta)$, while in $\partial_\eta F_i(x,\bar x;\eta)$ the arguments
$x$ and $\bar x$ are kept fixed. 
\vskip 1mm
\noindent
{\sf Exercise 21: Verify \eqref{eq:auxiliary-eta} by differentiating the second equation of 
\eqref{eq:dual-F0-Omega} with respect to $\eta$, keeping $x$ and $\bar x$ fixed.  Subsequently,
use \eqref{decom-tilde-om},
\eqref{eq:f-tilde} and \eqref{eq:holo-nonholo-Omega} to arrive at \eqref{eq:auxiliary-eta}.}
\noindent

Let us first consider \eqref{eq:auxiliary-eta}
in the case of a holomorphic function $F$, so that
$\Omega=0$. In that case \eqref{eq:auxiliary-eta} implies that the
derivative with respect to $x^i$ of $\partial_\eta \tilde
F- \partial_\eta F$ must vanish. Therefore it follows that
$\partial_\eta F$ transforms as a function under symplectic transformations
(possibly up to an $x$-independent expression, which is
irrelevant 
in view of the same argument that led to the equivalence
\eqref{eq:ambiguity}). 

When $\Omega\not=0$ one derives the following
result using \eqref{eq:auxiliary-eta},
\begin{equation}
  \label{eq:eta-der}
  \frac{\partial \tilde x^j}{\partial x^i}\,\partial_\eta\tilde F_j - 
  \frac{\partial \bar{\tilde x}^{\bar \jmath}}{\partial
    x^i}\,\partial_\eta (\overline{{{\tilde F}_{j}}} )
     = \partial_\eta F_i\,.  
\end{equation}
\vskip 1mm
\noindent
{\sf 
Exercise 22: Deduce \eqref{eq:eta-der} by suitably combining \eqref{eq:auxiliary-eta} with its complex
conjugate, and using the relation}
\begin{equation}
\bar{\cal Z}^{\bar \imath \bar \jmath} \, {\bar F}_{\bar \jmath k} \, [\hat{{\cal S}}^{-1} {\cal S}] ^k{}_l = [ 
\bar{\hat{\mathcal{S}}}^{-1}
{\bar {\cal S}} ]^{\bar \imath}{}_{\bar \jmath} \, \bar{\cal Z}^{\bar \jmath \bar k} \,
\bar{F}_{\bar k l} \;.
\end{equation}
\vskip 1mm

Next, we assume without loss of generality that the
dependence of $\tilde F$ on $\eta$ is entirely contained in $\tilde \Omega$.  Then, using 
\eqref{eq:rel-omega-bari_cond},
it follows that
\begin{equation}
\partial_\eta  (\overline{{{\tilde F}_{j}}} )
 = - \partial_\eta\tilde F_{\bar \jmath} \;,
 \label{cond_eta-omega}
\end{equation}
and the relation \eqref{eq:eta-der} simplifies.  Namely,
the left hand side of \eqref{eq:eta-der} becomes equal to $\partial 
( \partial_{\eta} {\tilde F} ) / \partial x^i$, where we used the existence of the new function $\tilde F$.
Thus, we obtain from \eqref{eq:eta-der},
\begin{equation}
\frac{\partial}{\partial x^i} \left( \partial_{\eta} \tilde F - \partial_{\eta} F \right) = 0 \;.
\label{eq:trafo-F-eta-tilde}
\end{equation}
This equation, together with its complex conjugate equation, implies that 
$\partial_\eta\tilde F-\partial_\eta F$ vanishes upon
differentiation with respect to $x$ and $\bar x$, so that
$\partial_\eta  F$ transforms as a function under symplectic transformations
(possibly up to an irrelevant term that is independent of $x$ and $\bar x$).

\section{The covariant derivative $\cal D_{\eta}$ \label{cov-der}}

The modified derivative \eqref{eq:cov-der-multiple} acts as a covariant derivative for symplectic transformations.
Here we verify this explicitly by showing that, given a quantity $G(x, \bar x; \eta)$ that transforms
as a function under symplectic transformations, also ${\cal D}_{\eta} G$ transforms
as a function.  

To establish this, we need the transformation law
of $\hat{N}^{ij}$ that enters in \eqref{eq:cov-der-multiple}.
Under symplectic transformations, $\hat{N}_{ij}$ given in \eqref{eq:hat-N-ij} transforms as 
\begin{align}
\tilde{\hat{N}}_{ij} =&\, 
[\hat {\cal S}^{-1}]^k{}_i \, [{\bar {\hat {\cal S}}}^{-1} ]^{\bar l}{}_{\bar \jmath}
\left[\hat{N}_{kl} + \mathrm{i} \, F_{k \bar m} \, {\bar {\cal Z}}^{\bar m \bar n} \, {\bar F}_{\bar n p } 
\left( \delta^p_l - {\cal Z}^{pq} \, F_{q \bar l}
\right) \right. \nonumber\\
&\, \left. \qquad \qquad \qquad \qquad \qquad 
- \mathrm{i} \,{\bar F}_{\bar k m} \, {\cal Z}^{mn} \, {F}_{n \bar p} \left( \delta^{\bar p}_{\bar l} - 
{\bar {\cal Z}}^{\bar p \bar q} \, {\bar F}_{\bar q l }
\right) 
\right] \nonumber\\
&\, +  \mathrm{i} \, [\hat {\cal S}^{-1}]^k{}_i \, [{\bar{\hat {\cal S}}}^{-1}]^{\bar l}{}_{\bar \jmath}
\, {\bar F}_{\bar k m} \, [{\cal S}^{-1} \, {\hat {\cal S}}]^m{}_l - \mathrm{i}
[{\bar {\hat {\cal S}}}^{-1} ]^{\bar k}{}_{\bar \imath} \, {\bar F}_{\bar k l} \, 
[{\cal S}^{-1}]^l{}_j \,,
\label{eq:N-til-multiple}
\end{align}
where ${\cal S}, \hat{\cal S}$ and ${\cal Z}$ are defined in \eqref{eq:Z-Z0}.
\vskip 1mm
\noindent
{\sf Exercise 23: Verify \eqref{eq:N-til-multiple} using \eqref{eq:f-tilde} and 
\eqref{eq:holo-nonholo-Omega}.}
\vskip 1mm

\noindent
Then, it follows that the inverse matrix $\hat{N}^{ij}$ transforms as
\begin{eqnarray}
\tilde{\hat{N}}^{ij}
= \left( {\cal S}^i{}_l - Z^{in}\, F_{n \bar l} \right) \, \hat{N}^{lk} \, 
\left( {\cal S}^j{}_k - Z^{jm}\, F_{m \bar k} \right)
- \mathrm{i} \, {\cal S}^i{}_k \, {\cal Z}^{kl} \,  {\cal S}^j{}_l \;.
\label{eq:transf-Ninv-ij}
\end{eqnarray}
Since the matrix 
${\cal Z} = {\cal S}^{-1} \, Z$ is symmetric \cite{deWit:1996ix}, so is 
$\tilde{\hat{N}}^{ij}$.  Observe that it can also be written as
\begin{eqnarray}
\tilde{\hat{N}}^{ij}
= \left( {\cal \bar S}^{\bar\imath}{}_{\bar l} - Z^{in}\, {\bar F}_{{\bar n} l} \right) \, \hat{N}^{lk} \, 
\left( {\cal S}^j{}_k - Z^{jm}\, F_{m \bar k} \right)
- \mathrm{i} \, Z^{il} \, Z^{jm} \, F_{l \bar m}
\;.
\label{eq:Ninv-transf}
\end{eqnarray}
Establishing the transformation behavior \eqref{eq:transf-Ninv-ij} turns out to be a 
tedious exercise, which we relegate to end of this appendix.

Now consider a quantity $G(x, {\bar x}; \eta)$ that 
transforms as a function
under symplectic transformations,
i.e. $G(x, {\bar x}; \eta) = {\tilde G}({\tilde x},
{\bar {\tilde x}}; \eta)$. We then calculate the behavior of
${\cal D}_{\eta} G$ under symplectic transformations.  First we establish
\begin{equation}
G_{\eta} = {\tilde G}_{\eta} + {\tilde G}_i \, Z^{ij} \,  F_{\eta  j}
+ {\tilde G}_{\bar\imath} \,Z^{ij} \, 
{\bar F}_{\eta {\bar\jmath}} \;,
\label{eq:rel-Gg-Ggdual}
\end{equation}
where, on the right hand side, the tilde quantities are differentiated with respect
to the tilde variables, while those without a tilde are differentiated with respect
to the original variables.  
Similarly,
\begin{equation}
G_i - G_{\bar\imath} = \left({\tilde G}_j - {\tilde G}_{\bar\jmath}\right) 
\left( {\cal S}^j{}_i - Z^{jk} \, F_{k \bar\imath }\right) + \mathrm{i} \, {\tilde G}_{\bar\jmath}
\, Z^{jk} \, \hat{N}_{ki} \;,
\label{eq:diff-G-G}
\end{equation}
as well as 
\begin{equation}
F_{\eta j} = {\tilde F}_{\eta i} \, {\cal S}^i{}_j + {\tilde F}_{\eta \bar\imath} \, Z^{ik} \, 
{\bar F}_{{\bar k} j} \;,
\label{eq:rel-Fg2-dual}
\end{equation}
where  we used that $F_{\eta}$ transforms as a symplectic function, as established in 
\eqref{eq:trafo-F-eta-tilde}.
\vskip 1mm
\noindent
{\sf Exercise 24: Verify \eqref{eq:rel-Gg-Ggdual} 
and \eqref{eq:diff-G-G}
using  $G(x, {\bar x}; \eta) = {\tilde G}({\tilde x},
{\bar {\tilde x}}; \eta)$.}
\vskip 1mm

\noindent
Then, inserting \eqref{eq:rel-Gg-Ggdual} and \eqref{eq:diff-G-G} into \eqref{eq:cov-der-multiple} yields,
\begin{equation}
{\cal D}_{\eta} G = 
{\tilde G}_{\eta} + \left({\tilde G}_i - {\tilde G}_{\bar\imath}\right) Z^{ij} \, F_{\eta j}
+ \mathrm{i} \, \hat{N}^{ij} \left( F_{\eta j} + {\bar F}_{\eta \bar\jmath} \right) 
\left({\tilde G}_k - {\tilde G}_{\bar k}\right) \left( {\cal S}^k{}_i - Z^{kl} \, F_{l \bar\imath }\right) \;.
\label{eq:cov-der-sympl}
\end{equation}
Next, using \eqref{eq:rel-Fg2-dual},
 we compute
\begin{align}
\left( F_{\eta j} + {\bar F}_{\eta \bar\jmath} \right) =&\, \left( {\tilde F}_{\eta k} + {\bar {\tilde F}}_{\eta\bar k} \right)
\left( {\cal S}^k{}_j - Z^{kl} \, F_{l \bar\jmath }\right) - \mathrm{i} \, {\bar {\tilde F}}_{\eta \bar l} \, Z^{lk}
\, \hat{N}_{kj} \nonumber\\
&\, +  \left( {\tilde F}_{\eta l} + {\bar {\tilde F}}_{\eta l} \right) Z^{lk} \, F_{k \bar j}
+  \left( {\tilde F}_{\eta \bar l} + {\bar {\tilde F}}_{\eta \bar l} \right)  Z^{lk} \, {\bar F}_{\bar k j} \;.
\label{eq:cov-der-dual}
\end{align}
Using that $\tilde F$ has the decomposition 
\begin{equation}
{\tilde F} ({\tilde x}, {\bar {\tilde x}};\eta) = {\tilde F}^{(0)} ({\tilde x}) + 2 \mathrm{i} \, 
{\tilde \Omega}({\tilde x}, {\bar {\tilde x}};\eta)
\end{equation}
with ${\tilde \Omega}$ real, 
it follows that the second line of \eqref{eq:cov-der-dual} vanishes.  Inserting the first line of 
\eqref{eq:cov-der-dual} into \eqref{eq:cov-der-sympl} and using $F_{i \bar\jmath} = - {\bar F}_{\bar\jmath i}$
as well as ${\cal S} \, Z^T = Z \, {\cal S}^T$, we obtain
\begin{equation}
{\cal D}_{\eta} G = {\tilde G}_{\eta} + \mathrm{i} \, {\tilde N}^{ij} \left( 
{\tilde F}_{\eta j} + \bar{\tilde{F}}_{\eta \bar\jmath} \right) \left(\tilde{G}_{i}
- \tilde{G}_{\bar\imath} \right) = \widetilde{\left( {\cal D}_{\eta} G \right)} \;,
\label{eq:cov-der-sympl-fin}
\end{equation}
which shows that ${\cal D}_{\eta} G$ transforms as a function under symplectic transformations.

Now we return to the transformation behavior of ${\hat{N}}^{ij}$ given in 
\eqref{eq:transf-Ninv-ij} and verify that it
is the inverse of \eqref{eq:N-til-multiple}, 
i.e. $\tilde{\hat{N}}^{-1} \, \tilde{\hat{N}} = \mathbb{I}$. We use the decomposition 
$F(x, \bar x; \eta) = F^{(0)}(x)  + 2 \mathrm{i} \Omega(x, \bar x ; \eta)$.
We find it useful to introduce the following matrix notation,
\begin{align}
{\bar {\cal S}}^{-1} \, {\cal S} =&\, \mathbb{I} + {\bar {\cal Z}} \left(F_{\cdot \cdot} - {\bar F}_{\---} \right) \;,
\nonumber\\
{\cal S}^{-1} \, {\hat {\cal S}} =&\, \mathbb{I} - X \;\;\;,\;\;\; X = {\cal Z} \, F_{\cdot\,\--} \, 
{\bar {\cal Z}} {\bar F}_{\-- \cdot}
= 4 \, {\cal Z} \, \Omega_{\cdot \,\--} \, {\bar {\cal Z}} \,
{\Omega}_{\-- \cdot}
 \;\;\;,\nonumber\\
{\hat {\cal S}}^{-1} {\cal S} =&\, \left(\mathbb{I} - X \right)^{-1} = \sum_{n=0}^{\infty} X^n \;,
\nonumber\\
{\bar {\hat {\cal S}}} =&\, {\cal S} \left[ \mathbb{I} - X - {\cal Z} \left( {\hat F}_{\cdot \cdot} - 
{\bar {\hat F}}_{\--- } \right) \right] =
\left[ \mathbb{I} - {\cal Z} \left(F_{\cdot \cdot } - {\bar F}_{\---}
\right) - 4 \,{\cal Z} \, \Omega_{\-- \cdot} \, {\cal Z} \, \Omega_{\cdot \--} \right] \;,
\nonumber\\
{\cal Z} - {\bar {\cal Z}} =&\, - {\bar {\cal Z}} \left(F_{\cdot \cdot } - {\bar F}_{\---} \right) {\cal Z} = 
- {\cal Z} \left(F_{\cdot \cdot } - {\bar F}_{\---} \right) {\bar {\cal Z}} \;,
\label{eq:relS-hatS}
\end{align}
where we assume that the power series expansion of  
${\cal S}^{-1} \, {\hat {\cal S}}$ is convergent.  Here $F_{\cdot \cdot } \;,\; F_{\---} \;,\; F_{\cdot\--}$ denote entries of the type
$F_{ij}, F_{\bar \imath \bar \jmath}, F_{i \bar \jmath}$, respectively.
Then, using \eqref{eq:N-til-multiple}, we compute
\begin{align}
{\cal S}^T \, \tilde{\hat{N}} \, {\bar {\hat {\cal S}}} =&\,
\sum_{n=0}^{\infty} \left(X^n\right)^T \left( \hat{N} 
+ 4 \mathrm{i} \, \Omega_{\cdot \--} \, {\bar {\cal Z}} \,
\Omega_{\-- \cdot} - 4 \mathrm{i} \Omega_{\-- \cdot} \, {\cal Z} \, \Omega_{\cdot\--}
+ 2 \Omega_{\cdot \--} \, {\bar X} + 2 \Omega_{\--\cdot}
\right)  \\
&\,   - 2 \left( {\bar {\cal S}}^{-1} \, {\cal S} \right)^T 
 \sum_{n=0}^{\infty} \left({\bar X}^n\right)^T \, \Omega_{\-- \cdot} 
 \left[ \mathbb{I} - {\cal Z} \left(F_{\cdot \cdot } - {\bar F}_{\---}
\right) - 4 \,{\cal Z} \, \Omega_{\-- \cdot} \, {\cal Z} \, \Omega_{\cdot \--} \right]\;. \nonumber
\end{align}
Multiplying this with $\tilde{\hat{N}}^{-1} \, {\cal S}^{-1,T}$ from the left and requiring the resulting
expression to equal ${\bar {\hat {\cal S}}}$ yields the relation
\begin{eqnarray}
&& 
\left[ \hat{N}^{-1} - 2 \mathrm{i} \, \hat{N}^{-1} \, \Omega_{\-- \cdot} \, {\cal Z} - 2 \mathrm{i} \,
{\cal Z} \, \Omega_{\cdot\--} \, \hat{N}^{-1} - 4 {\cal Z} \, \Omega_{\cdot \--} \, \hat{N}^{-1} \, \Omega_{\-- \cdot} 
\, {\cal Z} - \mathrm{i} {\cal Z} \right] \nonumber\\
&& \qquad 
\quad \left[ \sum_{n=0}^{\infty} \left(X^n\right)^T \left( \hat{N} + 4 \mathrm{i} \, \Omega_{\cdot \--} \, 
{\bar {\cal Z}} \,
\Omega_{\-- \cdot} - 4 \mathrm{i} \Omega_{\-- \cdot} \, {\cal Z} \, \Omega_{\cdot\--}
+ 2 \Omega_{\cdot \--} \, {\bar X} + 2 \Omega_{\-- \cdot}
\right) \right. \nonumber\\
&& \qquad \qquad \left. - 2 \left( {\bar {\cal S}}^{-1} \, {\cal S} \right)^T 
 \sum_{n=0}^{\infty} \left({\bar X}^n\right)^T \, \Omega_{\-- \cdot} 
 \left[ \mathbb{I} - {\cal Z} \left(F_{\cdot \cdot } - {\bar F}_{\---}
\right) - 4 \,{\cal Z} \, \Omega_{\-- \cdot} \, {\cal Z} \, \Omega_{\cdot \--} \right] \right]\nonumber\\
&& = 
\left[
\mathbb{I} - {\cal Z} \left(F_{\cdot \cdot} - {\bar F}_{\---}
\right) - 4 \,{\cal Z} \, \Omega_{\--  \cdot} \, {\cal Z} \, \Omega_{\cdot \--}  \right] \;.
 \label{eq:NtilNlhs}
\end{eqnarray}
Thus, checking $\tilde{\hat{N}}^{-1} \, \tilde{\hat{N}} = \mathbb{I}$ amounts to verifying the relation \eqref{eq:NtilNlhs}. To do so, 
we write \eqref{eq:NtilNlhs} as a power series in ${\cal Z}$ by converting $\bar{\cal Z}$ 
into ${\cal Z}$ using the last relation in \eqref{eq:relS-hatS}. 
Introducing the expressions
\begin{align}
\sigma =&\, 4 \, \Omega_{\-- \cdot} \, {\cal Z} \, \Omega_{\cdot \--} \, {\cal Z} \;, \nonumber\\
\Delta =&\, \sum_{n=1}^{\infty} \left[ \left(F_{\cdot \cdot } - {\bar F}_{\---} \right) {\cal Z} \right]^n \;,
\end{align}
we obtain
\begin{align}
{\bar X} =&\, 4 \, {\cal Z} \left(\mathbb{I}
 + \Delta \right) \, \Omega_{\-- \cdot} \, {\cal Z} \, \Omega_{\cdot\--} \;, \nonumber\\
{\bar X}^T \, \Omega_{\-- \cdot} =&\, \Omega_{\-- \cdot} \, X \;, \nonumber\\
\sum_{n=0}^{\infty} \left({\bar X}^n\right)^T \, \Omega_{\-- \cdot} =&\, \Omega_{\-- \cdot} \, \sum_{n=0}^{\infty} 
 X^n \;, \nonumber\\
 X^n =&\, 4 \, {\cal Z} \, \Omega_{\cdot\--} \, {\cal Z} \left[ \left(\mathbb{I} + \Delta\right) \sigma \right]^{n-1} 
 (\mathbb{I} + \Delta) \, \Omega_{\-- \cdot} \;\;\;,\;\;\; n \geq 1 \;, \nonumber\\
  \left(X^n\right)^T =&\, 4 \, \Omega_{\cdot\--} ( \mathbb{I} + \Delta^T) \left[ \sigma^T (\mathbb{I} + \Delta^T) \right]^{n-1} 
  {\cal Z} \, \Omega_{\-- \cdot} \, {\cal Z} \;\;\;,\;\;\; n \geq 1 \;,\nonumber\\
  \left( {\bar {\cal S}}^{-1} \, {\cal S} \right)^T =&\, \mathbb{I} + 
  \left(F_{\cdot \cdot} - {\bar F}_{\---} \right) {\cal Z}  \left( \mathbb{I} + \Delta \right) \;.
  \label{eq:exp-X-Delta}
  \end{align}
Then, \eqref{eq:NtilNlhs} becomes
\begin{eqnarray}
&& 
\left[ \mathbb{I} - 2 \mathrm{i}  \, \Omega_{\-- \cdot} \, {\cal Z} - 2 \mathrm{i} \, \hat{N} \, 
{\cal Z} \, \Omega_{\cdot \--} \, \hat{N}^{-1} - 4 \, \hat{N} \, {\cal Z} \, \Omega_{\cdot \--} \, \hat{N}^{-1} \, 
\Omega_{\-- \cdot} 
\, {\cal Z} - \mathrm{i} \, \hat{N} \, {\cal Z} \right] \nonumber\\
&& 
\left[ \sum_{n=0}^{\infty} \left(X^n\right)^T \left[ \hat{N} + 4 \mathrm{i} \, \Omega_{\cdot \--} \, 
{\cal Z} \,
\left( \mathbb{I} + \Delta \right)
\Omega_{\-- \cdot} - 4 \mathrm{i} \Omega_{\-- \cdot} \, {\cal Z} \, \Omega_{\cdot \--}
+ 8 \Omega_{\cdot \--} \, {\cal Z} \left(\mathbb{I} + \Delta \right) 
\Omega_{\-- \cdot} \, {\cal Z} \, \Omega_{\cdot \--}
 + 2 \Omega_{\-- \cdot}
\right] \right. \nonumber\\
&& 
\qquad 
\left. - 2 \left[\mathbb{I} + 
  \left(F_{\cdot \cdot} - {\bar F}_{\---} \right) {\cal Z}  \left( \mathbb{I} + \Delta \right)\right]
 \Omega_{\--\cdot} \, \sum_{n=0}^{\infty} X^n
 \left[ \mathbb{I} - {\cal Z} \left(F_{\cdot \cdot} - {\bar F}_{\---}
\right) - 4 \,{\cal Z} \, \Omega_{\-- \cdot} \, {\cal Z} \, \Omega_{\cdot \--} \right] \right]\nonumber\\
&& = \hat{N} \, \left[\mathbb{I} - {\cal Z} \left(F_{\cdot \cdot } - {\bar F}_{\---}
\right) - 4 \,{\cal Z} \, \Omega_{\-- \cdot} \, {\cal Z} \, \Omega_{\cdot  \--}  \right] \;,
 \label{eq:NtilNlhs-rew}
\end{eqnarray}
where $X^n$ (with $n\geq 1$) is expressed in terms of ${\cal Z}$ according to \eqref{eq:exp-X-Delta}.
Now we proceed to check that \eqref{eq:NtilNlhs-rew} is indeed satisfied, order by order in ${\cal Z}$.
Observe that the right hand side of  \eqref{eq:NtilNlhs-rew} is quadratic in ${\cal Z}$, so first we
check the cancellation of the terms up to order ${\cal Z}^2$.  Then we proceed to check the terms at order $n$
with $n\geq 3$.  Here we use the relations
\begin{align}
F_{\cdot \cdot } - {\bar F}_{\---} =&\,
 \mathrm{i} \hat{N} + 2 \mathrm{i} \Omega_{\cdot\--} + 2 \mathrm{i} \Omega_{\-- \cdot} \;,
\nonumber\\
\Delta^T \, {\cal Z} =&\, {\cal Z} \, \Delta \;, \nonumber\\
\left[ \sigma^T \left(\mathbb{I} + \Delta^T \right) \right]^n {\cal Z} =&\, {\cal Z} \left[\sigma \left(\mathbb{I} + \Delta \right) \right]^n \;,
\end{align}
and we organize the terms at order $n$ into those that end on either $N$ (introduced in 
\eqref{eq:hat-N-N}),
$\Omega_{\cdot \--}$ or $\Omega_{\-- \cdot}$.
It is then straightforward, but tedious, to check that at order $n$ in ${\cal Z}$ all these terms cancel out.
This establishes the validity of the transformation law \eqref{eq:transf-Ninv-ij}.


\section{The holomorphic anomaly equation in big moduli space\label{sec:top-string}}

The holomorphic anomaly equation \eqref{eq:holom-top-string}
of perturbative topological string theory 
\cite{Bershadsky:1993ta,Bershadsky:1993cx}
can be suscintly derived in the wave function approach \cite{Witten:1993ed} to the latter 
\cite{Dijkgraaf:2002ac,Verlinde:2004ck,Aganagic:2006wq,Gunaydin:2006bz}.  
In this approach, 
the topological string partition function $Z$
is represented by a wavefunction, 
\begin{equation}
 Z(t; t_B, {\bar t}_B ) = \int d\phi  \, {\rm e}^{-S(\phi,t; t_B, {\bar t}_B)/\hbar} \, 
Z(\phi) \;,
\label{eq:overlap-int}
\end{equation}
where $S(\phi,t; t_B, {\bar t}_B)$ denotes the generating function \eqref{eq:Sf-back} of canonical transformations\footnote{We use the conventions of section 
 \ref{sec:hesse-top} and suppress the superscript of $F^{(0)}$.}.
We take the background dependent constant
$c(t_B, {\bar t}_B)$ appearing in $S$ to be given by \cite{Dijkgraaf:2002ac,Verlinde:2004ck,Aganagic:2006wq,Gunaydin:2006bz}
\begin{equation}
 c(t_B, {\bar t}_B) = - \frac{\hbar}{2} \,  \ln \det N_{IJ}(t_B, {\bar t}_B) \;, 
 \label{eq:c-ver-f1}
 \end{equation}
with $N_{IJ}$ as in \eqref{eq:def-N}.

Differentiating \eqref{eq:overlap-int} with respect to the background field ${\bar t}_B$ on the one hand, 
and with respect
to the fluctuations $t$ on the other hand, yields the relation \cite{Verlinde:2004ck},
\begin{eqnarray}
\label{eq:anomal-new}
 \frac{\partial Z(t; t_B, {\bar t}_B)}{\partial {\bar t}_B^L} =  \frac{\hbar }{2} \, {\bar F}_{\bar L}{}^{IJ} \, 
\frac{\partial}{\partial t^I}  \frac{\partial}{\partial t^J} Z(t; t_B, {\bar t}_B)\;.
\end{eqnarray}
Here  ${\bar F}_{\bar L}{}^{IJ}$ is evaluated on the background, and 
is given by
${\bar F}_{\bar L}{}^{IJ} = 
{\bar F}_{\bar L \bar M \bar O} N^{MI} N^{OJ}$.  Assigning scaling dimension $1$ to both $t_B$ and $t$ (and to
their complex conjugates)
and scaling dimension $2$ to $\hbar$, we see that \eqref{eq:anomal-new} has scaling dimension $-1$.
Setting
\begin{equation}
Z(t; t_B, {\bar t}_B) = {\rm e}^{W(t; t_B, {\bar t}_B)/\hbar} \;,
\end{equation}
we obtain from \eqref{eq:anomal-new}
\begin{equation}
\frac{ \partial W(t; t_B, {\bar t}_B)}{\partial {\bar t}_B^L} = \tfrac12 \, {\bar F}_{\bar L}{}^{IJ} \, 
\left( \hbar \, \frac{\partial^2 W }{\partial t^I \, \partial t^J} 
+ \frac{\partial W}{\partial t^I}  \, \frac{\partial W}{\partial t^J} 
\right)
\;,
\label{eq:anomal-W}
\end{equation}
which has scaling dimension $1$.
The BCOV-solution \cite{Bershadsky:1993cx} 
is obtained by making the ansatz \cite{Grimm:2007tm}
\begin{eqnarray}
W = \sum_{g=0,n=0}^{\infty} \frac{\hbar^g}{n!} \, 
C^{(g)}_{I_1 \dots I_n} (t_B, {\bar t}_B)  \, t^{I_1} \dots t^{I_n} \;,
\label{eq:ansatz-W}
\end{eqnarray}
with
\begin{equation}
 C^{(g)}_{I_1 \dots I_n} = 0 \;\;\;,\;\;\; 2g - 2 + n \leq 0 \;.
 \label{eq:cond_C}
 \end{equation}
The $ C^{(g)}_{I_1 \dots I_n}$ are symmetric in $I_1, \dots, I_n$ and have scaling dimension $2 - 2g -n$.
Inserting the ansatz \eqref{eq:ansatz-W} into \eqref{eq:anomal-W}, 
equating the terms of order $\hbar^g$ for $g \geq 2$ and setting $t=0$ gives,
\begin{equation}
 \partial_{\bar L} C^{(g)}(t_B, {\bar t}_B) = \tfrac12 \, {\bar F}_{\bar L}{}^{IJ} \, 
\left( C^{(g-1)}_{IJ}
+ \sum_{r=1}^{g-1} C^{(r)}_{I} \, C^{(g-r)}_{J}
\right)
\;\;\;,\;\;\; g \geq 2 \;.
\label{eq:anomal-W-C}
\end{equation}
\vskip 1mm
\noindent
{\sf Exercise 25: Verify \eqref{eq:anomal-W-C}.}
\vskip 1mm
\noindent
Now we set \cite{Grimm:2007tm}
\begin{eqnarray}
C^{(g)}_{I_1 \dots I_n} = D_{I_1} \dots D_{I_n} F^{(g)} \;\;\;, \;\;\; g\geq 1 \;,
\label{eq:sol-g-geq-2}
\end{eqnarray}
where $D_L$ is given by 
\begin{equation}
D_L V_M = \partial_L V_M + \mathrm{i} \,  N^{PI} F_{ILM}
V_P \;.
\end{equation}
$D_L$ acts as a covariant derivative for symplectic reparametrizations $V_M \rightarrow \left({\cal S}_0^{-1}
\right)^P{}_M \, V_P$, since $N^{IJ}$ transforms as $N^{IJ} \rightarrow 
[{\cal S}_0  \, N^{-1} \, {\cal S}_0 ]^{IJ} - \mathrm{i} [{\cal S}_0 \, {\cal Z}_0 \, {\cal S}_0]^{IJ}$
(see \eqref{eq:transf-Ninv-ij}).
The $F^{(g)}$ have scaling dimension $2 - 2g$ and transform as functions under symplectic
transformations.  Inserting \eqref{eq:sol-g-geq-2} into \eqref{eq:anomal-W-C}
yields the
holomorphic anomaly equation in big moduli space \cite{Grimm:2007tm},
\begin{equation}
 \partial_{\bar L} F^{(g)}(t_B, {\bar t}_B) = \tfrac12 \, {\bar F}_{\bar L}{}^{IJ} \, 
\left( D_I \partial_J F^{(g-1)}
+ \sum_{r=1}^{g-1} \partial_I F^{(r)}\, \partial_J F^{(g-r)}
\right)
\;\;\;,\;\;\; g \geq 2 \;.
\label{eq:non-holo-Fg-big}
\end{equation}

As an example, consider solving \eqref{eq:non-holo-Fg-big} for $g=2$.
We need $F_I^{(1)} = 
\partial_I F^{(1)}(t_B, \bar t_B)$, which is non-holomorphic and given by\footnote{$F^{(1)}$ contains an additional term proportional
to the K\"ahler potential \eqref{eq:spec-K-pot}, but this term drops out of \eqref{eq:non-holo-Fg-big} due to the
special geometry relation $\bar F_{\bar I \bar J \bar K} \ {\bar t}^{\bar K} = 0$.}
\begin{eqnarray}
\partial_I F^{(1)}(t_B, \bar t_B) = \partial_I f^{(1)}(t_B) + \tfrac12 \, \mathrm{i} \, F_{IJK} \, N^{JK} \;.
 \label{eq:F1-nh-hat}
\end{eqnarray}
Then, solving \eqref{eq:non-holo-Fg-big} for
$F^{(2)}$ yields
\cite{Aganagic:2006wq,Grimm:2007tm}
\begin{align}
 F^{(2)}(t_B, \bar t_B) =&\, f^{(2)}(t_B) 
 + \tfrac12 \, \mathrm{i} \, 
{N}^{IJ} \left(D_I {F}_J^{(1)} +  {F}^{(1)}_I  {F}^{(1)}_J \right) \nonumber\\
&\, + \tfrac12  {N}^{IJ} {N}^{KL} \left( \tfrac14  F_{IJKL} 
+ \tfrac13 \mathrm{i} \, N^{MN} F_{IKM} F_{JLN} + F_{IJK} {F}^{(1)}_L
\right) \;.
\label{eq:G2-exp}
\end{align}
In this expression, all the terms are evaluated on the background $(t_B, \bar t_B)$.
\\[1mm]
{\sf Exercise 26: Verify that \eqref{eq:G2-exp} solves \eqref{eq:non-holo-Fg-big}.}
\\[1mm]
Observe that 
\eqref{eq:F1-nh-hat}
transforms covariantly under symplectic transformations, provided that
$f^{(1)}$
transforms as $f^{(1)} \longrightarrow f^{(1)} - \tfrac12  \ln \det \mathcal{S}_0$ in order to compensate
for the transformation behavior $N_{IJ} \longrightarrow N_{KL} \, [ \bar{\mathcal{S}_0}^{-1}]^K{}_I \, [ \mathcal{S}_0^{-1} ]^L{}_J$
\cite{deWit:1996ix}, so that
\begin{eqnarray}
f_I^{(1)} &\longrightarrow& \left(f_J^{(1)}- \tfrac12 {\cal Z}_0^{PQ} \, F_{PQJ} \right) \left({\cal S}_0^{-1}
\right)^J{}_I \;, \nonumber\\
f_{IJ}^{(1)} &\longrightarrow& 
\left({\cal S}_0^{-1}
\right)^Q{}_J \, \partial_Q \left[
\left(f_L^{(1)}- \tfrac12  {\cal Z}_0^{PQ} \, F_{PQL} \right) \left({\cal S}_0^{-1}
\right)^L{}_I \right]\;.
\end{eqnarray}
{\sf Exercise 27: Determine the transformation behavior of $f^{(2)}(t_B)$ under symplectic
transformations \eqref{eq:duality-new-Y} that ensures that 
$F^{(2)}(t_B, \bar t_B)$ transforms as a function.
A useful transformation law is
\begin{align}
F_{IJKL} \longrightarrow &\, \left({\cal S}_0^{-1}
\right)^M{}_I \, \partial_M \left[ F_{NOP} 
\left({\cal S}_0^{-1}
\right)^N{}_J \, \left({\cal S}_0^{-1}
\right)^O{}_K \,\left({\cal S}_0^{-1}
\right)^P{}_L \right] \nonumber\\
&\, 
 =  \left(\mathcal{S}_0^{-1}\right)^M{}_I \left(\mathcal{S}_0^{-1}\right)^N{}_J \left(\mathcal{S}_0^{-1}\right)^O{}_K  \left(\mathcal{S}_0^{-1}\right)^P{}_L 
 \Big[F_{MNOP} \nonumber\\
 & \, \qquad -  F_{MPS} {\cal Z}_0^{SR} F_{RNO} -  F_{OPS} {\cal Z}_0^{SR} F_{RMN} -  F_{NPS} {\cal Z}_0^{SR} F_{ROM}
 \Big] \;.
 \end{align}
 }

\section{The functions $\mathcal{H}^{(a)}_i$ for $a\geq2$    }
\label{sec:funct-H-a-i-geq2}
\setcounter{equation}{0}
Here we collect the explicit results for the various functions
$\mathcal{H}^{(a)}_i$ (with $a \geq 2$) that appear in \eqref{eq:Hesse-decomp}. These
functions can be determined by iteration.  We present the functions 
up to order $\mathcal{O}(\Omega^4)$. We use the notation $(N \Omega)^I = N^{IJ} \Omega_J \,,\, 
(N \bar\Omega)^I = N^{IJ} \Omega_{\bar J}$.   The symmetrization 
$F_{R(IJ} N^{RS} F_{KL)S}$ is defined with a symmetrization factor
$1/(4!)$.
\begin{align}
  \label{eq:H-from-2}
  \mathcal{H}^{(2)}=&\,8\, N^{IJ}\Omega_I \Omega_{\bar J}
  -16\,\big[\Omega_{IJ} 
  (N\bar\Omega)^I (N\Omega)^J +\,\Omega_{I\bar J} (N\bar\Omega)^I
  (N\bar\Omega)^J + {\rm h.c.} \big] \nonumber\\
  &\, - 8 \mathrm{i} \,\big[F_{IJK} (N\bar\Omega)^I (N\Omega)^J
  (N\Omega)^K - {\rm h.c.} \big] \nonumber\\
  &\, + \tfrac{16}3 \mathrm{i} \big[ \left( F_{IJKL} + 3 \mathrm{i}
    F_{IJR} N^{RS} F_{SKL} \right) \, (N\Omega)^I (N\Omega)^J
  (N\Omega)^K (N \bar \Omega)^L - {\rm h.c.} \big]
  \nonumber\\
  &\, + 16\, \big[  \Omega_{IJK} \,  (N\Omega)^I (N\Omega)^J (N \bar
  \Omega)^K + {\rm h.c.} \big] \nonumber\\ 
  & \, + 16\, \big[ \left( \Omega_{IJ\bar K} + \mathrm{i} F_{IJP} N^{PQ}
    \Omega_{Q \bar K} \right) \, \left( (N\Omega)^I (N\Omega)^J (N
    \Omega)^K + 2 (N\Omega)^I (N \bar \Omega)^J (N \bar \Omega)^K
  \right)   + {\rm h.c.}    \big] \nonumber\\
  &\, + 32\, \Big[\Omega_{IQ} \, N^{QR} \, \Omega_{RJ} \, (N \Omega)^I (N
  \bar \Omega)^J + {\rm h.c.} \Big]
  \nonumber\\
  &\, + 32\, \Omega_{IQ} \, N^{QR} \, {\Omega}_{\bar R \bar J} \, (N
  \Omega)^I (N \bar \Omega)^J
  \nonumber\\
  & \, + 16 \mathrm{i} \Big[ F_{IJK} \, N^{KP} \, \Omega_{PQ} \, \left(
    (N \Omega)^I (N \Omega)^J (N \bar \Omega)^Q + 2 (N \Omega)^Q (N
    \Omega)^I (N \bar \Omega)^J \right) - {\rm h.c.} \Big]
  \nonumber\\
  & \, + 16 \mathrm{i} \Big[ F_{IJK} \, N^{KP} \, \Omega_{\bar P \bar
    Q} \, (N \Omega)^I (N \Omega)^J (N \bar \Omega)^Q - {\rm h.c.}
  \Big]
  \nonumber\\
  & \, +8\, (N \Omega) ^I \, (N \Omega)^J \, F_{IJQ} N^{QR} {\bar
    F}_{\bar R \bar K \bar L} (N \bar \Omega)^K (N \bar \Omega)^L \nonumber  \\
   & \, + 32 \Big[(N \Omega)^I \, \Omega_{IJ} \, N^{JK} \Omega_{K \bar L} (N \Omega)^L + {\rm h.c.} \Big] \nonumber\\
      & + 32 \Big[(N \bar \Omega)^I \, \Omega_{IJ} \, N^{JK} \Omega_{K \bar L} (N \bar \Omega)^L + {\rm h.c.} \Big] \nonumber\\
      & + 32 \Big[(N \Omega)^I \, \Omega_{IJ}\,  N^{JK} \Omega_{\bar K L} (N \Omega)^L + {\rm h.c.} \Big] \nonumber\\
      & + 16 \mathrm{i} \Big[ (N\Omega)^I (N\Omega)^J F_{IJK} N^{KL} \Omega_{\bar L P}  (N\Omega)^P - {\rm h.c.} \Big]
      \nonumber\\
            &\, + 32 \Big[(N \Omega)^I  \Omega_{I\bar J} \,  N^{JK} \, 
   \Omega_{\bar K L} 
   \, (N \bar \Omega)^L + {\rm h.c.} \Big] \nonumber\\
 &\, + 32 \Big[(N \Omega)^I  \Omega_{I\bar J} \,  N^{JK} \, 
   \Omega_{K \bar L} 
   \, (N \bar \Omega)^L  \Big]   \;,      
\\[.4ex]
  \mathcal{H}^{(3)}_1=&\,- \tfrac83\mathrm{i} F_{IJK}
  (N\bar\Omega)^I(N\bar\Omega)^J(N\bar\Omega)^K \nonumber\\
  &\, +8 \mathrm{i}\, F_{IJK} (N\bar\Omega)^J (N\bar\Omega)^K \, N^{IP} \left[2
    \Omega_{\bar P \bar Q} (N \bar{\Omega})^Q + 2 \Omega_{\bar P Q} (N
    \Omega)^Q - \mathrm{i} \bar{F}_{\bar P \bar Q \bar R} (N
    \bar{\Omega})^Q (N \bar{\Omega})^R \right]  \,,
   \\[.4ex]
  \mathcal{H}^{(3)}_2=&\, 
  8\, \left(\Omega_{IJ} + \mathrm{i} F_{IJK} (N
    \Omega)^K \right) (N \bar{\Omega})^I (N \bar{\Omega})^J 
      \nonumber\\
     &\, - \tfrac43 \mathrm{i}  \left(F_{IJKL} + 3 \mathrm{i}
       F_{R(IJ} N^{RS} F_{KL)S} \right)    
          \left(
            6 (N \Omega)^I (N \Omega)^J (N \bar \Omega)^K (N \bar
            \Omega)^L
          \right. \nonumber\\
          & \, \left. \qquad \qquad \qquad \qquad \qquad \qquad \qquad
            - 4 (N \bar \Omega)^I (N \bar \Omega)^J (N \bar \Omega)^K
            (N \Omega)^L \right) \Big] \nonumber\\
          &\, - \tfrac{16}3 \, \Omega_{IJK} \left( 3\,(N
              \bar{\Omega})^I (N \bar{\Omega})^J (N \Omega)^K - (N
              \bar{\Omega})^I (N \bar{\Omega})^J (N \bar \Omega)^K
            \right)  \nonumber\\
          &\, - 16\,\Omega_{IJ\bar K} (N \bar{\Omega})^I (N
            \bar{\Omega})^J (N \bar \Omega)^K \nonumber\\
          &\ - 16 \mathrm{i} \,F_{IJK} N^{KP} \,\Omega_{PQ}
          \big[
          - (N \bar \Omega)^I ( N \bar \Omega)^J (N \bar \Omega)^Q
          \nonumber\\ 
          & \, \qquad \qquad \qquad \qquad \qquad + (N \bar \Omega)^I
          ( N \bar \Omega)^J (N \Omega)^Q + 2 (N \bar \Omega)^I ( N
          \Omega)^J (N \bar \Omega)^Q \big]  \nonumber\\
          & \, - 16 \, (N \bar \Omega)^P \, \Omega_{PQ} \, N^{QR}
          \Omega_{RK} \, (N \bar \Omega)^K \nonumber\\
          & \, - 32 \, (N \Omega)^I \, \left( \Omega_{IJ} +
            \mathrm{i} F_{IJP} (N \Omega)^P \right) \, N^{JK} \left(
            \Omega_{\bar K \bar L} - \mathrm{i} {\bar F}_{\bar K
              \bar L \bar M} (N \bar \Omega)^M \right) \, (N \Omega)^L
          \nonumber\\
          & \, + 16 \mathrm{i} (N \Omega)^I (N \Omega)^J
            F_{IJP}N^{PK} \left({\Omega}_{\bar K \bar L} -
              \mathrm{i} {\bar F}_{ \bar K \bar L \bar Q} (N \bar
              \Omega)^{\bar Q} \right) (N \Omega)^L 
          \nonumber\\
          & \, - 16\, (N \Omega)^P \Omega_{\bar P Q} N^{QR}
          \Omega_{R \bar K} (N \Omega)^K  \nonumber\\
          & \, - 32 \, (N \bar \Omega)^I \left( \Omega_{IJ} + \mathrm{i} F_{IJK} (N \Omega)^K  \right) N^{JL} 
            \Omega_{\bar L M} (N \Omega)^M \nonumber\\
            & - 16 \mathrm{i} \,( N \bar \Omega)^I \, (N \bar \Omega)^J F_{IJK} N^{KP} \Omega_{P \bar Q} (N \bar 
                      \Omega)^Q  \;,
                         \\[.4ex]
  \mathcal{H}^{(3)}_3=&\,16\,\Omega_{I\bar J} (N\bar\Omega)^I (N
  \Omega)^J  \nonumber\\ 
   &\,- 16\,\Big[ 2 (N \bar{\Omega})^K (N \Omega)^L \left(\Omega_{KM} 
     N^{MN} \Omega_{N \bar L}  + \Omega_{K \bar L Q} (N \Omega)^Q
   \right) \nonumber\\ 
  &\, \qquad + (N \bar{\Omega})^K \Omega_{K \bar L} N^{LP}
  \left( \mathrm{i} F_{PMN} (N \Omega)^M (N \Omega)^N + 2 \Omega_{PJ}
    (N \Omega)^J + 2 \Omega_{P \bar J} (N \bar{\Omega})^J \right)
  \nonumber\\
   & \qquad + 2 \mathrm{i} (N \bar \Omega)^I (N \Omega)^J F_{IJK} N^{KP} \Omega_{P \bar Q} (N \Omega)^Q \nonumber\\
  & \, \qquad + {\rm h.c.} \Big]\,,  \\[.4ex]
  \mathcal{H}^{(4)}_1=&\, 32 \, (N \bar \Omega)^I \left(
    \Omega_{IJ} + \mathrm{i} {F}_{IJK} (N \Omega)^K \right)
  N^{JP} \left( \Omega_{\bar P \bar Q} - \mathrm{i} {\bar
      F}_{\bar P \bar Q \bar R} (N \bar \Omega)^R
  \right) (N \Omega)^Q \,,  \\[.4ex]
  \mathcal{H}^{(4)}_2=&\, 32 \, (N \Omega)^P \,
  \Omega_{\bar{P} Q} \, N^{QR} \, \Omega_{\bar{R} K}
  \, (N \bar \Omega)^K  \\[.4ex]
  \mathcal{H}^{(4)}_3=&\, 8 \, F_{IJR} N^{RS} {\bar F}_{\bar S
    \bar K \bar L} \, (N \bar \Omega)^I (N \bar \Omega)^J (N
  \Omega)^K (N \Omega)^L \,, \\[.4ex]
  \mathcal{H}^{(4)}_4=&\, - \tfrac43 \mathrm{i} \left(
    F_{IJKL} + 3 \mathrm{i} \, F_{IJR} N^{RS} F_{SKL} \right)
  (N \bar \Omega)^I (N \bar \Omega)^J (N \bar \Omega)^K (N
  \bar \Omega)^L \,,  \\[.4ex]
  \mathcal{H}^{(4)}_5=& \, - 16 \mathrm{i} \, F_{IJK} N^{KL} \Omega_{\bar L Q} \, 
   (N   \bar \Omega)^Q (N \bar \Omega)^I (N \bar \Omega)^J \,,
  \\[.4ex] 
  \mathcal{H}^{(4)}_6=&\, - 16 \mathrm{i}\, F_{IJK} N^{KP}
  \left( \Omega_{\bar P \bar Q} - \mathrm{i} {\bar
      F}_{\bar P \bar Q \bar R} (N \bar \Omega)^R \right)
  (N \bar \Omega)^I (N \bar \Omega)^J (N  \Omega)^Q \,,
  \\[.4ex] 
  \mathcal{H}^{(4)}_7=&\, 16 \,\left( \Omega_{IJ\bar K} +
    \mathrm{i} F_{IJP} \, N^{PQ} \, \Omega_{Q \bar K}\right)
  \, (N \bar \Omega)^I (N \bar \Omega)^J (N \Omega)^K \,,
  \\[.4ex] 
    \mathcal{H}^{(4)}_8 =&\,
32 \, (N \bar \Omega)^I 
\left( \Omega_{IJ} + \mathrm{i} {F}_{IJK}  (N \Omega)^K \right) N^{JP} 
 \Omega_{\bar P  Q} \,  (N \bar \Omega)^Q \,,
  \\[.4ex] 
   \mathcal{H}^{(4)}_9  =&\,
 - 16 \mathrm{i} \, (N  \bar \Omega)^I  (N  \bar \Omega)^J F_{IJK} N^{KL} \Omega_{\bar L P} (N  \bar \Omega)^P
\;.
\end{align}

\section{Transformation laws by iteration }
\label{sec:transf-der-om}
\setcounter{equation}{0}

The Hesse potential in section \ref{sec:hesse-top} depends on $\Omega$, whose 
behavior under symplectic transformations can be determined by iteration.  
Here we summarize the result for the 
transformation behavior of derivatives of $\Omega$
(expressed in terms of the covariant variables of section \ref{sec:new-var}),
up to a certain order.
  We use the conventions of section 
 \ref{sec:hesse-top} and suppress the superscript of $F^{(0)}$.
 \begin{align}
  \label{eq:transf-Omega3}
    \tilde\Omega_I=&\, [\mathcal{S}_0^{-1}]^J{}_I \Big[ \Omega_J
  +\mathrm{i} F_{JKL} \,\left( \mathcal{Z}_0 \Omega\right)^K\,(\mathcal{Z}_0
  \Omega)^L -2\mathrm{i} \Omega_{JK} (\mathcal{Z}_0 \Omega)^K
  +2\mathrm{i} \Omega_{J\bar K} (\bar{\mathcal{Z}}_0
  \bar{\Omega})^{\bar K} \nonumber\\
  &\,+ \tfrac23 F_{JKLP} (\mathcal{Z}_0 \Omega)^K  (\mathcal{Z}_0 \Omega)^L  (\mathcal{Z}_0 \Omega)^P
  + 2 F_{KLP} (\mathcal{Z}_0 \Omega)^K{}_J  (\mathcal{Z}_0 \Omega)^L  (\mathcal{Z}_0 \Omega)^P \nonumber\\
  &\, + 4 F_{JKL} (\mathcal{Z}_0 \Omega)^K (\mathcal{Z}_0 \Omega)^L{}_{P}  (\mathcal{Z}_0 
  {\Omega})^{P}
   - 4 F_{JKL} (\mathcal{Z}_0 \Omega)^K (\mathcal{Z}_0 \Omega)^L{}_{\bar P}  (\mathcal{\bar Z}_0 
  {\bar \Omega})^{\bar P} \nonumber\\
 & \, - 2 F_{JKL} \mathcal{Z}_0^{LP} F_{PQS} (\mathcal{Z}_0 \Omega)^K  (\mathcal{Z}_0 \Omega)^Q
 (\mathcal{Z}_0 \Omega)^S 
 + 2 {\bar F}_{\bar K \bar L \bar P} (\bar{\mathcal{Z}}_0 \bar \Omega)^{\bar K}{}_J
 (\bar{\mathcal{Z}}_0 \bar \Omega)^{\bar L} (\bar{\mathcal{Z}}_0 \bar \Omega)^{\bar P} \nonumber\\
 &\, - 2 \Omega_{JKL} (\mathcal{Z}_0 \Omega)^K (\mathcal{Z}_0 \Omega)^L- 4 \Omega_{KL} 
 (\mathcal{Z}_0 \Omega)^K{}_J (\mathcal{Z}_0 \Omega)^L - 2 \Omega_{J \bar K \bar L} 
 (\bar{\mathcal{Z}}_0 \bar \Omega)^{\bar K} (\bar{\mathcal{Z}}_0 \bar \Omega)^{\bar L} \nonumber\\
 &\,  - 4 \Omega_{\bar K \bar L}  (\bar{\mathcal{Z}}_0 \bar \Omega)^{\bar K}{}_J 
 (\bar{\mathcal{Z}}_0 \bar \Omega)^{\bar L} + 4 \Omega_{JK \bar L} (\mathcal{Z}_0 \Omega)^K
 (\bar{\mathcal{Z}_0 } \bar{\Omega})^{\bar L} + 4 \Omega_{K \bar L} (\mathcal{Z}_0 \Omega)^K{}_J 
  (\bar{\mathcal{Z}_0 } \bar{\Omega})^{\bar L} \nonumber\\
 & \,  + 4 \Omega_{K \bar L} (\mathcal{Z}_0 \Omega)^K  (\bar{\mathcal{Z}_0 } \bar{\Omega})^{\bar L}{}_J \Big]
    +\mathcal{O}(\Omega^4)\,,\nonumber \\
  \tilde\Omega_{IJ}=&\, [\mathcal{S}_0^{-1}]^K{}_I
  [\mathcal{S}_0^{-1}]^L{}_J  \Big[ \Omega_{KL} -F_{KLM}
  \,\mathcal{Z}_0^{MN} \Omega_N  \nonumber\\
  & \,  - \mathrm{i} F_{KLP} {\cal Z}_0^{PM} F_{MQR} ({\cal Z}_0 \Omega)^{Q} ({\cal Z}_0 \Omega)^R 
  + 2 \mathrm{i} F_{KLP} ({\cal Z}_0 \Omega)^{P}{}_{Q} ({\cal Z}_0 \Omega)^Q
   - 2 \mathrm{i} F_{KLP} ({\cal Z}_0 \Omega)^{P}{}_{\bar Q} (\bar{\cal Z}_0 {\bar \Omega})^{\bar Q}
  \nonumber\\
   & \,  + \mathrm{i} F_{KLMN} (\mathcal{Z}_0 \Omega)^M (\mathcal{Z}_0 \Omega)^N \nonumber\\
  & \, +  \mathrm{i} F_{KMN}  (\mathcal{Z}_0 \Omega)^M{}_L  (\mathcal{Z}_0 \Omega)^N
  + \mathrm{i} F_{KMN}  (\mathcal{Z}_0 \Omega)^N{}_L  (\mathcal{Z}_0 \Omega)^M \nonumber\\
  & \, 
   - 2 \mathrm{i} F_{KMN} \mathcal{Z}_0^{MP} F_{PQL}  (\mathcal{Z}_0 \Omega)^Q  (\mathcal{Z}_0 \Omega)^N 
   \nonumber\\
   & - 2 \mathrm{i} \Omega_{KLP} (\mathcal{Z}_0 \Omega)^P -  2 \mathrm{i} \Omega_{KP} (\mathcal{Z}_0 \Omega)^P{}_L
   + 2 \mathrm{i} \Omega_{KP} \mathcal{Z}_0^{PQ} F_{QLS}   (\mathcal{Z}_0 \Omega)^S \nonumber\\
   & \,+  2 \mathrm{i} \Omega_{KL \bar P} (\bar{\mathcal{Z}}_0 \bar{\Omega})^{\bar P}
   +  2 \mathrm{i} \Omega_{K  \bar P} (\bar{\mathcal{Z}}_0 \bar{\Omega})^{\bar P}{}_L
  \Big] +\mathcal{O}(\Omega^3)
  \,,\nonumber\\ 
  \tilde\Omega_{I\bar J}=&\, [\mathcal{S}_0^{-1}]^K{}_I
  [\bar{\mathcal{S}}_0^{-1}]^{\bar L}{}_{\bar J} \,\Big[ \Omega_{K\bar
    L}  + 2 \mathrm{i} F_{KMN} (\mathcal{Z}_0 \Omega)^M{}_{\bar L} (\mathcal{Z}_0 \Omega)^N
    - 2 \mathrm{i} {\bar F}_{\bar L \bar P \bar N} (\bar{\mathcal{Z}}_0 \bar{\Omega})^{\bar N}{}_{K} 
    (\bar{ \mathcal{Z}}_0 \bar \Omega)^{\bar P} \nonumber\\
     & \, - 2 \mathrm{i} \Omega_{KM{\bar L}} (\mathcal{Z}_0 \Omega)^M 
    - 2 \mathrm{i} \Omega_{KM} (\mathcal{Z}_0 \Omega)^M{}_{\bar L} + 
    2 \mathrm{i} \Omega_{K \bar L \bar M}  (\bar{\mathcal{Z}}_0 \bar \Omega)^{\bar M}
    + 
    2 \mathrm{i} \Omega_{K \bar M }  (\bar{\mathcal{Z}}_0 \bar \Omega)^{\bar M}{}_{\bar L} 
    \Big]
    \nonumber\\
    & \, +\mathcal{O}(\Omega^3) \,, \nonumber\\
     \tilde\Omega_{IJL}=&\, [\mathcal{S}_0^{-1}]^M{}_I [\mathcal{S}_0^{-1}]^N{}_J [\mathcal{S}_0^{-1}]^K{}_L
     \left[ \Omega_{MNK} - F_{MNKP} ({\cal Z}_0 \Omega)^P \right. \nonumber\\
      & \, \left. 
     - F_{MNP} ({\cal Z}_0 \Omega)^P{}_K 
     - F_{KMP} ({\cal Z}_0 \Omega)^P{}_N - F_{NKP} ({\cal Z}_0 \Omega)^P{}_M 
     \right.
     \nonumber\\
     & \, \left. + 
     F_{MNP} {\cal Z}_0^{PQ} F_{KQR} ({\cal Z}_0 \Omega)^R + 
     F_{KMP} {\cal Z}_0^{PQ}    F_{QNR} ({\cal Z}_0 \Omega)^R + 
       F_{NKP}  {\cal Z}_0^{PQ}
       F_{QMR} ({\cal Z}_0 \Omega)^R \right] \nonumber\\ 
      & \, +\mathcal{O}(\Omega^2) \,, \nonumber\\
      \tilde\Omega_{IJ \bar K}=&\, [\mathcal{S}_0^{-1}]^M{}_I [\mathcal{S}_0^{-1}]^N{}_J [\bar{\mathcal{S}}_0^{-1}]^{\bar L}{}_{\bar K} \left[ \Omega_{M N \bar L} - F_{MNQ} ({\cal Z}_0 \Omega)^Q{}_{\bar L}
      \right] +\mathcal{O}(\Omega^2) \,,       
\end{align}
where $(\mathcal{Z}_0 \Omega)^M = \mathcal{Z}_0^{MN}  \Omega_N $, 
$(\bar{\mathcal{Z}}_0 \bar{\Omega})^{\bar M} = \bar{\mathcal{Z}}_0^{\bar M \bar N} {\Omega}_{\bar N} \,,\,
 (\mathcal{Z}_0 \Omega)^M{}_{\bar L} = \mathcal{Z}_0^{MN}  \Omega_{N \bar L} \,,\,
(\bar{\mathcal{Z}}_0 \bar{\Omega})^{\bar P}{}_L = \bar{\mathcal{Z}}_0^{\bar P \bar N} 
{\Omega}_{\bar N L}$, $ ({\mathcal{Z}}_0 \Omega)^L{}_{\bar P} = {\mathcal{Z}}_0^{LK} \Omega_{K {\bar P}}$, 
$(\bar{\mathcal{Z}}_0 \bar{\Omega})^{\bar P}{}_{\bar L} = \bar{\mathcal{Z}}_0^{\bar P \bar N} 
{\Omega}_{\bar N \bar L}$.

\providecommand{\href}[2]{#2}\begingroup\raggedright\endgroup


\end{document}